\documentclass[letter]{aa}  

\usepackage{graphicx}
\usepackage{media9}
\everymath={\rm}
\usepackage{txfonts}
\usepackage{natbib}
\usepackage{footmisc}
\bibpunct{(}{)}{,}{a}{}{,}

\newcommand{\kmps}{km\,s$^{-1}$}
\newcommand{\kmpsb}{km\,s$^{-1}$ }

\newcommand{\odho}{$^{16}$O$^{18}$O}

\newcommand{\dzcob}{$^{12}$CO }

\newcommand{\htp}{H$_3^+$}
\newcommand{\htpb}{H$_3^+$ }

\newcommand{\cc}{cm$^{-3}$}

\newcommand{\jkk}{\emph{J}$_\mathrm{KK\arcmin}$}

\newcommand{\jyb}{Jy\,beam$^{-1}$}

\newcommand{\pdix}[1]{$\times$~10$^{#1}$}
\newcommand{\pdixb}[1]{$\times$~10$^{#1}$ }

\newcommand{\paperIt}{\citetalias{Pagani:2017fx}}
\newcommand{\paperIp}{\citepalias{Pagani:2017fx}}

\begin{document} 

    \title{The complexity of \object{Orion}: an ALMA view\thanks{This paper makes use of the following ALMA data: ADS/JAO.ALMA\#2013.1.00533.S. ALMA is a partnership of 
    ESO (representing its member states), NSF (USA) and NINS (Japan), together with NRC (Canada), NSC and ASIAA (Taiwan), and KASI (Republic of Korea), in cooperation with 
    the Republic of Chile. The Joint ALMA Observatory is operated by ESO, AUI/NRAO and NAOJ.}}
 
    \subtitle{III. The explosion impact}
 
   \author{L. Pagani           \inst{1}
          \and
                E. Bergin \inst{2}
                \and
                P. F. Goldsmith \inst{3}
                \and
                G. Melnick \inst{4} 
          \and
                R. Snell \inst{5}
                \and
            C. Favre\inst{6}
         }

   \offprints{L.Pagani}

\institute{ LERMA \& UMR8112 du CNRS, Observatoire de Paris, PSL  University, Sorbonne Universit\'es, CNRS, F- 75014 Paris, France\\
\email{laurent.pagani@obspm.fr}
\and
Department of Astronomy,
       University of Michigan,
       311 West Hall, 1085 S. University Ave,
       Ann Arbor, MI 48109, USA
\and
JPL, Pasadena, California, USA
\and
Harvard-Smithsonian Center for Astrophysics, Cambridge, Massachusetts, USA
\and
Department of Astronomy, University of Massachusetts, Amherst, MA, 01003, USA
\and
Univ. Grenoble Alpes, CNRS, IPAG, F-38000 Grenoble, France
  }

   \date{Received 14/02/2019; accepted 16/03/2019}

 
 \abstract{The chemistry of complex organic molecules in interstellar dark clouds is still highly uncertain in part because of the
 lack of constraining observations. Orion is the closest massive star-forming region, and observations making use of ALMA allow us to
 separate the emission regions of various complex organic molecules (COMs) in both velocity and space. Orion also benefits from an
 exceptional situation, in that it is the site of a powerful explosive event that occurred $\sim$550 years ago. We show that the
 closely surrounding Kleinmann-Low region has clearly been influenced by this explosion; some molecular species have been pushed away
 from the densest parts while others have remained in close proximity. This dynamical segregation reveals the time dependence of
 the chemistry and, therefore  allows us to better constrain  the formation sequence of COMs and other species, including deuterated molecules.
 }


\titlerunning{The explosion impact}
\maketitle
%
\defcitealias{Pagani:2017fx}{Paper I}

\section{Introduction} 
  Though Orion is a well-studied region and has been explored with a wide variety of instruments, including the NOEMA (former
  Plateau de Bure) Interferometer, the Berkeley-Illinois-Millimeter-Array (BIMA), the Combined Array for Research in
  Millimeter-wave Astronomy (CARMA), and the Submillimeter Array (SMA), the arrival of the Atacama Large Millimeter Array (ALMA)
  holds the promise of new discoveries thanks to its higher angular resolution and sensitivity. The ALMA instrument provides high-velocity
  resolution while maintaining a high dynamic range that enables the detection of previously unseen structures, leading to a
  better understanding of the source structure and evolution.
  
  Based on the search for the lowest  observable rotational transition of \odho\ at 234 GHz, we performed deep observations of
  this source during Cycle 2 with 37--39 antennas, surveying 16 GHz in ALMA band 6, and improving the sensitivity by a factor
  $\sim$5 compared to the Cycle 0 Science Verification  (SV0) observations for the frequencies in common (\citealt{Pagani:2017fx},
  hereafter Paper I). \paperIt\ presents a more detailed history of the recent work on Orion. In \paperIt, we present the data and first results including the detection of several species not previously seen in
  Orion (n-- and i--propyl cyanide, C$_3$H$_7$CN,...) as well as several vibrationally excited levels of cyanoacetylene (HC$_3$N)
  and of its $^{13}$C isotopologues. A companion paper \citep{Favre:2017ch} presents the first detection  of gGg$'$ ethylene
  glycol (gGg$'$ (CH$_2$OH)$_2$) and of acetic acid (CH$_3$COOH\footnote{Though not explicitly announced  nor discussed in their
  paper, acetic acid was already identified in \citet{Cernicharo:2016fm}.}) in Orion.
  
  One remarkable feature present in the central region of Orion is an explosive event that occurred {550 $\pm$ 25} years
  ago {(J.Bally, priv. communication)} and was revealed by  the three runaway stars BN, n, and I
  \citep{2005ApJ...635.1166G,Rodriguez:2005fy,Rodriguez:2017bi}, and by the CO and H$_2$ fingers
  \citep[e.g.,][]{1993Natur.363...54A,Zapata:2009ho,2012A&A...540A.119N, Youngblood:2016jr,Bally:2017cy}.  {However,
  \citet{Luhman:2017fe} showed that the object n is no longer a runaway member because its real proper motion is much lower than
  previously estimated; but conversely another object, named x, is moving away at high speed from the same explosion center. J.
  Bally  (priv. comm.) confirms the fast proper motion of x and the absence of movement of n from ground-based H$_2$ images 14
  years apart. X is further out having passed our 20 \% beam coupling mark (see Fig. 1 of Paper I), and therefore does not appear in
  the figures presented in this Letter.} \citet{Zapata:2011ez} and \citet{OrozcoAguilera:2017cg} in their follow-up work with ALMA
  proposed that the hot core (HC) is externally heated despite its high temperature, and that the heating source could be the
  nearby explosion. Similarly, \citet{Blake1987}, \citet{Wang:2011fr}, and \citet{Favre:2011ha} advocated that the Compact Ridge is
  also externally heated, although the heating source should not be the same since we presented evidence in \paperIt\ that the
  Compact Ridge has not yet been affected by the impact of the explosion. A possibility could be the outflow from source I hitting
  the Compact Ridge \citep{2002ApJ...576..255L}. In \paperIt, we also presented  evidence for an interaction between the explosive
  event and the main components of the \object{Orion KL} region including the HC, several infrared (IR) components
  \citep{Rieke1973}, and methyl formate (CH$_3$OCHO; hereafter MF) peaks \citep{Favre:2011ha}. We showed that the IRc6/MF5 and
  IRc20/MF4 sources, west of the explosion center,  display  emission lines of various species having only red wings, while
  sources on the east and south sides display emission lines having only blue wings. We also confirmed that excited emission lines
  are found preferentially surrounding the explosion center and  that complex organic molecules (COMs) rich in oxygen (O-COMs) do
  not occupy the same volumes as CN rich COMs (CN-COMs). We identified the ethylene glycol peak (EGP) to be coincident with a hollow
  sphere of material, which we interpreted to have originated from the impact of  a ``bullet''  launched from the explosion center
  \citep{Favre:2017ch,Wright:2017jk}. We also proposed that the Compact Ridge (MF1) is sufficiently far away from the rest of the
  KL region to have not yet been perturbed by the explosion, the evidence being the absence of asymmetric emission line wings and
  the narrowness of the lines themselves ($\sim$1 \kmps).
In this Letter, we study further the  interaction of the explosion blowout with the surrounding gas and dense sources. 

\section{Observations} The observations have been described in detail in \paperIt, and we give here only a short summary.
Observations were carried out in December 2014, during the Cycle 2 period, with 37 antennas, and 39 antennas for the last run. Sixteen
GHz of discontinuous bandwidth were covered in the frequency range 215 to 252 GHz. One band, centered on \odho\ at 234 GHz, was
observed all the time (2.5 hours), reaching a sensitivity of $\sim$2 mJy/beam while the other bands were observed only in one
setup each (30 minutes duration), with 488 kHz {channel spacing }and reaching a typical noise of 5 mJy/beam (see Table 1 in \paperIt). As in
\paperIt, the part of the data discussed in this work are centered on  ra$_{ J2000}$:  05$^h$35$^m$14.160$^s$, dec$_{ J2000}$:
-05$\degr$22$'$31.504$''$. Since we do not yet have total power observations to include to the data set, this paper concentrates on the kinematics and spatial positioning of the sources rather than on discussions of column density.

\section{Channel maps}

\begin{figure*}[h!]
\centering 
\includegraphics[scale = 0.245,trim=0 20 49 0,clip]{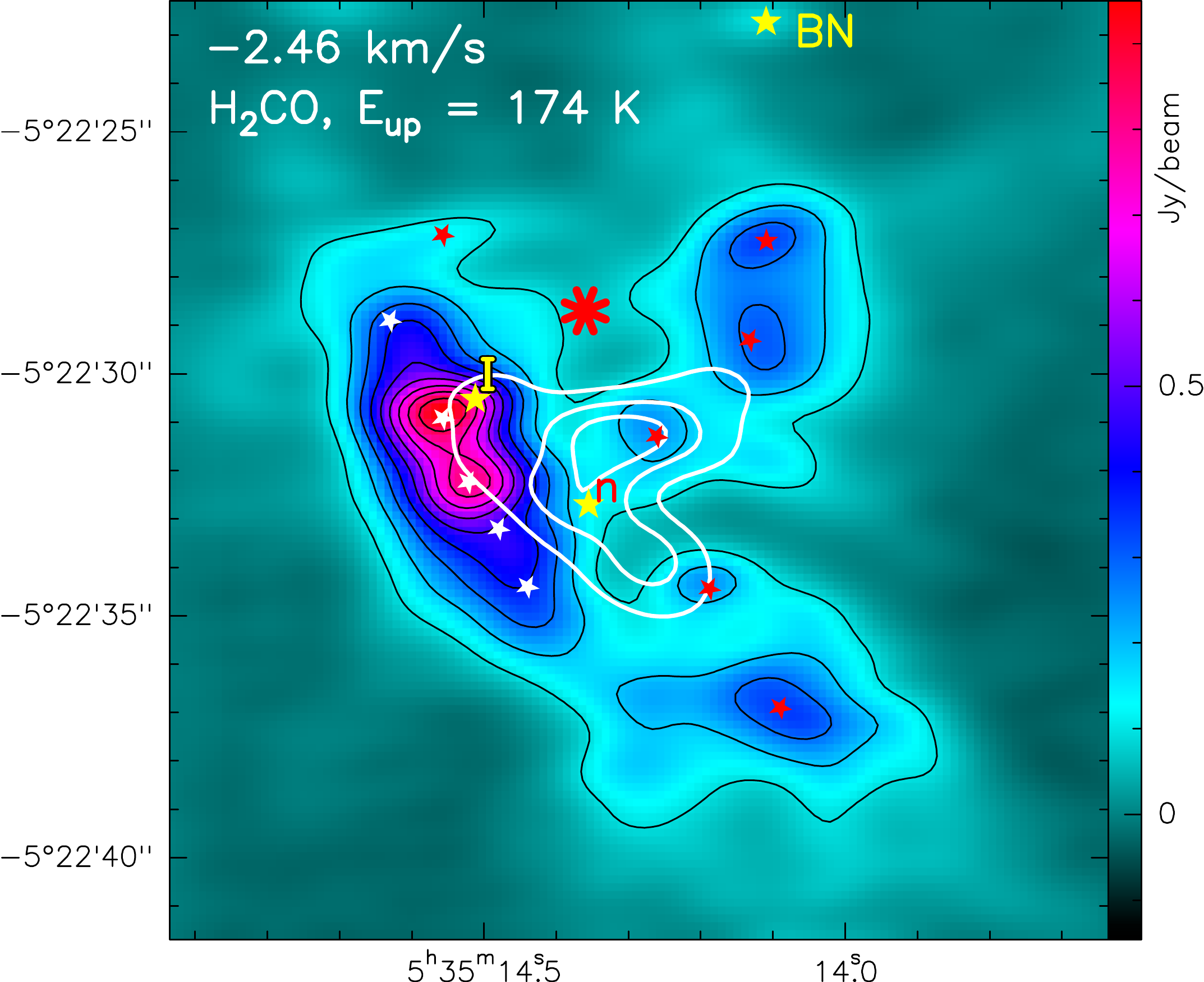}
\includegraphics[scale = 0.245, trim=85 20 48 0,clip]{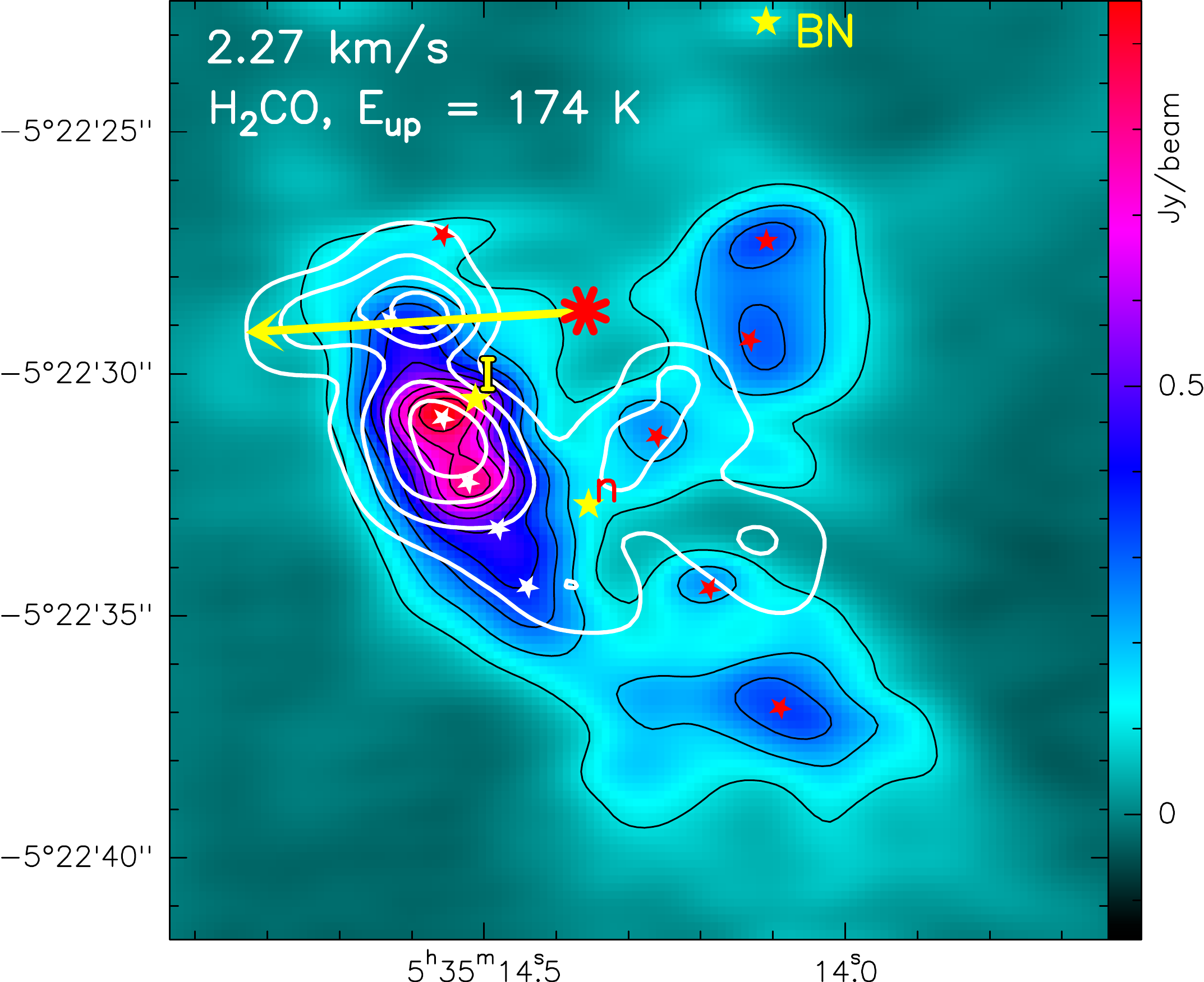} 
\includegraphics[scale = 0.245, trim=85 20 48 0,clip]{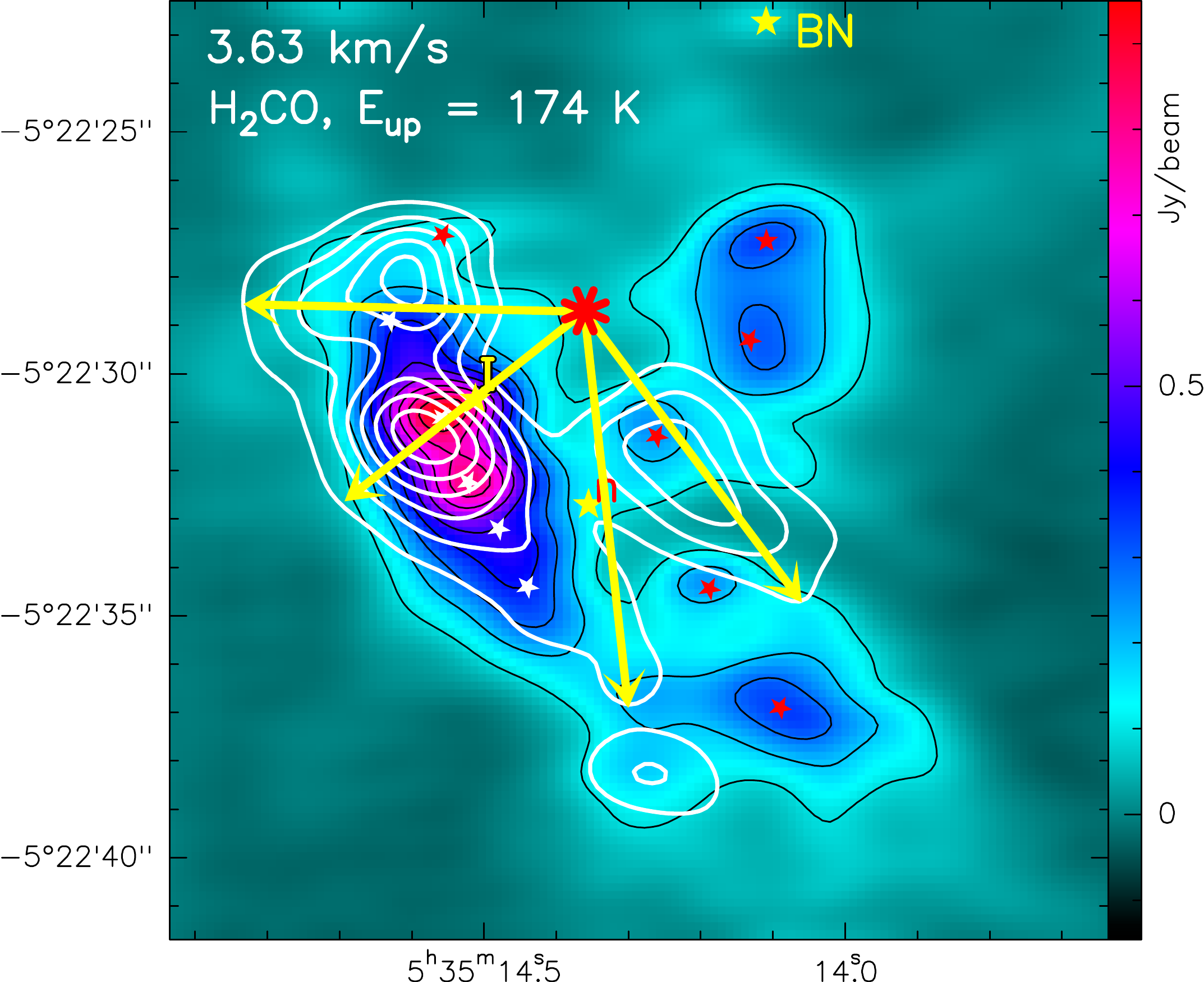} 
\includegraphics[scale = 0.245, trim=85 20 0 0,clip]{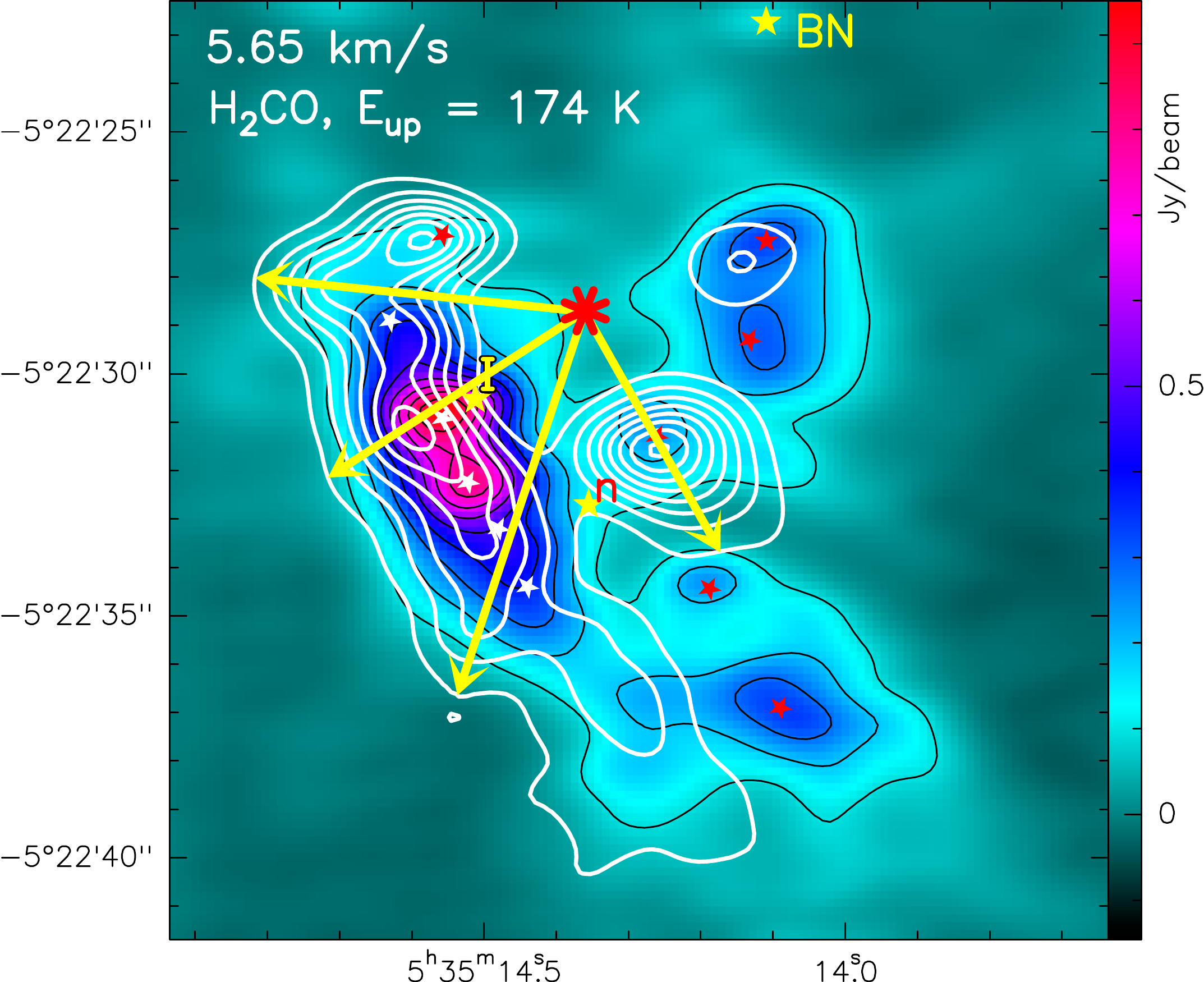}
\includegraphics[scale = 0.245, trim=0 0 49 0,clip]{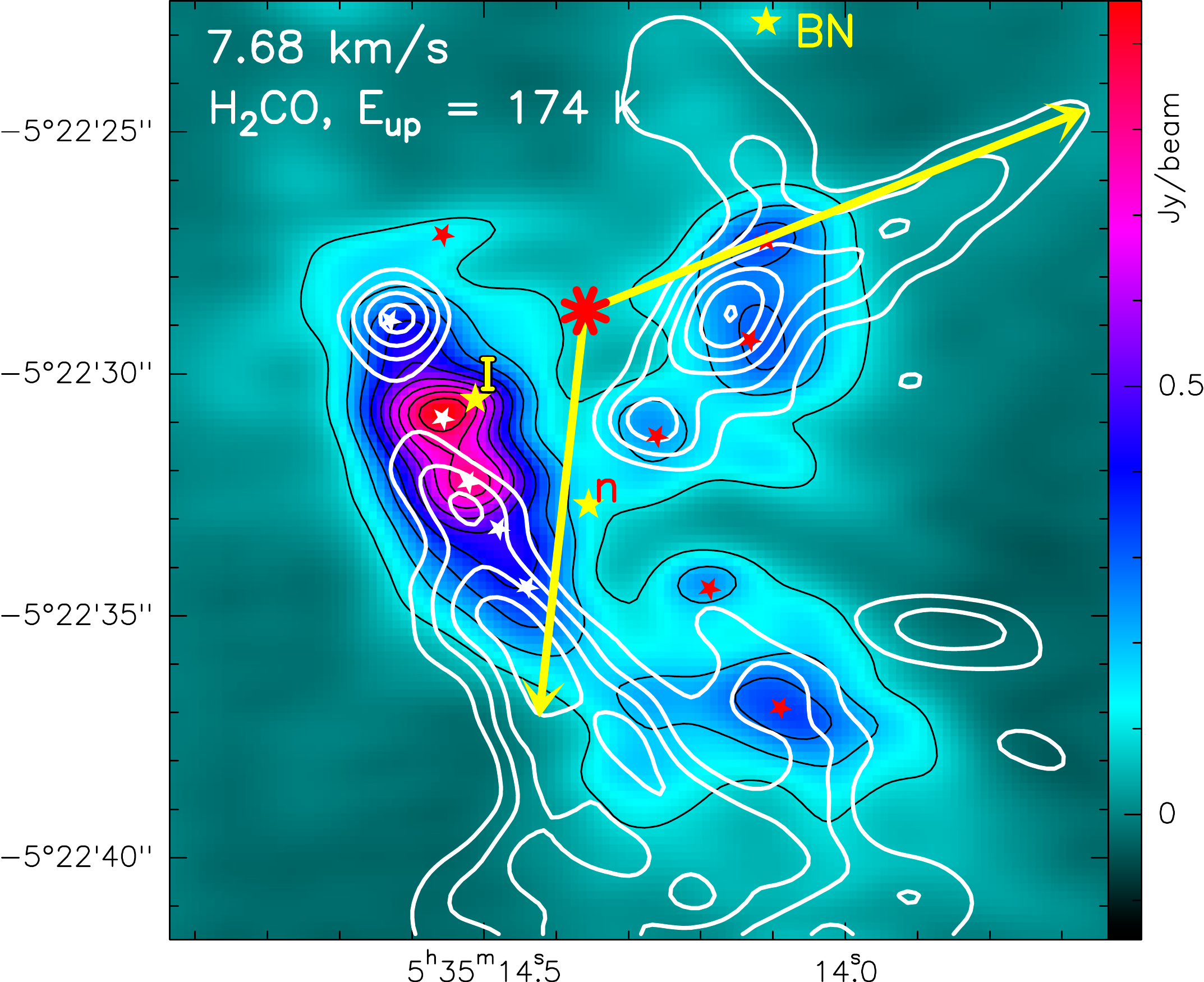} 
\includegraphics[scale = 0.245, trim=85 0 48 0,clip]{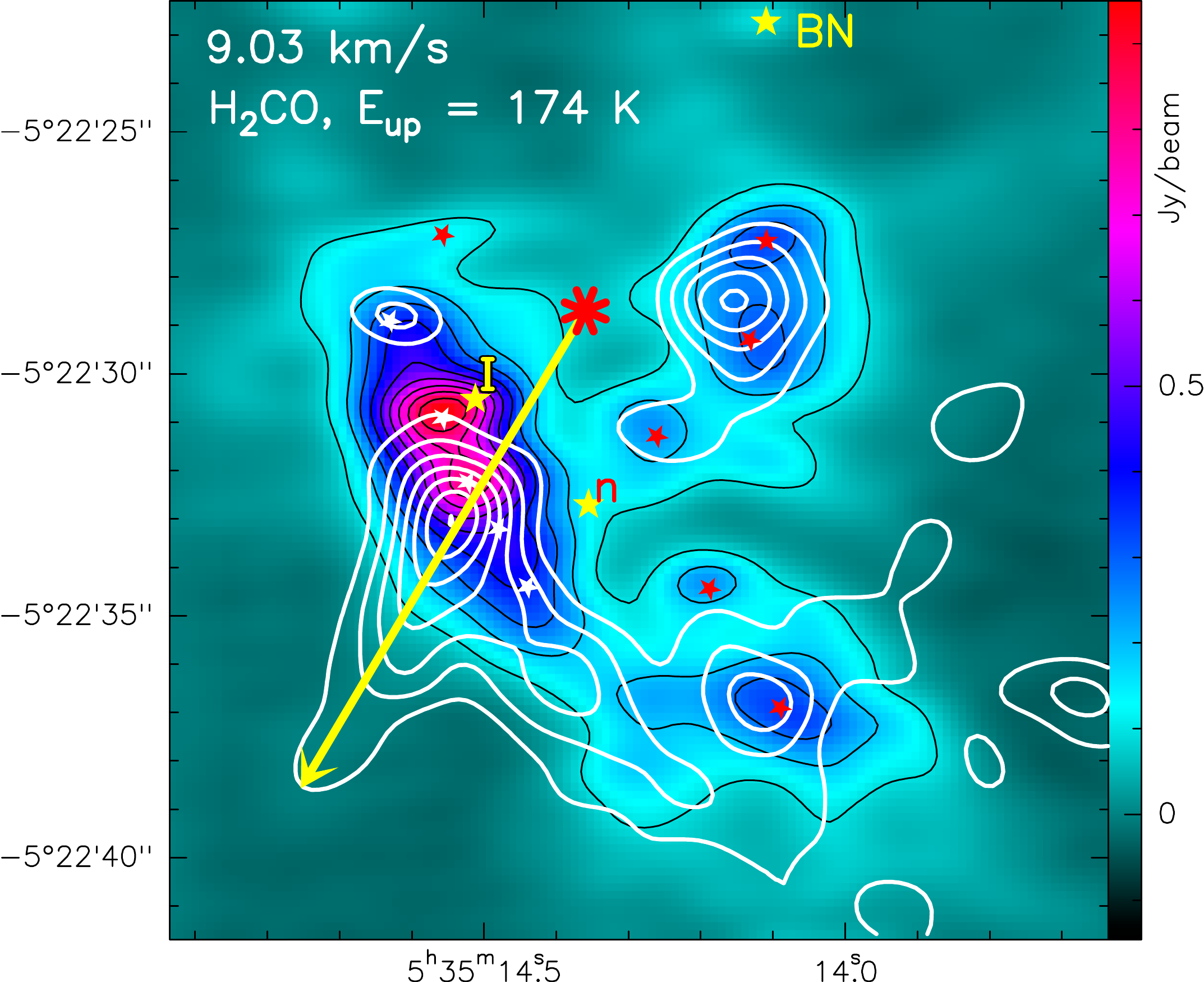} 
\includegraphics[scale = 0.245, trim=85 0 48 0,clip]{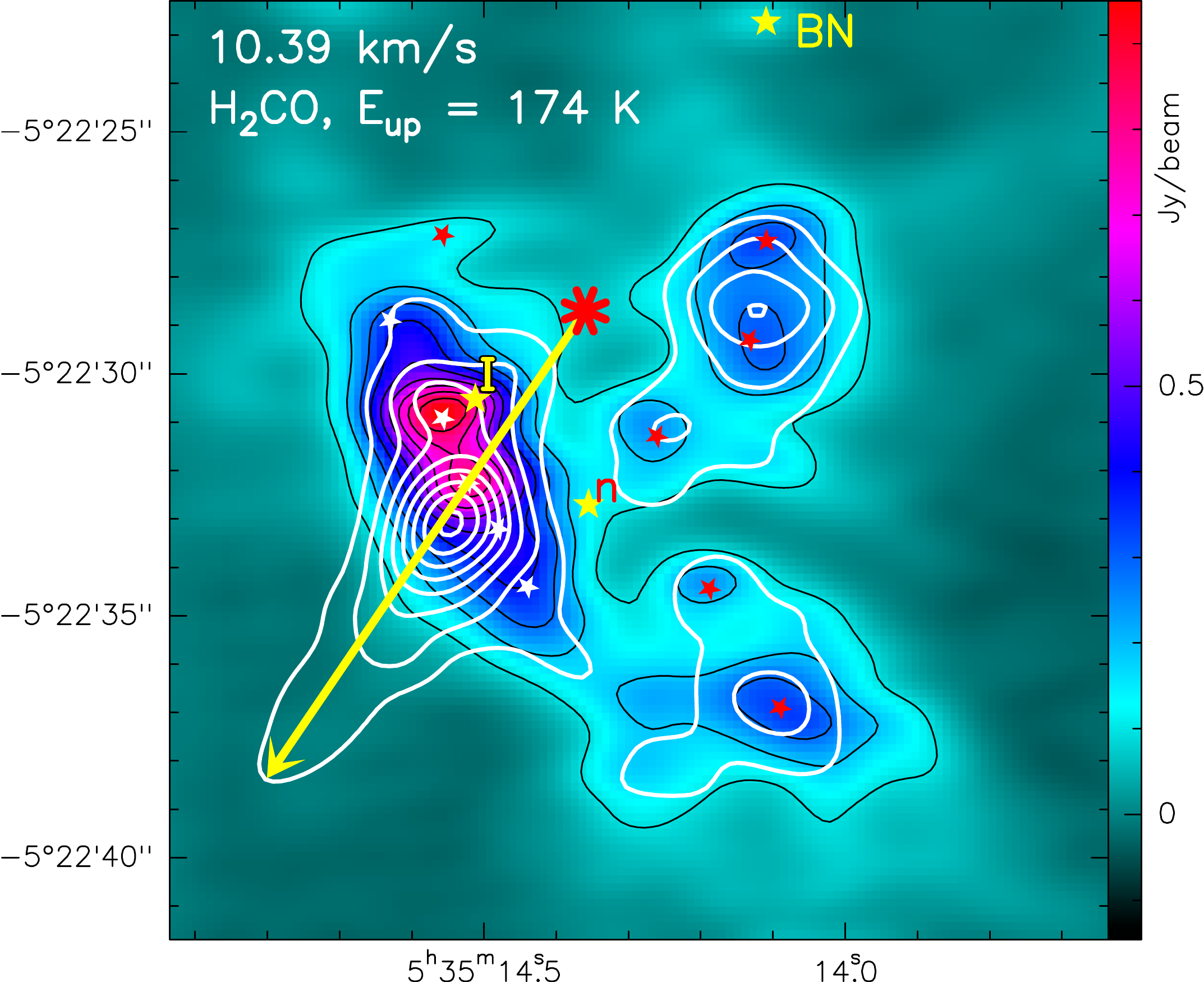}
\includegraphics[scale = 0.245, trim=85 0 0 0,clip]{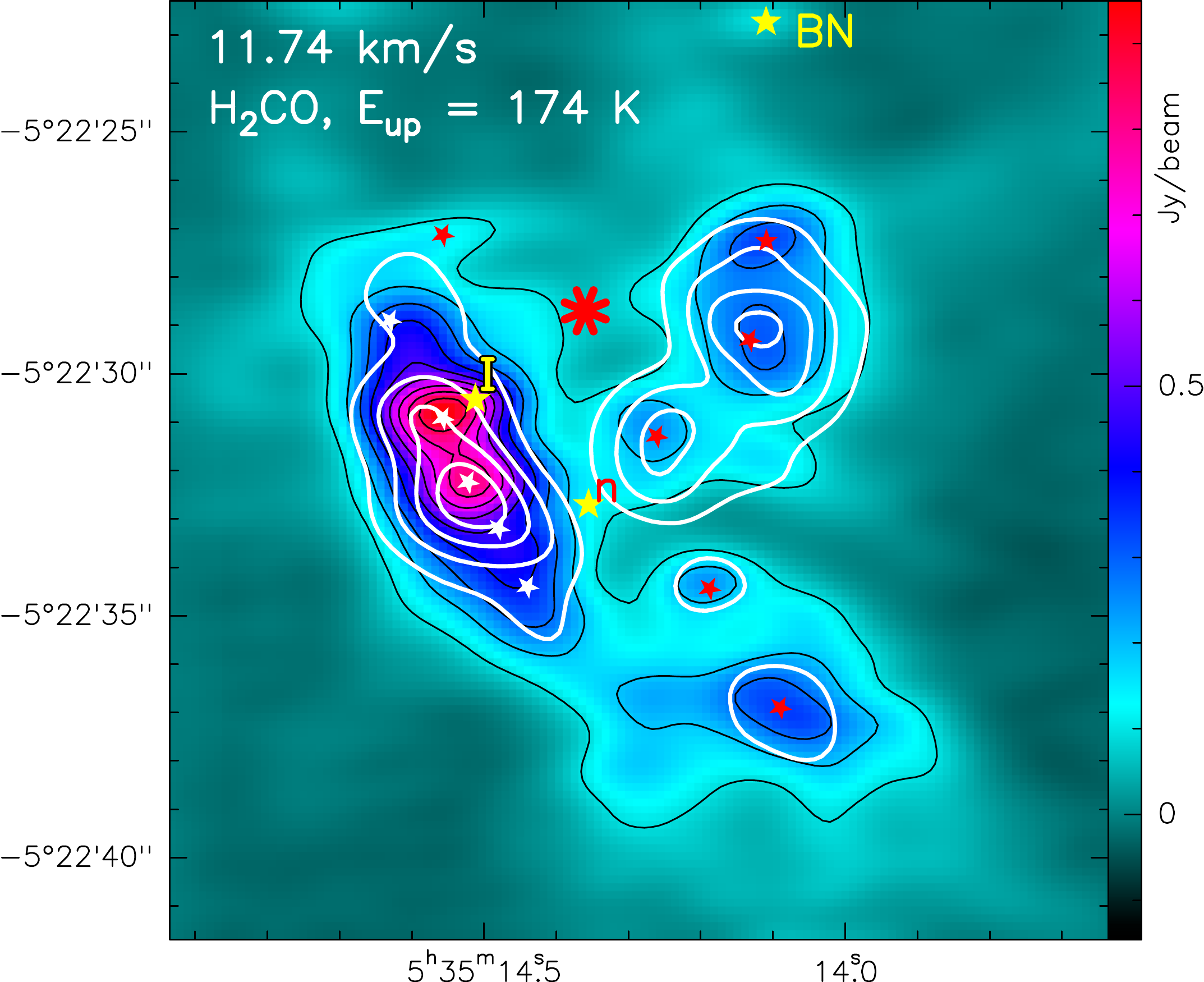} 
\caption{Selection of  formaldehyde  (H$_2$CO, \jkk:9$_{18}$--9$_{19}$, at
216568.651 MHz) 
channel maps {($\delta$V = 0.68\,\kmps)}.  The top left corner indicates the channel velocity, the species, and its upper energy level. White contours
are at 10 to 90\,\% of the peak emission of the strongest channel for that species. Yellow arrows starting from the explosion center 
(red eight-pointed star) suggest possible displacement of gas linked to the explosive event, which occurred {550\,$\pm$\,25} years ago.
Five-pointed stars denote the position of the 10 sources studied in \paperIt\ ( Fig. 1, see also Fig.\,\ref{fig:bnklcontinuummapcontours} here). 
Underlying color map {(enhanced by  black contours)} in this and subsequent figures is the 1.2 mm continuum emission \paperIp.}
\label{fig:H2COchannelmaps}
\end{figure*}

\begin{figure*}[h!]
\centering 
\includegraphics[scale = 0.245,trim=0 20 49 0,clip]{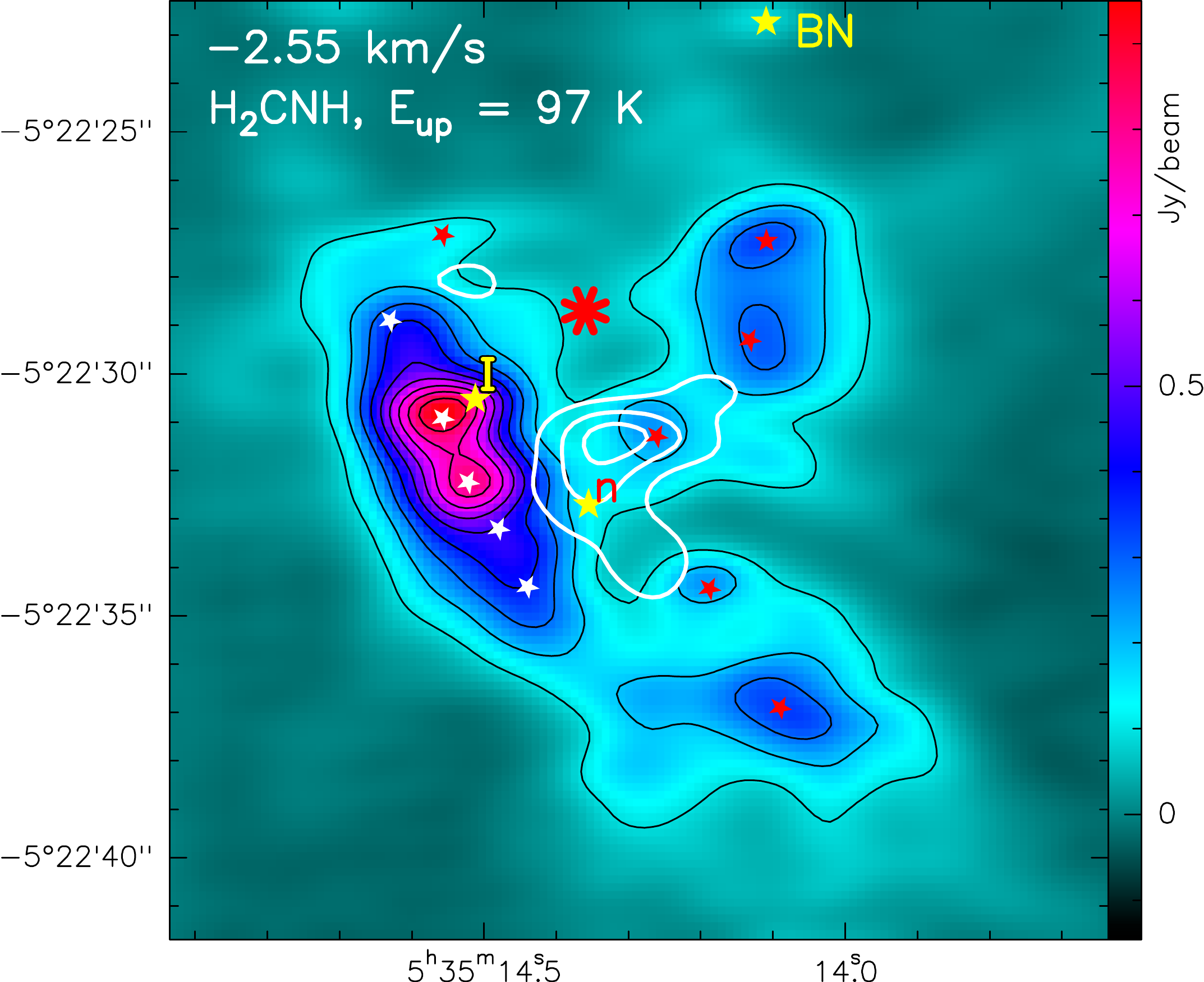}
\includegraphics[scale = 0.245, trim=85 20 48 0,clip]{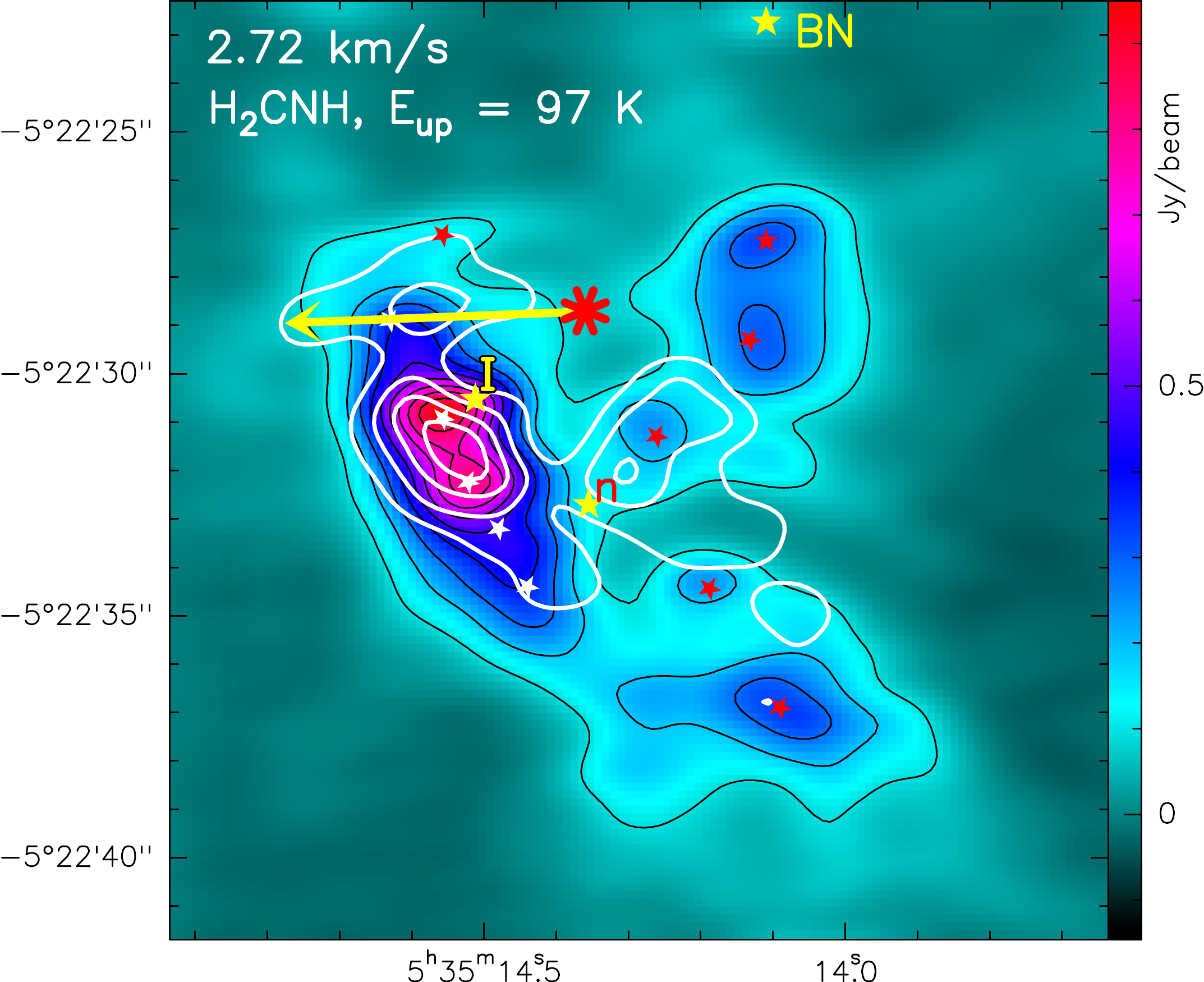}
\includegraphics[scale = 0.245, trim=85 20 48 0,clip]{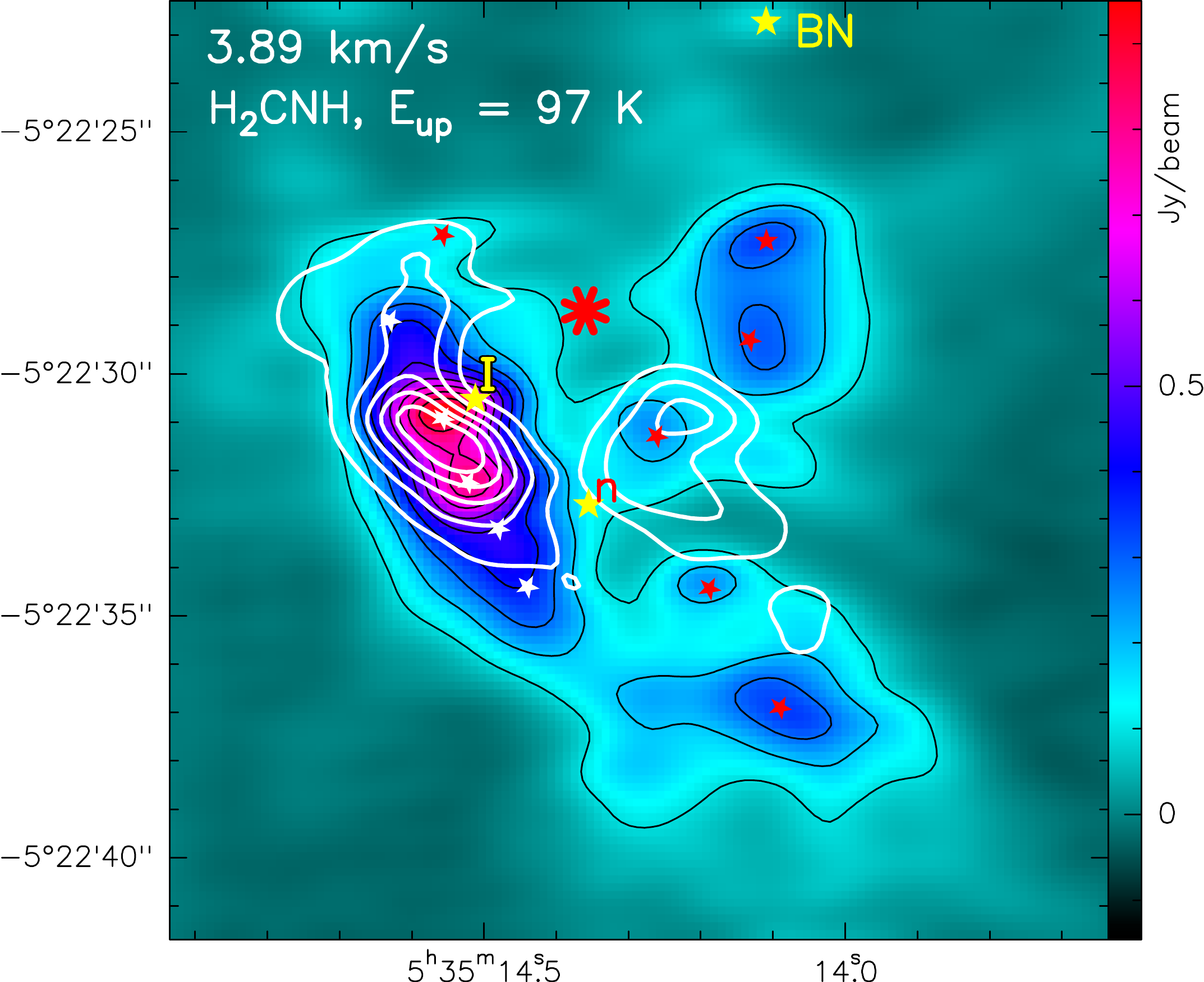}
\includegraphics[scale = 0.245, trim=85 20 0 0,clip]{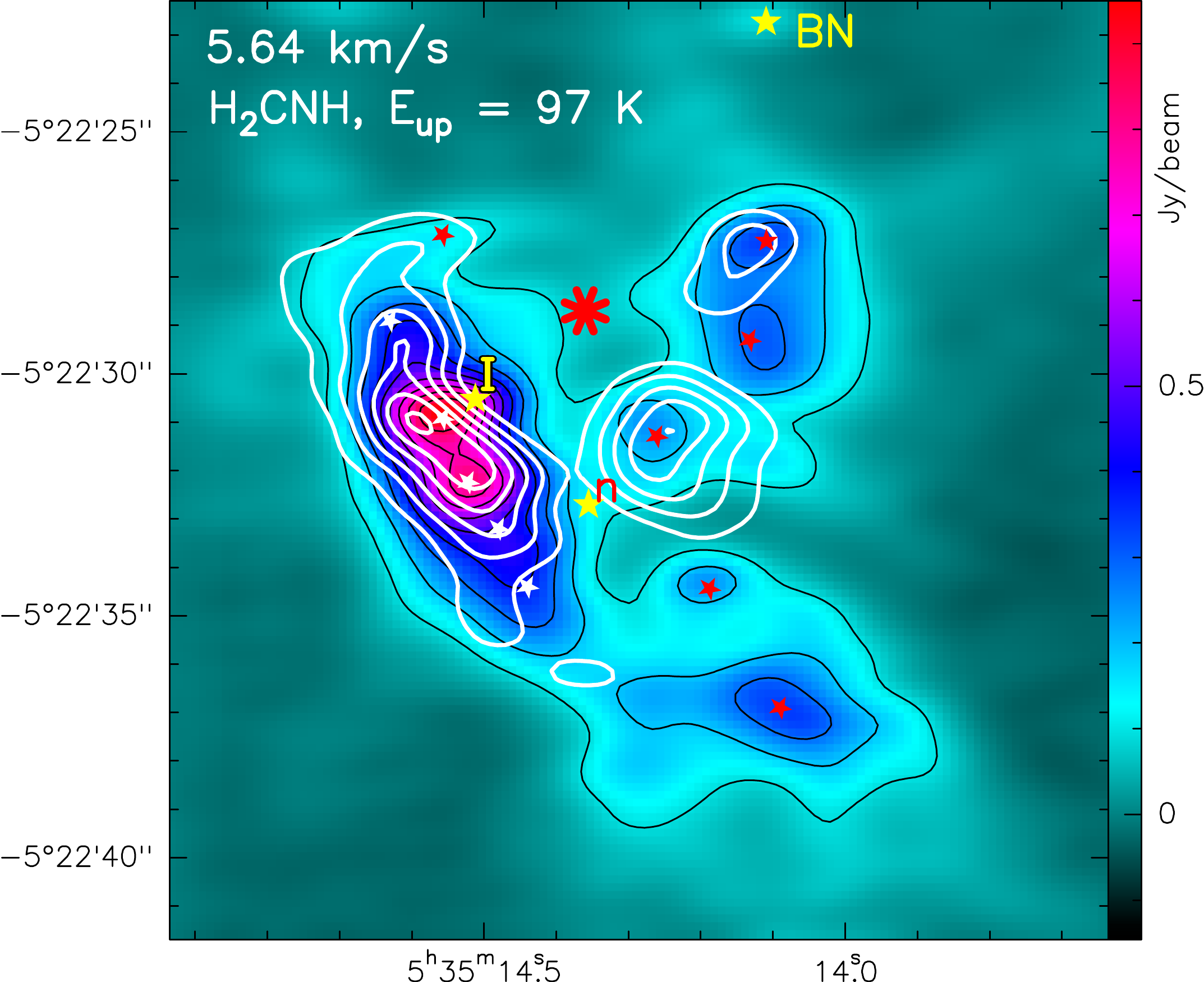}
\includegraphics[scale = 0.245, trim=0 0 49 0,clip]{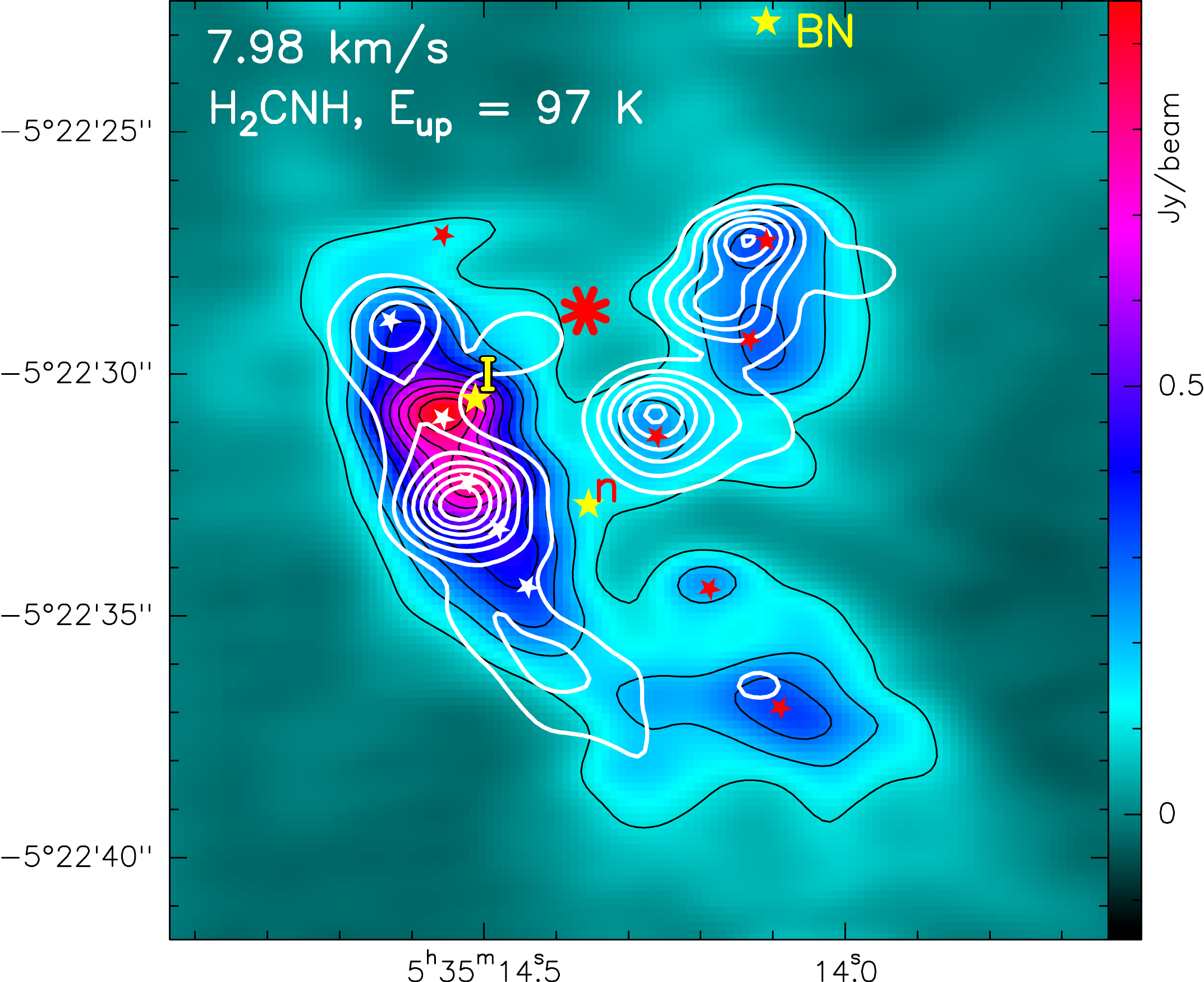}
\includegraphics[scale = 0.245, trim=85 0 48 0,clip]{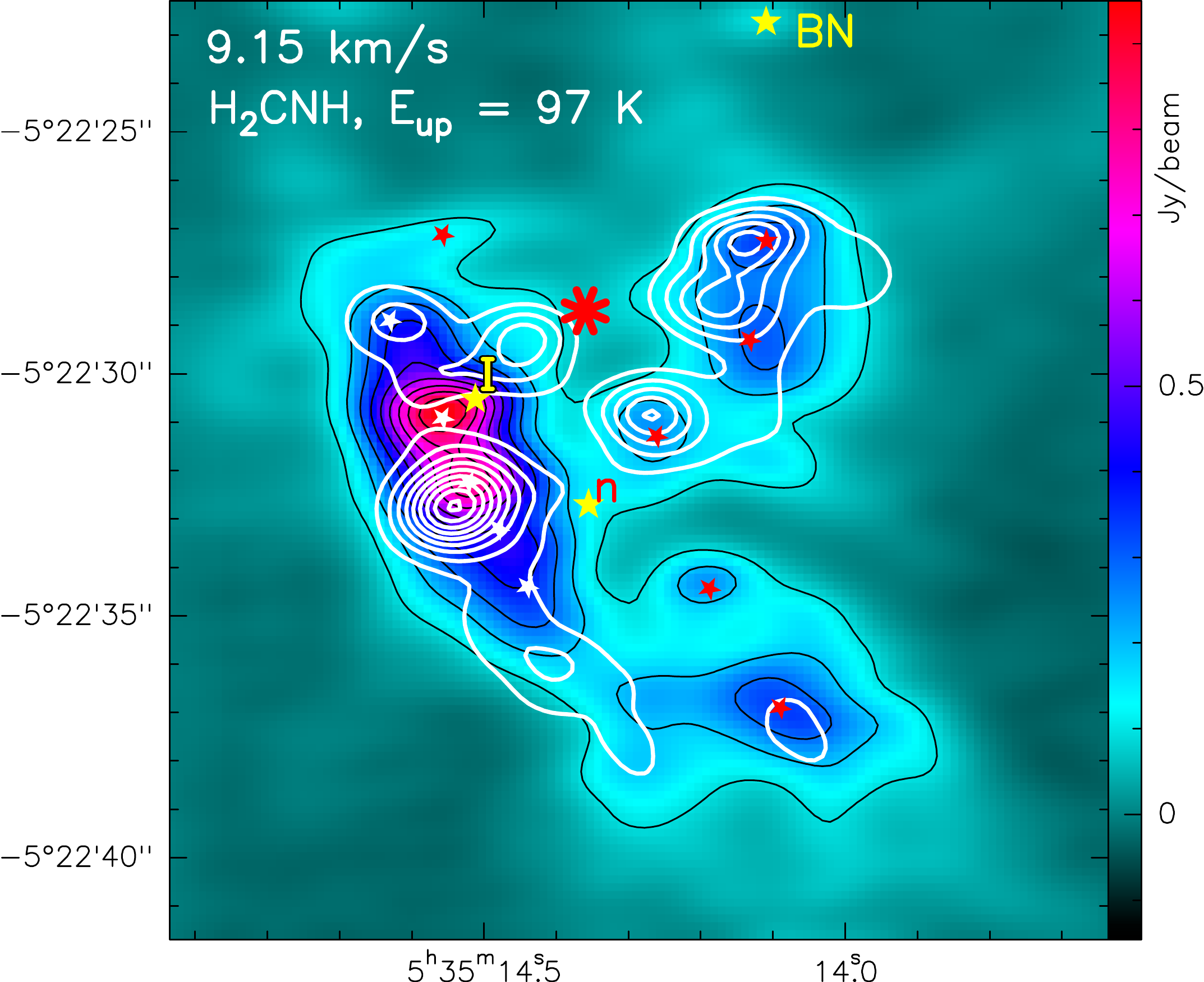}
\includegraphics[scale = 0.245, trim=85 0 48 0,clip]{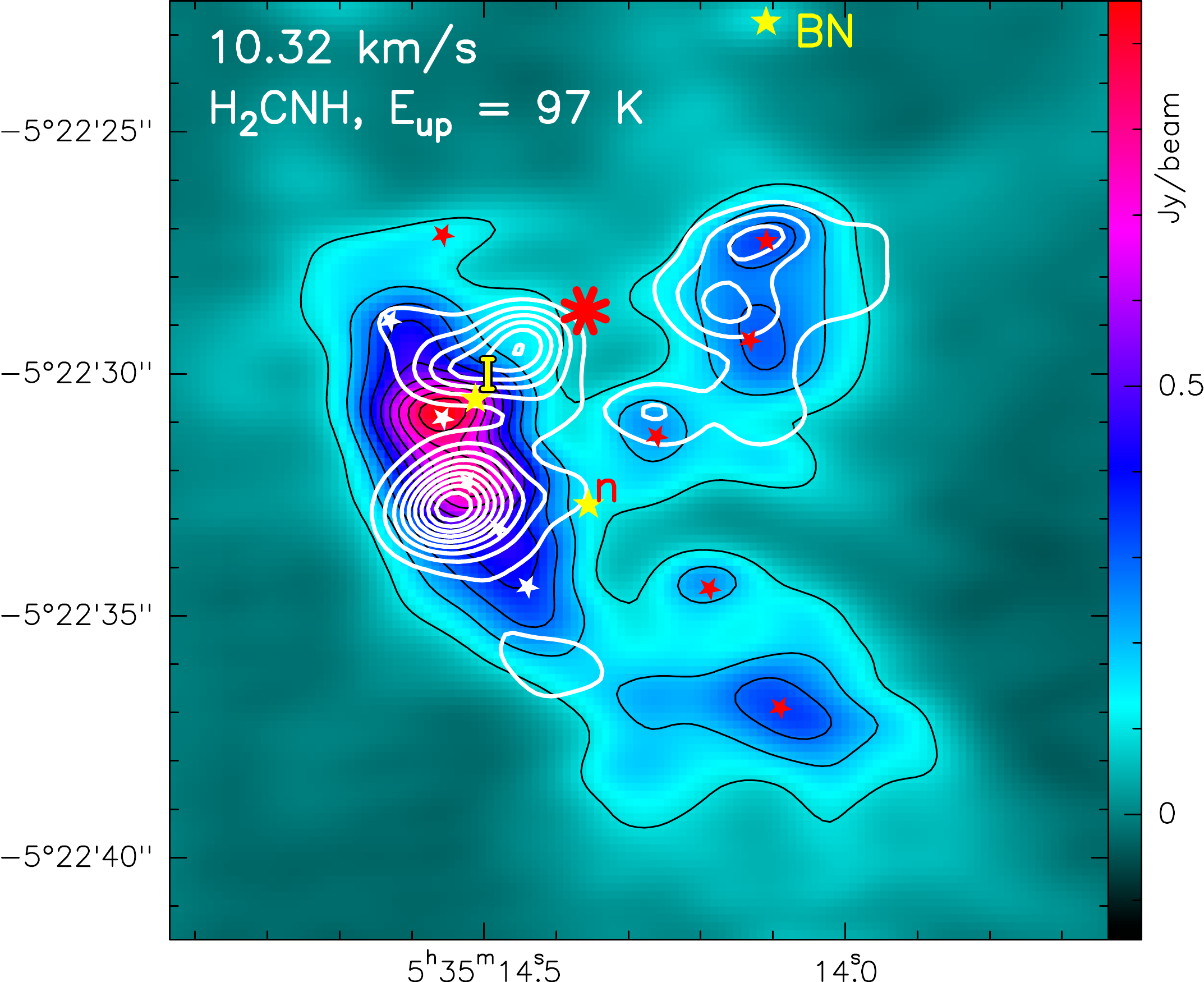}
\includegraphics[scale = 0.245, trim=85 0 0 0,clip]{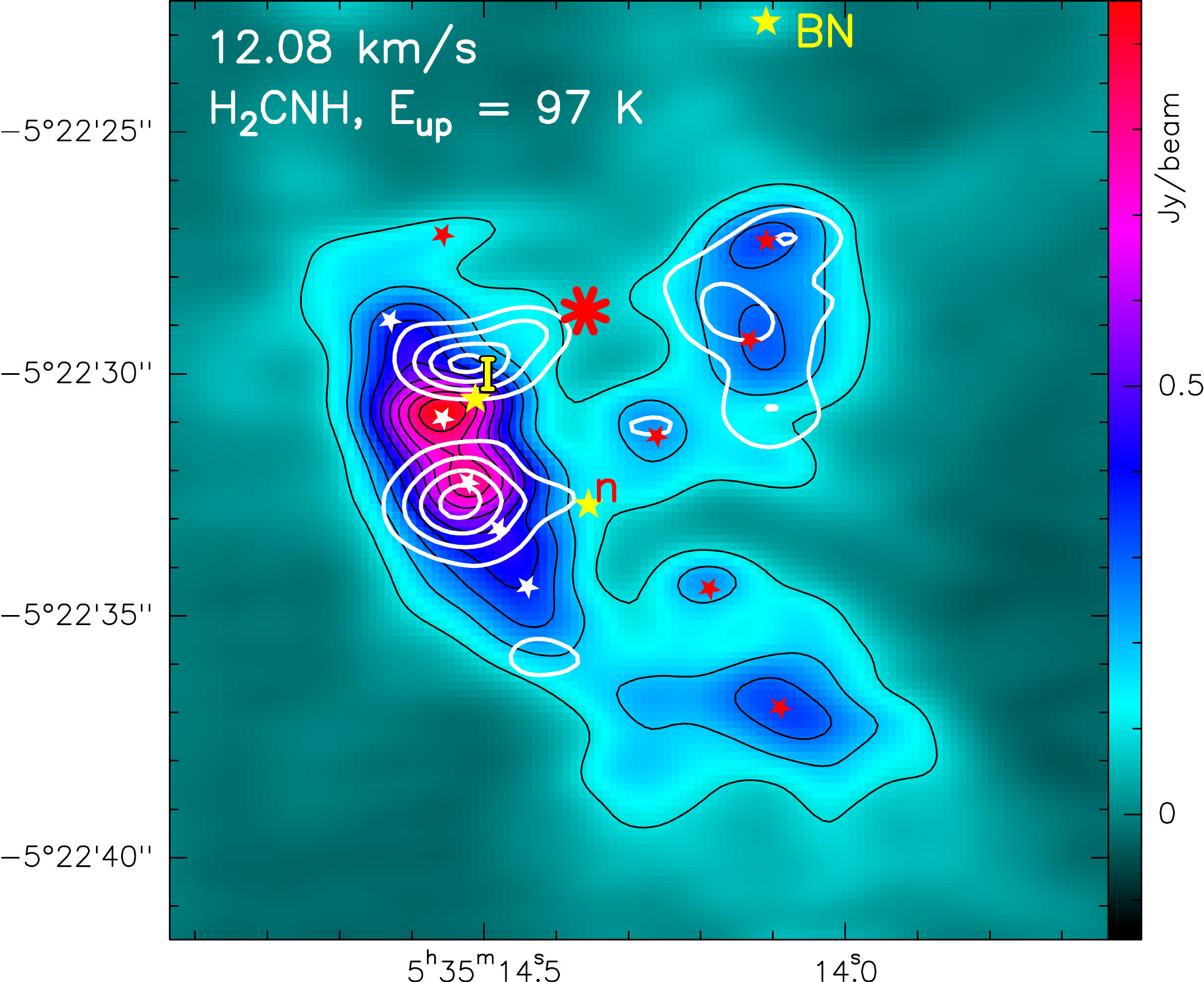}
 \caption{Same as Fig. \ref{fig:H2COchannelmaps} for a selection of  methanimine (H$_2$CNH, \jkk:7$_{16}$--7$_{07}$, at 250161.775 MHz) 
  channel maps {($\delta$V = 0.59 \kmps)}.}
 \label{fig:H2CNHchannelmaps}
\end{figure*}

It is interesting to examine channel maps rather than integrated intensity maps when the signal is strong enough. Figures
\ref{fig:H2COchannelmaps} and \ref{fig:H2CNHchannelmaps} show a selection of  channels for H$_2$CO (\jkk:9$_{18}$--9$_{19}$, at
216568.651 MHz) and H$_2$CNH (\jkk:7$_{16}$--7$_{07}$, at 250161.775 MHz), respectively. For all channel maps, the white contours
represent 10 to 90\% of the peak emission of the strongest channel for that species (which may not be one of the displayed
channels) in steps of 10\%. This allows us to estimate the importance of the emission at the given velocity. These two species
were chosen to be representative of two types of molecules: those showing evidence of a strong interaction with the wind of
the explosion and those that do not.  The first category is illustrated by H$_2$CO, the channel maps of which show elongations
that can be traced back to the 550-year-old explosion center as suggested by the yellow arrows (Fig. \ref{fig:H2COchannelmaps}).
Not only do the lowest contours form convex structures oriented away from the explosion center but the peak emission is always
displaced outward with respect to the local continuum peaks. The offset is especially strong southeastward in the 9 and 10
\kmpsb channel maps where the maximum emission from MF6 and the  EGP is clearly displaced (sources are
labeled in Fig.\,1 of \paperIt\ and in Fig. \ref{fig:bnklcontinuummapcontours}).  There is also a northwest elongation past MF5
between 7 and -8 \kmps. This feature is also clearly seen in OCS and D$_2$CO, and  in other species such as NO but at a much lower
intensity level ($\sim$ 0.15\,\jyb\ for NO). Similarly, the emission is neatly stretching away from the IRc7 spot at velocities
3--6 \kmps.  The elongated emission north of the HC  east from the explosion center, typically at 0 to 3 \kmps,  is seen (Fig.
\ref{fig:2.5kmps}) in all species that are emitting strongly enough in that velocity range. 

{Spatial velocity cuts along
the two most prominent northeast and southwest features of the H$_2$CO maps reveal essentially constant velocity ($<$ 1\,\kmps
shift) over their $\sim$5--10$\arcsec$ (0.01--0.02 pc) extent. The dynamical time of these features cannot be inferred from their
line-of-sight velocity since these movements are close to the plane of the sky with relative velocities of only one to a few \kmps. Conversely, the longest H$_2$CO elongation, 
$\sim$10$\arcsec$ can be attained in 550 years if the gas is moving at $\sim$40\,\kmps. This
is much lower than the maximum \dzcob clump velocities noticed in various data sets including our own, on the order of 140\,\kmpsb
\citep{Zapata:2009ho,Bally:2017cy}.} 

It is remarkable that a number of species do not show these various elongations (except that at $\sim$2\,\kmps) and remain close to the dust emission peaks. This is the case of the  H$_2$CNH shown (Fig.
\ref{fig:H2CNHchannelmaps}) but also of, for example, CH$_3$OCH$_3$, CH$_3$OCHO, NH$_2$CHO, and NH$_2$D. A few others, like ethyl cyanide and vinyl cyanide, show only
limited displacements.
{We  do not see any elongations related to the BN, I, and x runaway stars either. This probably indicates that the stars
are moving independently of  the gas.}

In Figs.\,\ref{fig:2.5kmps}--\ref{fig:9.0kmps}, we compare all species that are strong enough to be plotted channel-wise, at three representative
velocities of 2.5, 7.5, and 9.0 \kmps. The white contours have the same identification as in Figs. \ref{fig:H2COchannelmaps} and
\ref{fig:H2CNHchannelmaps}. We did not include  species such as CO, SO, and SO$_2$ that  extend beyond the BN-KL region and for
which zero-spacing data are missing to the point that the line profiles are totally distorted. We allowed three exceptions:
methanol, for which we selected a relatively high upper energy level transition (E$_{up}$/k = 231 K) to keep its spatial extent
within the limits of  efficient primary beam coupling; and H$_2$CO and OCS, which are mildly distorted but still provide
information on the small scale components in the KL region. At 2.5 km/s, the D$_2$CO and NH$_2$D maps are missing owing to line blending.
Depending on the species, the elongations are present or not and when absent, the peak emission is generally centered on the
infrared peaks or at least along the continuum ridges. In a few cases, and  mostly around the IRc7 peak,  the local maximum emission is on
the inner side of the infrared peak, i.e., displaced toward the explosion center while the extended emission on the other edge of
the continuum peak is still present. This is the case for CH$_3$COCH$_3$, CH$_3$NCO, c-C$_2$H$_4$O, D$_2$CO, and without extension
for NH$_2$D and H$_2$CNH,   at $\sim$7.5 and $\sim$9 \kmps (Figs.\,\ref{fig:7kmps} \& \ref{fig:9.0kmps}).

\section{Discussion}

From Figs. \ref{fig:H2COchannelmaps}, \ref{fig:H2CNHchannelmaps}, and \ref{fig:2.5kmps}--\ref{fig:9.0kmps}, we see clear evidence for the effect of
the explosion in many species. This was already mentioned in Paper I, in which  we revealed the existence of  blue wings for the
sources east and south of the explosion center and of only red wings for the western sources (MF4 and MF5). We also showed that 
systemic velocities are below 7 \kmpsb for the eastern sources and above 7 \kmpsb for the western sources. We explained these
features in terms of the sources expanding away from the explosion center.  In this picture the eastern sources are in front of the center
and the western sources are behind (Fig. 30 of Paper I). The asymmetrical wings were explained as gas being accelerated by the
explosion blowout. \citet{Wright:2017jk} and ourselves \citep{Favre:2017ch} also presented evidence for a bullet having hit the
dense ridge near MF6. This bullet was  most probably  launched by the explosion.That the HC has no internal source and is
heated from outside, most probably by the explosion, has  been advocated by \citet{Zapata:2011ez} and \citet{
OrozcoAguilera:2017cg}.  This is particularly obvious when 12 $\mu$m maps of the region
are superimposed \citep{2004ApJ...605L..57G,2005AJ....129.1534R} on our 1.2 mm continuum map (see Fig. \ref{fig:bnklcontinuummapcontours}). The
IRc2 spot marked by \citet{2004ApJ...605L..57G} is halfway between the explosion center and the HC while the latter is isible
at 12 $\mu$m.
 
With the channel maps, we see some direct evidence of the fast expansion as many species are being driven  away from the explosion center at
the same velocities: from -2 to 3 \kmpsb eastward, through MF10; from 6 to 11 \kmpsb southeastward through MF6; from 6 to 9
\kmpsb northwestward through MF5; from 7 to 14 \kmpsb in between MF4 and MF5 and then through MF4 (especially visible in H$_2$CO,
D$_2$CO, and OCS, Fig.\,\ref{fig:7kmps}); and from 2 to 6 (or 8 depending on the species) \kmpsb southward through IRc7. These
offsets are  due to  the winds from the explosion center that pass near the dense dust cores,  collect the molecules recently
released from the grains, and  push these molecules away. The different velocities  possibly reflect the direction of propagation if we
suppose the winds to be spherically expanding. For example, the long southeast extension crossing MF6 probably passes behind
the dust ridge since its velocity range (from 6 to 11 \kmps)  is globally higher than the HC-MF6 range (4 -- 6 \kmps).
\begin{figure}
\centering
\includegraphics[width=\linewidth]{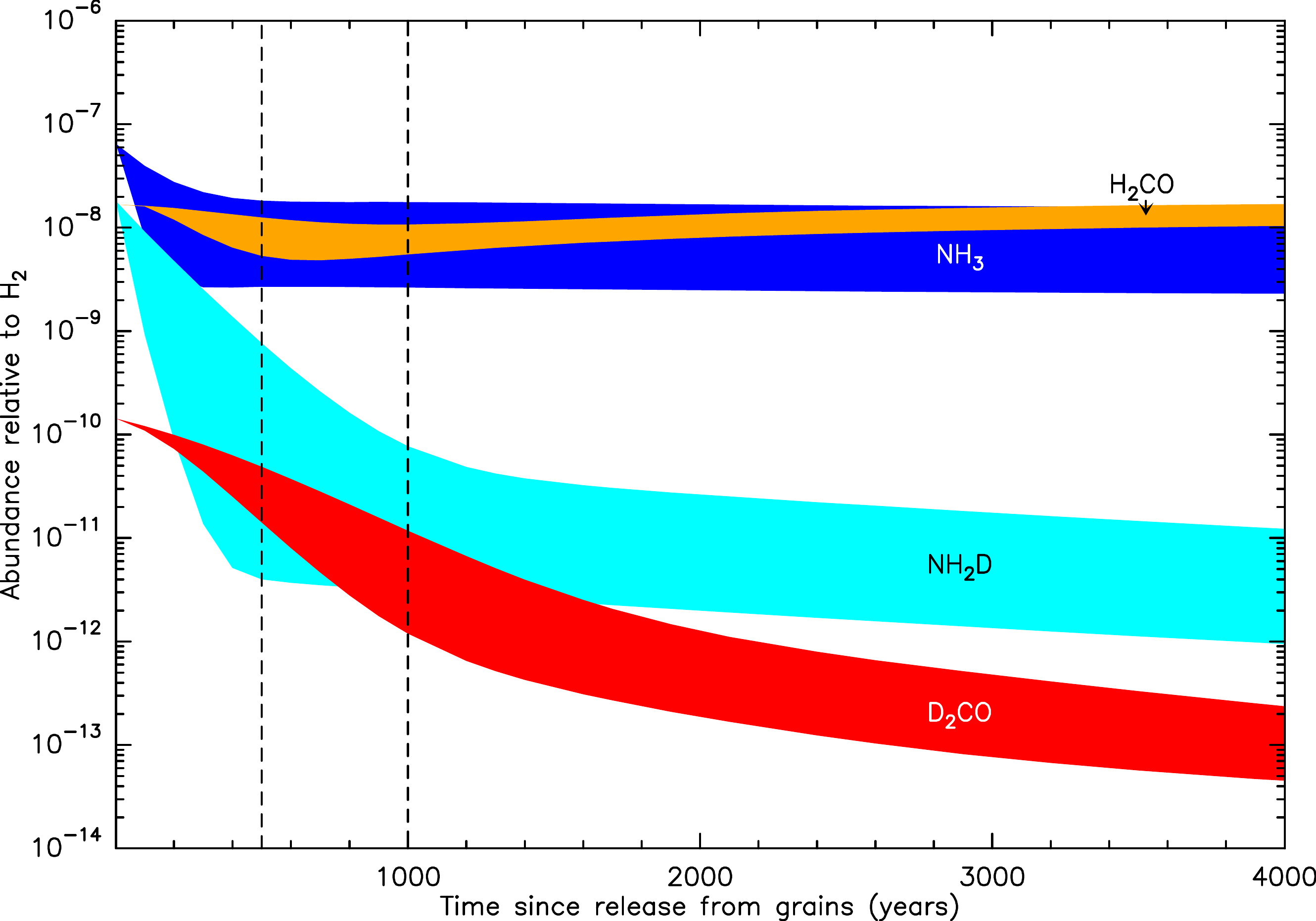}
\caption{Abundance variation with time of deuterated species following their release from grains. The range of values covers a factor of 10 in UV illumination variation.
Vertical dashed lines indicate 500 and 1000 years.}
\label{fig:destructionnh3h2coorion}
\end{figure}
In summary, we see that some molecular species show emission elongated to significant distances all pointing away from the
explosion center, while other species do not. A number of these species are predominantly formed on grain surfaces, and these
channel maps may be understood as due to the evolution of these species after they are released from the dust grains and are driven
outward by the explosion. Those molecules that do not show elongations, such as H$_2$CNH and NH$_2$D, are probably rapidly
destroyed after their release in the gas phase.  In the case of NH$_2$D, the tentative detection of NH$_3$D$^+$ by
\citet{Cernicharo:2013gn}, later confirmed by laboratory measurements of \citet{2016A&A...593A..56S}, could suggest an
explanation. NH$_2$D can react with H$_3^+$ to form NH$_3$D$^+$ in two-thirds of the cases, and NH$_4$+ + HD in one-third of the cases. Although
the probability of dissociative recombination of NH$_3$D$^+$ to form NH$_3$ is statistically three times less probable than the
formation of NH$_2$D,  the rapid destruction of NH$_2$D is still expected since 50\,\% of the reactions with H$_3^+$ eventually  lead to NH$_3$ and in this warm environment (100 -- 300 K), deuteration is not expected to occur in the gas phase to
replenish NH$_2$D.  Similarly, D$_2$CO extends away from the explosion center but less than H$_2$CO. D$_2$CO is  converted back to HDCO
and next to H$_2$CO by repeated reactions with \htp, followed by dissociative recombinations. To test this idea, we ran  our
gas-phase chemical model including our deuteration network with the spin states of H$_2$ and \htpb isotopologues
\citep{Pagani2011}. Starting from a gas rich in deuterated species (D$_2$CO/H$_2$CO $\approx$ 0.01 and NH$_2$D/NH$_3$ $\approx$
0.3) at a density of 1\,\pdixb{5} \cc, a temperature of 150 K, low extinction to the ISRF and a harsh environment (approximated
by a cosmic ionization rate $\zeta$ of 1\, \pdixb{-15} s$^{-1}$), we found that the D$_2$CO/H$_2$CO ratio drops by a factor of 3
to 40 in 500 to 1000 years, the abundance of D$_2$CO itself drops by a factor 3 to 100 in the same time,  while the
NH$_2$D/NH$_3$ ratio drops by a factor 6 to 200 in 500 years, and the abundance of NH$_2$D itself decreases by a factor 20 to 5000
(Fig. \ref{fig:destructionnh3h2coorion}). The decrease of NH$_2$D is much more rapid than the decrease of D$_2$CO, which is
illustrated by the fact that NH$_2$D does not expand at all away from the icy dust evaporation sites, unlike D$_2$CO.
It would be interesting to map H$_2$COH$^+$ (2 and 3 mm transitions of this ion have been detected in a number of
sources, including Orion KL; \citealt{Ohishi_1996}) and NH$_3$D$^+$ in the source at a similar resolution and make quantitative
statements about the abundance of each species in each velocity channel, but this is beyond the scope of this letter.

Species such as ethyl cyanide and vinyl cyanide are only slightly displaced outward. They probably survive longer than NH$_2$D or
H$_2$CNH before being destroyed or converted to other species.

It seems therefore that the closeness of Orion, combined with the ongoing expansion and availability of high spatial resolution
observations opens a new window on the time evolution of different species soon after their release in the gas phase in a
way similar to  a time-of-flight experiment. This will allow us to select the important chemical paths between species and possibly
constrain the chemistry of COMs. Recently, \citet{Tercero:2018ba} have suggested segregation of various O-bearing COMs based on
spatial variation inside BN-KL. Including the channel information presented in this work, combined with quantitative estimates, which
require the inclusion of zero-spacing data that is not yet available, will allow us to follow the abundance variations of each species
along the expanding paths and better understand the chemistry in hot cores, both in the gas phase and on grain surfaces.
\vspace{-0.5cm}
   \section{Conclusions}
   
We show that thanks to the high quality of the ALMA data (large signal/noise ratio, good sampling of the UV plane), we can study
channel by channel the fate of numerous molecules, and separate the components both spatially and in velocity. The expansion of
each molecular species due to the 550-year-old explosion is comparable to a time-of-flight experiment and allows us to put constraints on the
way the chemistry proceeds in the cloud. This is an exceptional situation not met in other sources in which all species are mixed
spatially and therefore their temporal evolution cannot be discerned. Hence, Orion KL should  be the prime benchmark for chemical
models to treat the formation and evolution of COMs in hot cores. To quantify these constraints, zero-spacing data should be
added to the present set and other species (DCO$^+$, DNC, HNC, HCO$^+$, NH$_3$D$^+$ , N$_2$H$^+$, to name a few) should be observed
with at least similar sensitivity to better constrain the models of chemistry in general and of this source in particular.

   \begin{acknowledgements}
        We thank the referee, J. Bally, for the important corrections concerning sources x and n and for the suggested improvements to 
        the Letter. We thank the IRAM ARC center for their hospitality and coaching for the ALMA data reduction process, and ESO for
        the financial support during the visit to the ARC (MARCUs funding). This work made use of the SIMBAD resource offered by CDS
        Strasbourg, France, of the CDMS and JPL molecular line databases offered by the K\"oln University and by NASA, respectively. LP
        acknowledges funding from AF ALMA-NOEMA, Paris Observatory. CF work is supported by the French National Research Agency in the
        framework of the Investissements d'Avenir program (ANR-15-IDEX-02), through the funding of the "Origin of Life" project of the
        Univ. Grenoble-Alpes. This work was carried out in part at the Jet Propulsion Laboratory, which is operated for NASA by the
        California Institute of Technology. EAB acknowledges  support from the National Science Foundation grant AST-1514670. We thank
        E. Roueff for interesting discussions.
 \end{acknowledgements}
   \bibliographystyle{aa}
   
   \bibliography{/Users/laurent/bibtex/references,/Users/laurent/bibtex/papers}
\appendix

\section{Mid-infrared and millimeter continuum map comparison} \label{douze_mic}
\begin{figure}[!h]
\centering
\includegraphics[width=\linewidth]{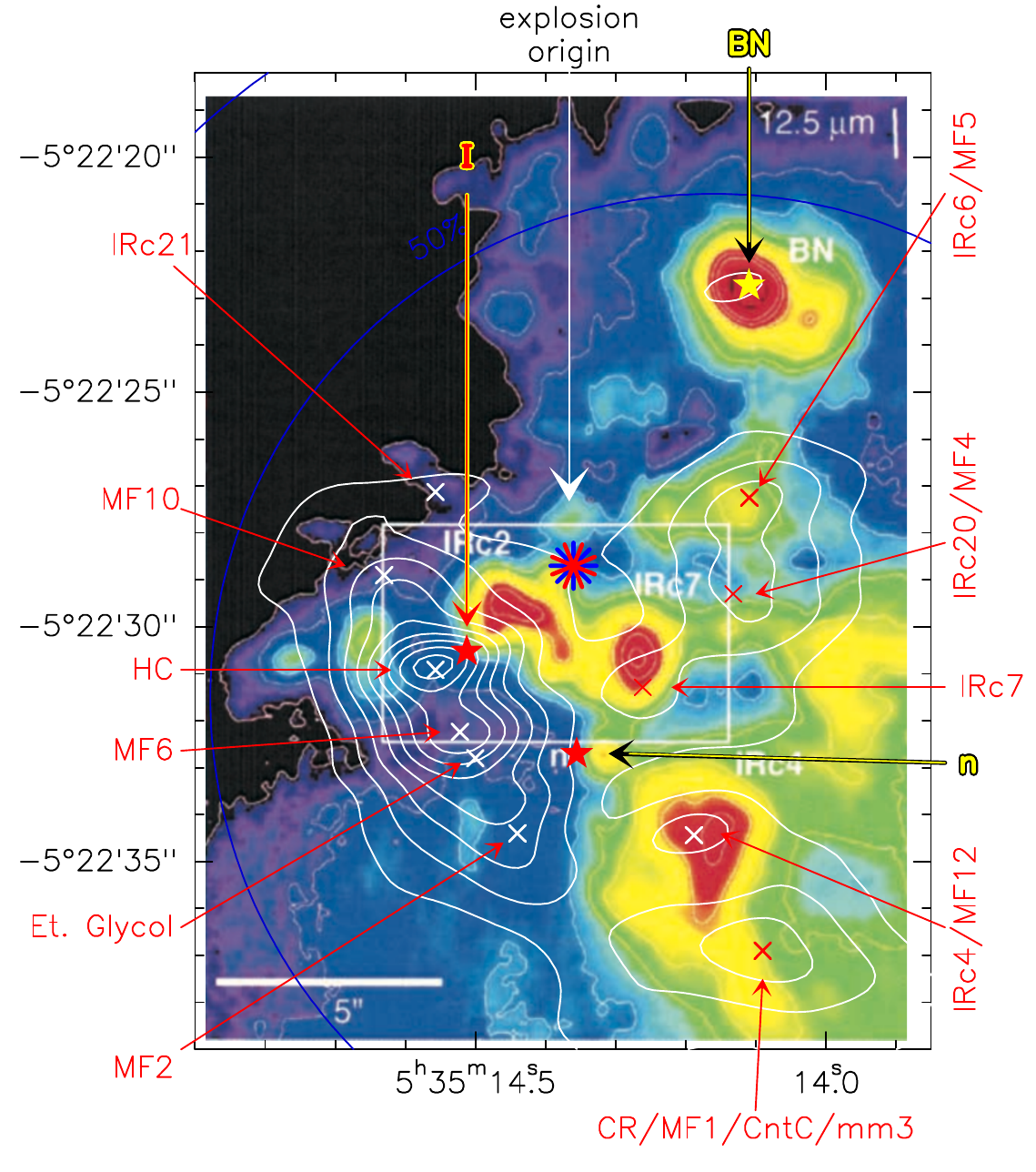}
\caption{1.2 mm continuum emission (white contours) superimposed on the 12.5\,$\mu$m  map from \citet[shown in colors]{2004ApJ...605L..57G}. {The two runaway stars 
(BN and I) plus the quasi-static object n }are 
indicated by five-pointed stars (yellow or red) while the 10 sources studied in \paperIt\ are identified and denoted by crosses (white or red; see \paperIt\ for more details). The 
presumed position of the origin of the explosion is denoted by a multi-pointed red and blue star.}
\label{fig:bnklcontinuummapcontours}
\end{figure}
 
 We superimposed our 1.2 mm ALMA continuum map on the \citet{2004ApJ...605L..57G} 12.5 $\mu$m map. The hot sources at 12.5 
 $\mu$m do not coincide with our 1.2 mm strongest peaks. Alternately, the emission peaks coincide toward both BN and IRc4. The
 HC itself is particularly weak at 12.5 $\mu$m, while there is a strong mid-infrared zone  between the explosion center and HC (known as
 IRc2) and a weaker mid-infrared zone on the other side of the HC ridge. This confirms the claim of \citet{Zapata:2011ez} of the absence of an
 internal heating source for the HC but of  external heating possibly due to the explosion.  {From the \citet{Feng:2015cl} data, we find 
 that the HC does not seem to be gravitationally bound. The effect of the explosion on the overall stability of the HC is yet to be determined.}
 
 
 \section{Multi-species comparisons at three remarkable velocities} 
  \begin{table*}
 \caption{List of molecular species shown in Figs. \ref{fig:2.5kmps}--\ref{fig:9.0kmps}.}
\label{Tab:molfreq}
 \begin{tabular}{llcccl}
 \hline 
  Name&  Transition& E$_{up}$/k\tablefootmark{a}  &Frequency  & A$_{ul}$\tablefootmark{b}& Database \\ 
  &  &  (K)&(MHz)&  s$^{-1}$&  \\ 
 \hline 
  C$_2$H$_3$CN& 23$_{4,19}$--22$_{4,18}$ &  160.4& 218\,615.092 &8.40(-4)&cdms  \\ 
  C$_2$H$_5$CN&  26$_{3,23}$--25$_{3,22}$ & 162.2 &237\,170.450  &1.11(-3)& cdms \\ 
  c-C$_2$H$_4$O  & 8$_{0,8}$--7$_{1,7}$ &~~52.4&235\,105.022 &  2.33(-4)& cdms\\ 
  & 8$_{1,8}$--7$_{0,7}$ &~~52.4&235\,105.055  &  2.33(-4)& cdms\\ 
  C$_2$H$_5$OH&  &130.6  &216\,415.624  & 9.07(-5)  &jpl\\ 
  CH$_3$CN&  12$_4$--11$_4$& 183.1 &220\,679.287  &  8.21(-4) &cdms\\ 
  CH$_3$COCH$_3$& 22$_{0,22}$--21$_{0,21}$  &123.9& 220\,361.881 & 3.93(-4) & jpl \\ 
 & 22$_{0,22}$--21$_{1,21}$  &123.9& 220\,361.881 & 1.61(-4) &  jpl \\ 
 & 22$_{1,22}$--21$_{0,21}$  &123.9& 220\,361.881 & 1.61(-4) &  jpl\\ 
 & 22$_{1,22}$--21$_{1,21}$  &123.9& 220\,361.881 & 3.93(-4) &  jpl \\ 
    CH$_3$NCO&  24$_{3,0}$--23$_{3,0}$, v$_b$=0 &  191.0&217\,701.086  & 4.40(-4) &cdms\\ 
  CH$_3$OCH$_3$& 22$_{4,19,3}$--21$_{3,20,3}$  &253.4  &  217\,189.668&5.43(-5)&cdms  \\ 
  & 22$_{4,19,5}$--21$_{3,20,5}$  &253.4  &  217\,189.669&5.43(-5)&cdms  \\
  & 22$_{4,19,1}$--21$_{3,20,1}$  &253.4  &  217\,191.400&5.43(-5)&cdms  \\
  & 22$_{4,19,0}$--21$_{3,20,0}$  &253.4  &  217\,193.132&5.43(-5)&cdms  \\
  CH$_3$OCHO& 17$_{3,14,0}$--16$_{3,13,0}$  &~~99.7  &218\,297.890  & 1.51(-4)&jpl  \\ 
  CH$_3$OH&12$_{3,9}$--12$_{2,10}$, v$_t$=0   & 230.8 &  250\,635.207&  8.29(-5)&cdms\\ 
  D$_2$CO& 4$_{0,4}$--3$_{0,3}$ & ~~27.9 & 231\,410.234 &3.47(-4)  &cdms\\ 
  DCN& 3$_3$--2$_3$ & ~~20.9 &217\,236.999  & 5.08(-5) &cdms\\ 
  & 3$_2$--2$_1$ & ~~20.9 &217\,238.300  & 3.84(-4) &cdms\\ 
  & 3$_3$--2$_2$ & ~~20.9 &217\,238.555  & 4.07(-4) &cdms\\ 
  & 3$_4$--2$_3$ & ~~20.9 &217\,238.612  & 4.57(-4) &cdms\\ 
  & 3$_2$--2$_3$ & ~~20.9 &217\,239.079  & 2.03(-6) &cdms\\ 
  & 3$_2$--2$_2$ & ~~20.9 &217\,240.622  & 7.12(-5) &cdms\\
  H$_2$CNH&7$_{1,6,6}$--7$_{0,7,6}$& ~~97.2 & 250\,161.110 & 1.47(-4) & cdms\\
  &7$_{1,6,8}$--7$_{0,7,8}$& ~~97.2 & 250\,161.373 & 1.48(-4) & cdms\\
  &7$_{1,6,7}$--7$_{0,7,6}$& ~~97.2 & 250\,161.945 & 2.65(-6) & cdms\\
  &7$_{1,6,6}$--7$_{0,7,7}$& ~~97.2 & 250\,161.972 & 3.06(-6) & cdms\\
  &7$_{1,6,7}$--7$_{0,7,8}$& ~~97.2 & 250\,161.990 & 2.66(-6) & cdms\\
  &7$_{1,6,8}$--7$_{0,7,7}$& ~~97.2 & 250\,162.190 & 2.34(-6) & cdms\\
  &7$_{1,6,7}$--7$_{0,7,7}$& ~~97.2 & 250\,162.807 & 1.45(-4) & cdms\\ 
  H$_2$CO& 9$_{1,8}$--9$_{1,9}$  & 174.0 & 216\,568.651 & 7.22(-6)&cdms  \\ 
  HC$_3$N& 24$_{1}$--23$_{-1}$, v$_7$ = 1 & 452.3 & 219\,173.757 & 8.30(-4) &cdms \\ 
  HNCO&10$_{1,10,9}$--9$_{1,9,9}$& 101.1 & 218\,980.248 & 1.63(-6) & jpl\\
  &10$_{1,10,9}$--9$_{1,9,10}$& 101.1 & 218\,981.031 & 4.10(-9) & jpl\\
  &10$_{1,10,9}$--9$_{1,9,8}$& 101.1 & 218\,981.170 & 1.46(-4) & jpl\\
  &10$_{1,10,10}$--9$_{1,9,9}$& 101.1 & 218\,981.170 & 1.46(-4) & jpl\\
  &10$_{1,10,11}$--9$_{1,9,10}$& 101.1 & 218\,981.170 & 1.48(-4) & jpl\\
  &10$_{1,10,10}$--9$_{1,9,10}$& 101.1 & 218\,981.900 & 1.48(-6) & jpl\\
  NH$_2$CHO & 11$_{3,9}$--10$_{3,8}$ & ~~94.1 & 233\,897.318&  8.62(-4)&  jpl\\ 
  NH$_2$D&3$_{2,2,0,2}$--3$_{1,2,1,2}$& 119.6 & 216\,562.487 & 5.44(-5) & jpl\\
  &3$_{2,2,0,3}$--3$_{1,2,1,2}$& 119.6 & 216\,562.489 & 4.86(-6) & jpl\\
  &3$_{2,2,0,4}$--3$_{1,2,1,4}$& 119.6 & 216\,562.621 & 5.74(-5) & jpl\\
  &3$_{2,2,0,3}$--3$_{1,2,1,4}$& 119.6 & 216\,562.622 & 4.92(-6) & jpl\\
  &3$_{2,2,0,2}$--3$_{1,2,1,3}$& 119.6 & 216\,563.000 & 6.80(-6) & jpl\\
  &3$_{2,2,0,4}$--3$_{1,2,1,3}$& 119.6 & 216\,563.001 & 3.83(-6) & jpl\\
  &3$_{2,2,0,3}$--3$_{1,2,1,3}$& 119.6 & 216\,563.002 & 5.15(-5) & jpl\\
  NO\tablefootmark{c} &3$_{-1,3,4}$--2$_{1,2,3}$& ~~19.3 & 250\,796.436 & 1.85(-6) & jpl\\
  &3$_{-1,3,3}$--2$_{1,2,2}$& ~~19.3 & 250\,815.594 & 1.55(-6) & jpl\\
  &3$_{-1,3,2}$--2$_{1,2,1}$& ~~19.3 & 250\,816.954 & 1.39(-6) & jpl\\
  OCS&  18--17, v=0&~~99.8 & 218\,903.356 & 3.04(-5)& cdms \\ 
 \hline 
 \end{tabular} 
\tablefoot{ 
\tablefoottext{a}{Energy expressed in Kelvins (k = Boltzmann's constant)}\\
\tablefoottext{b}{Einstein spontaneous emission coefficient. a(b) = a\pdix{b}.}\\
\tablefoottext{c}{To avoid contamination by strong nearby lines, NO maps are a combination of the 250.796 GHz transition 
for velocities from 6 to  19 \kmps, and of the 250.81 lines for velocities from 
-5 to 11 \kmps; the overlapping velocities are a sanity check.}
}
 \end{table*}

 \begin{figure*}[h!]
 \centering 
 \includegraphics[scale = 0.248,trim=0 20 49 0,clip]{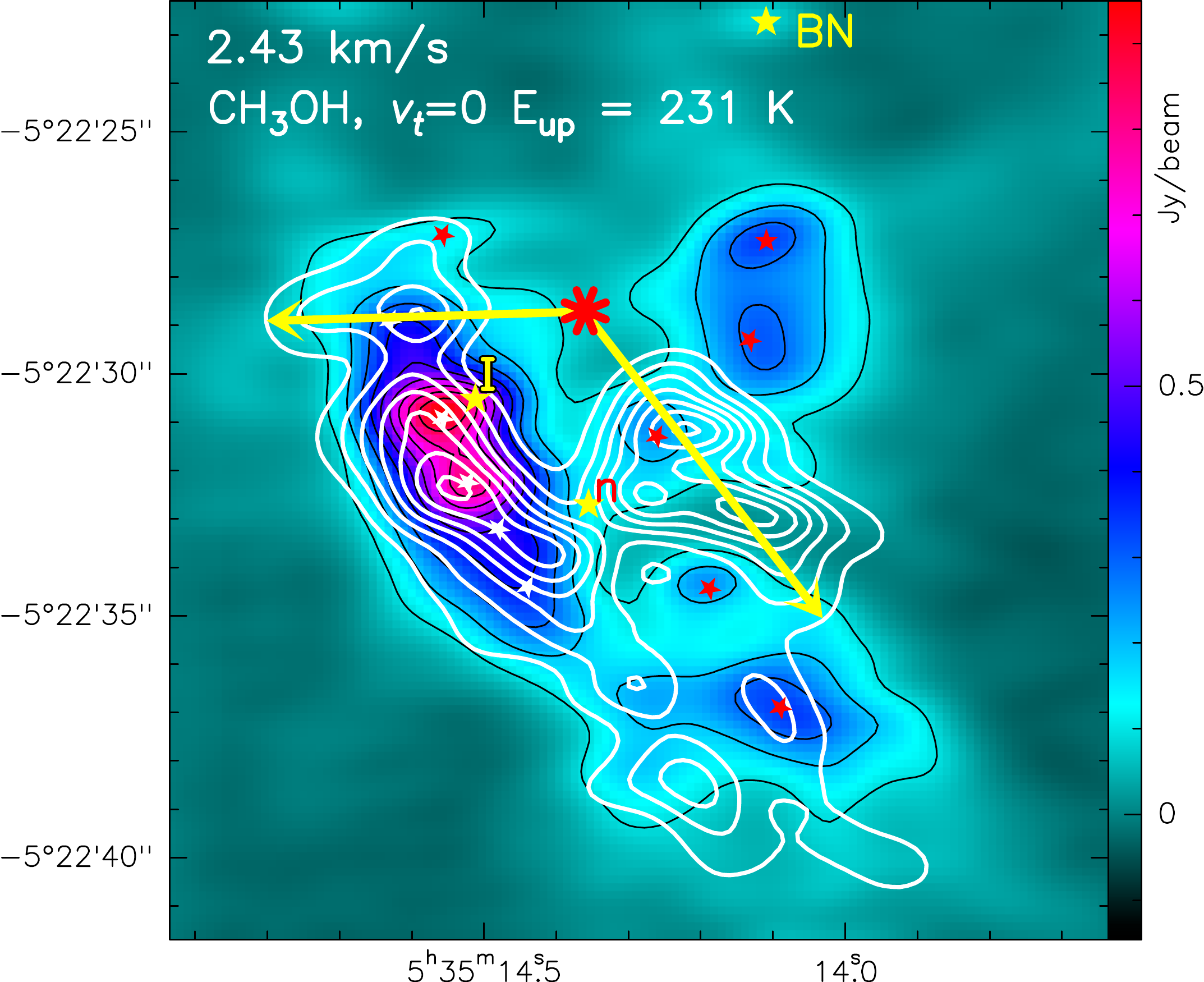}
 \includegraphics[scale = 0.248, trim=85 20 48 0,clip]{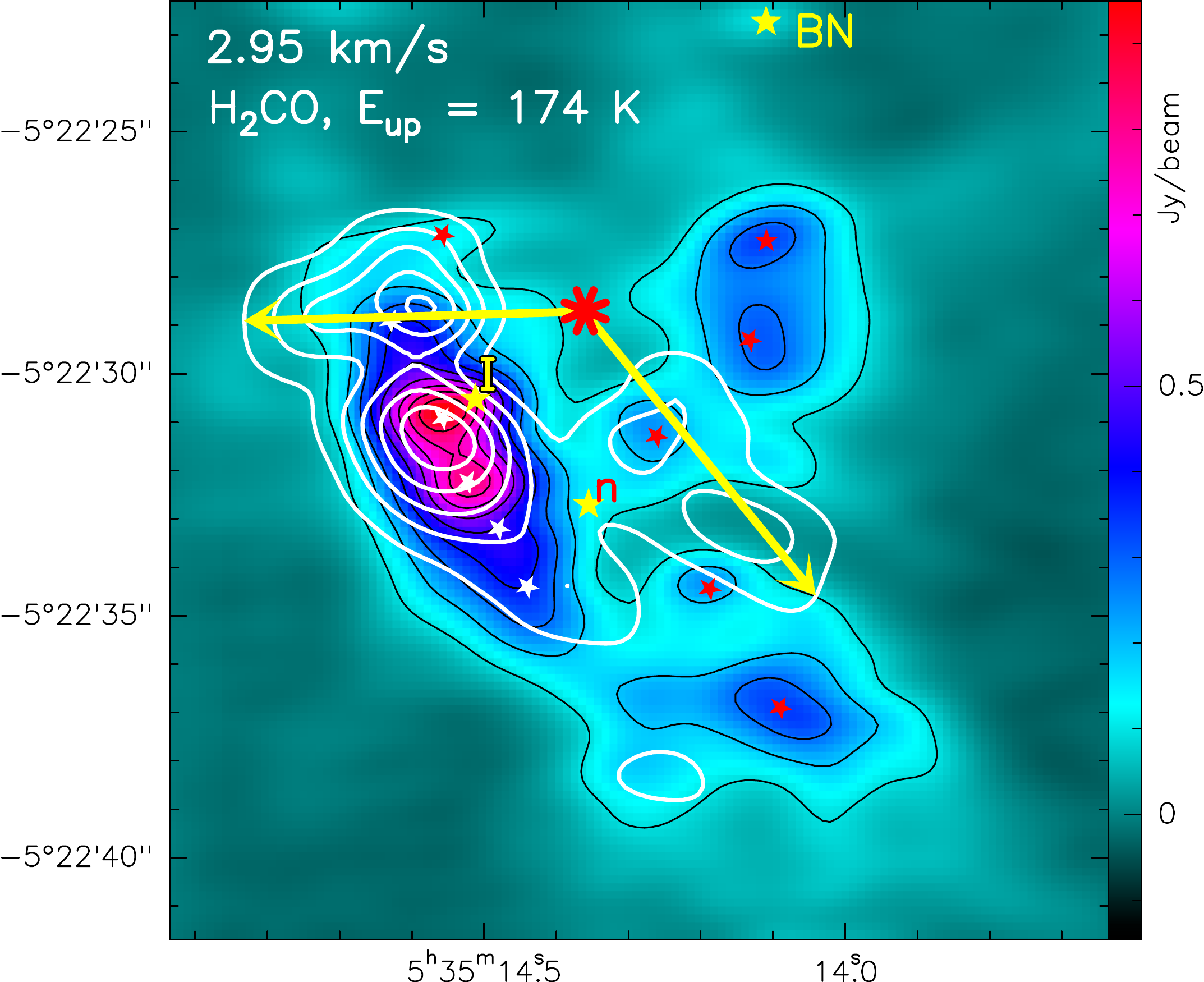}
 \includegraphics[scale = 0.248, trim=85 20 48 0,clip]{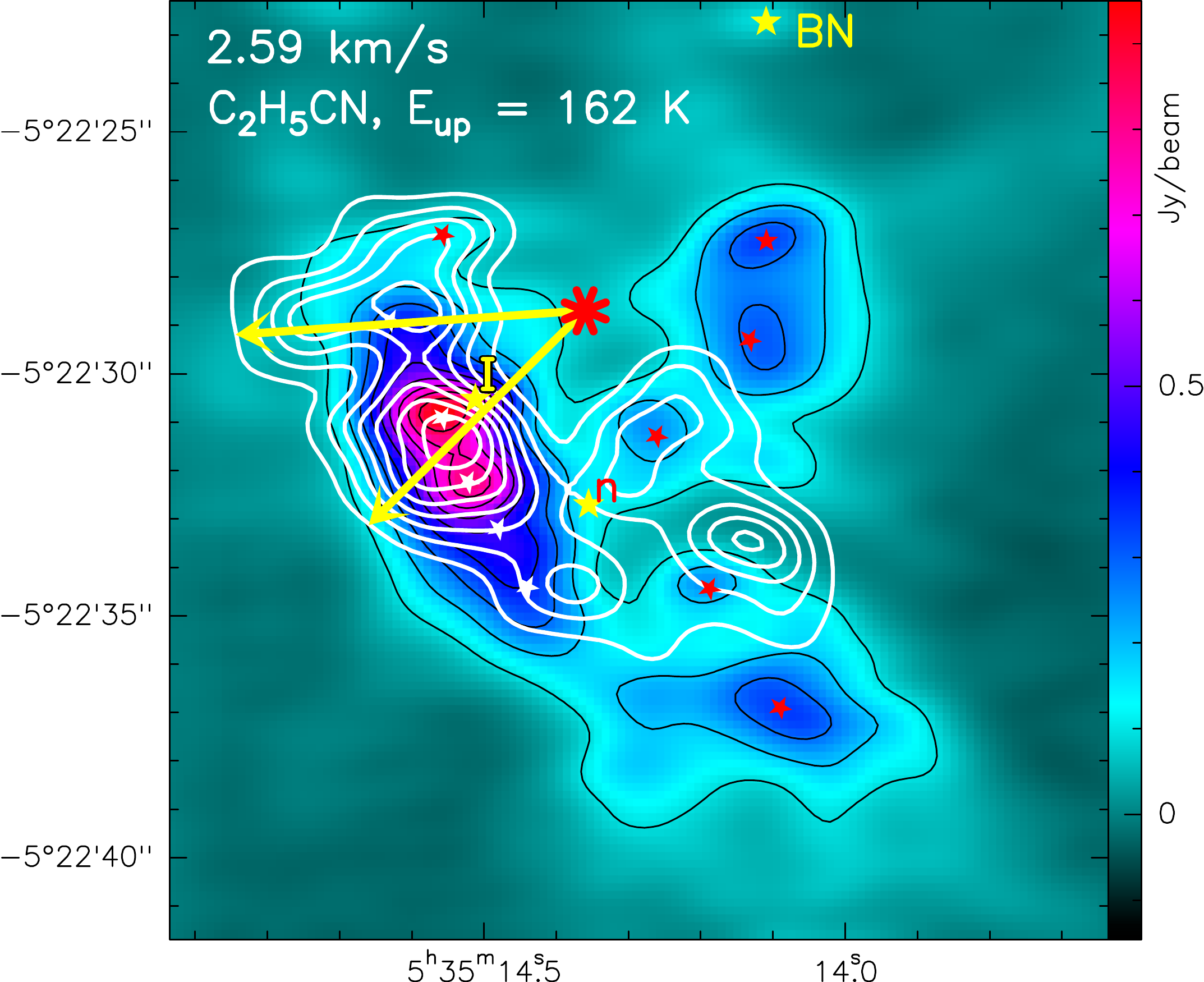}
 \includegraphics[scale = 0.248, trim=85 20 0 0,clip]{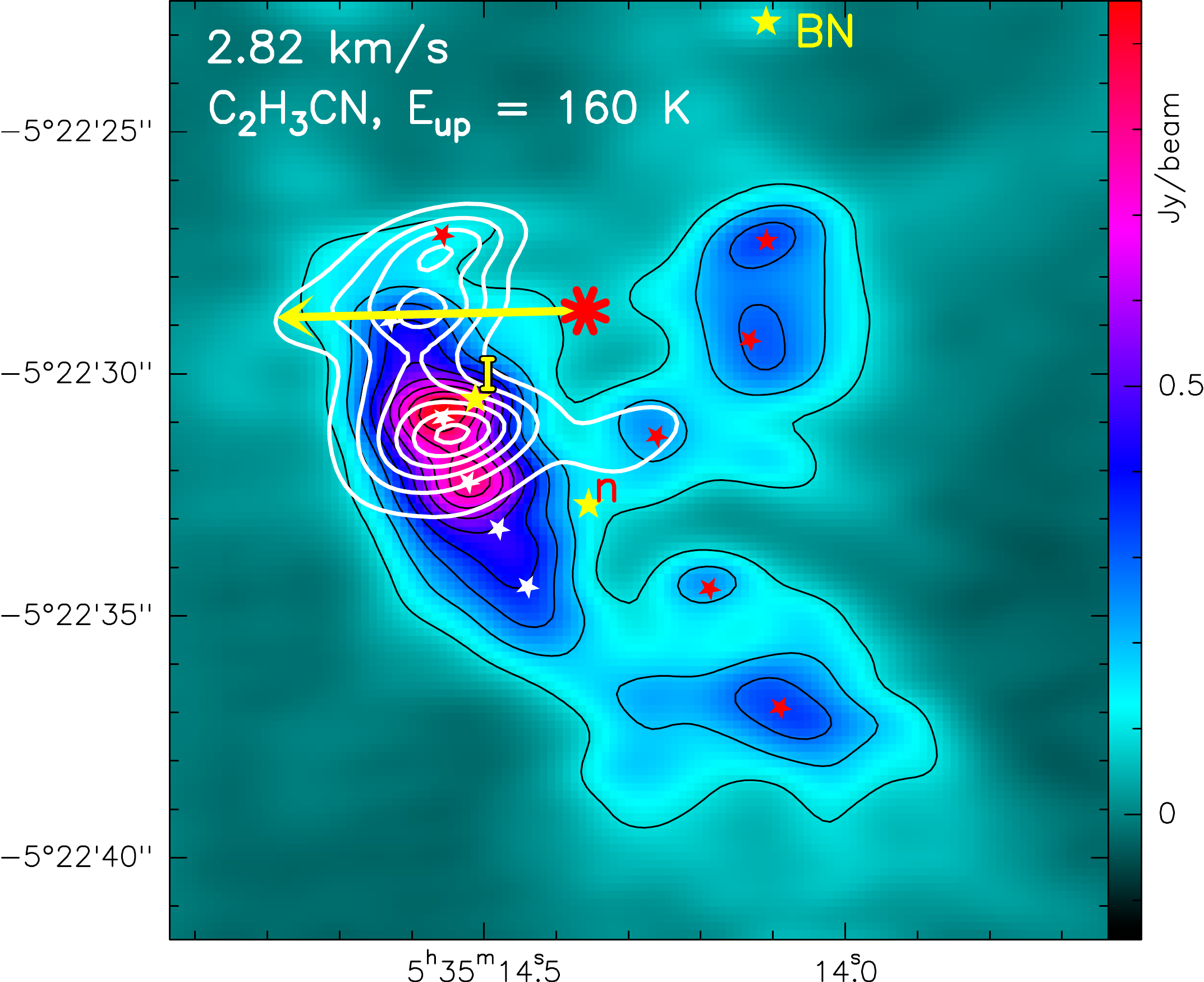}
 
  \includegraphics[scale = 0.248, trim=0 20 49 0,clip]{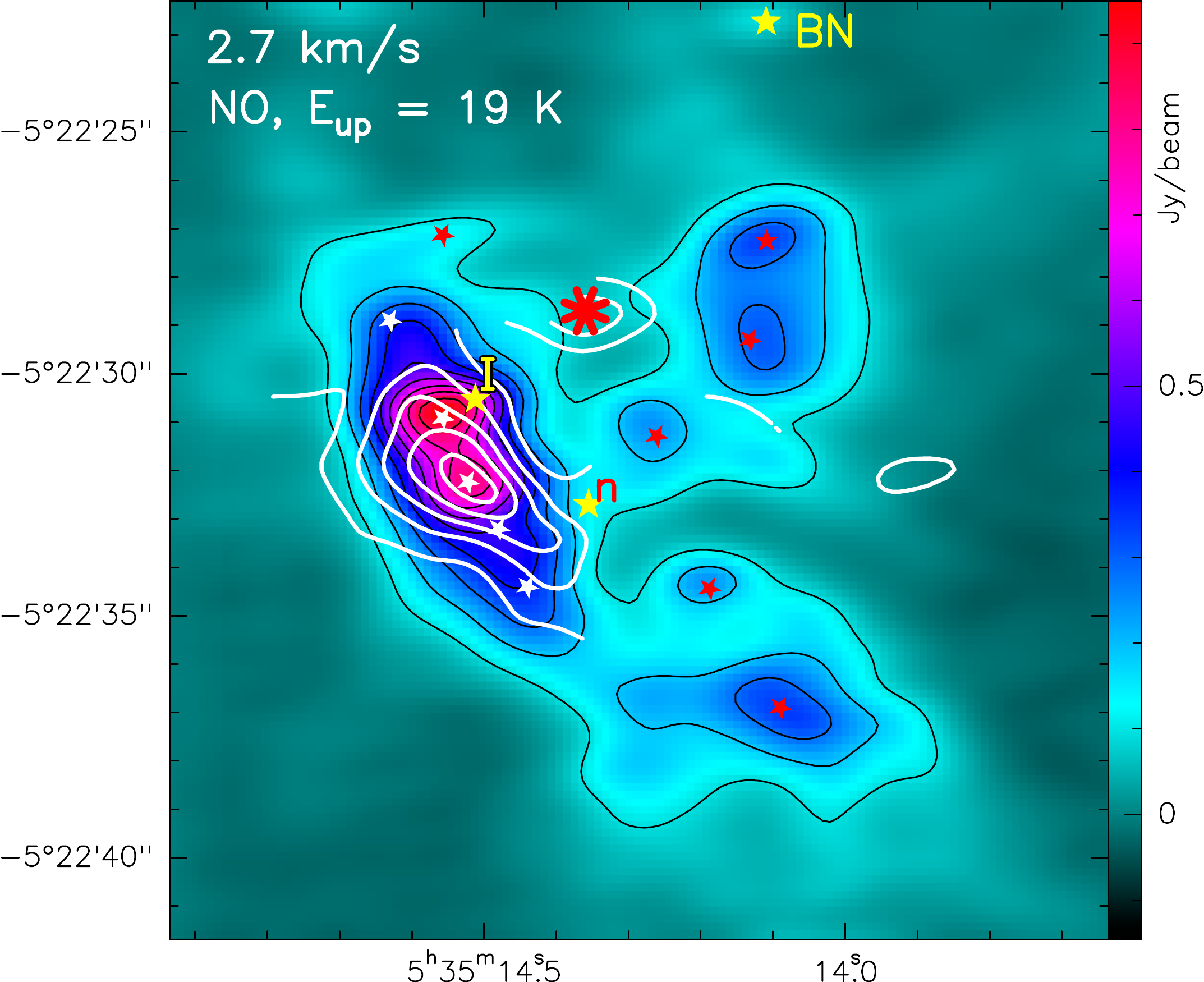}
  \includegraphics[scale = 0.248, trim=85 20 49 0,clip]{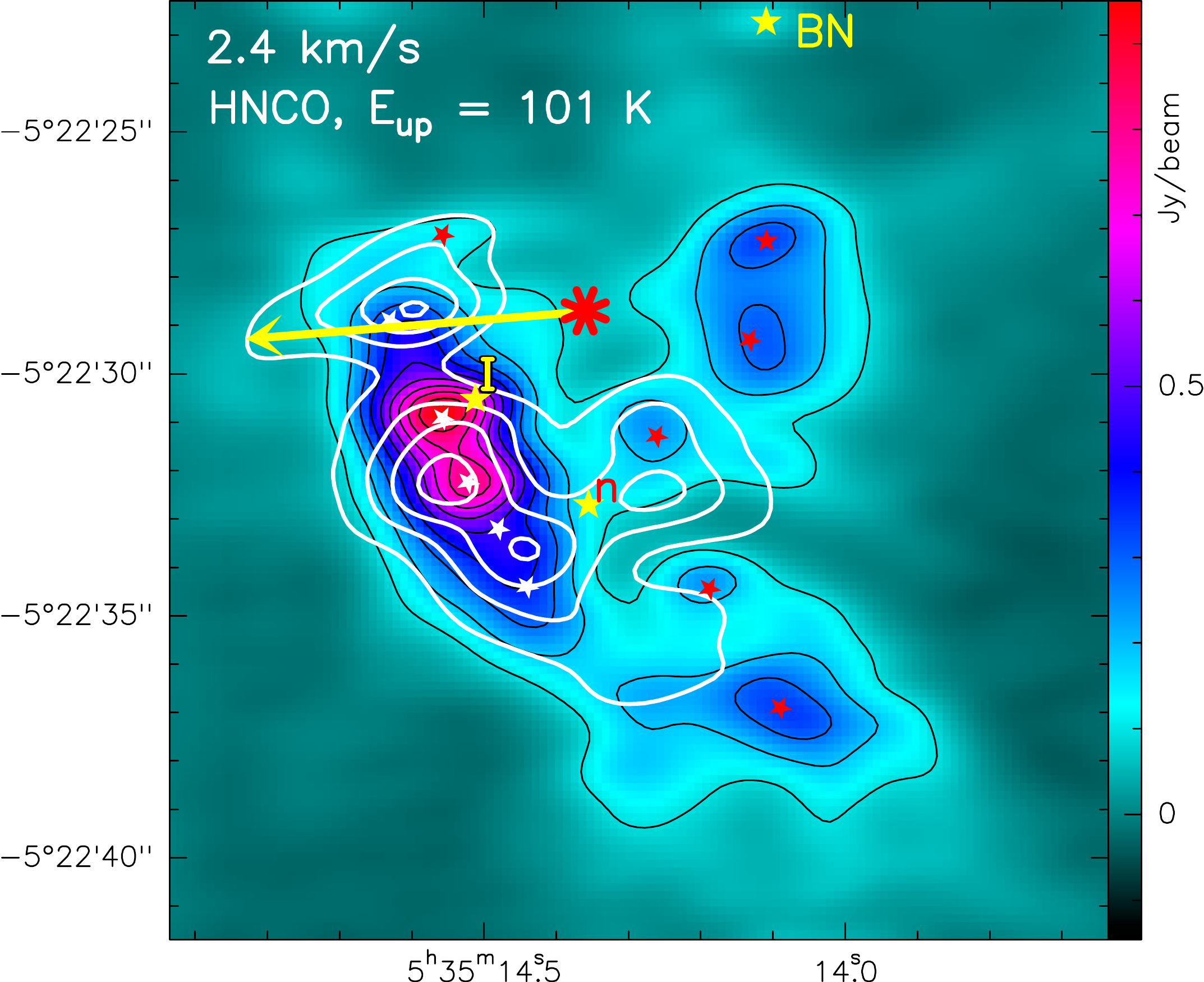}
 \includegraphics[scale = 0.248, trim=85 20 48 0,clip]{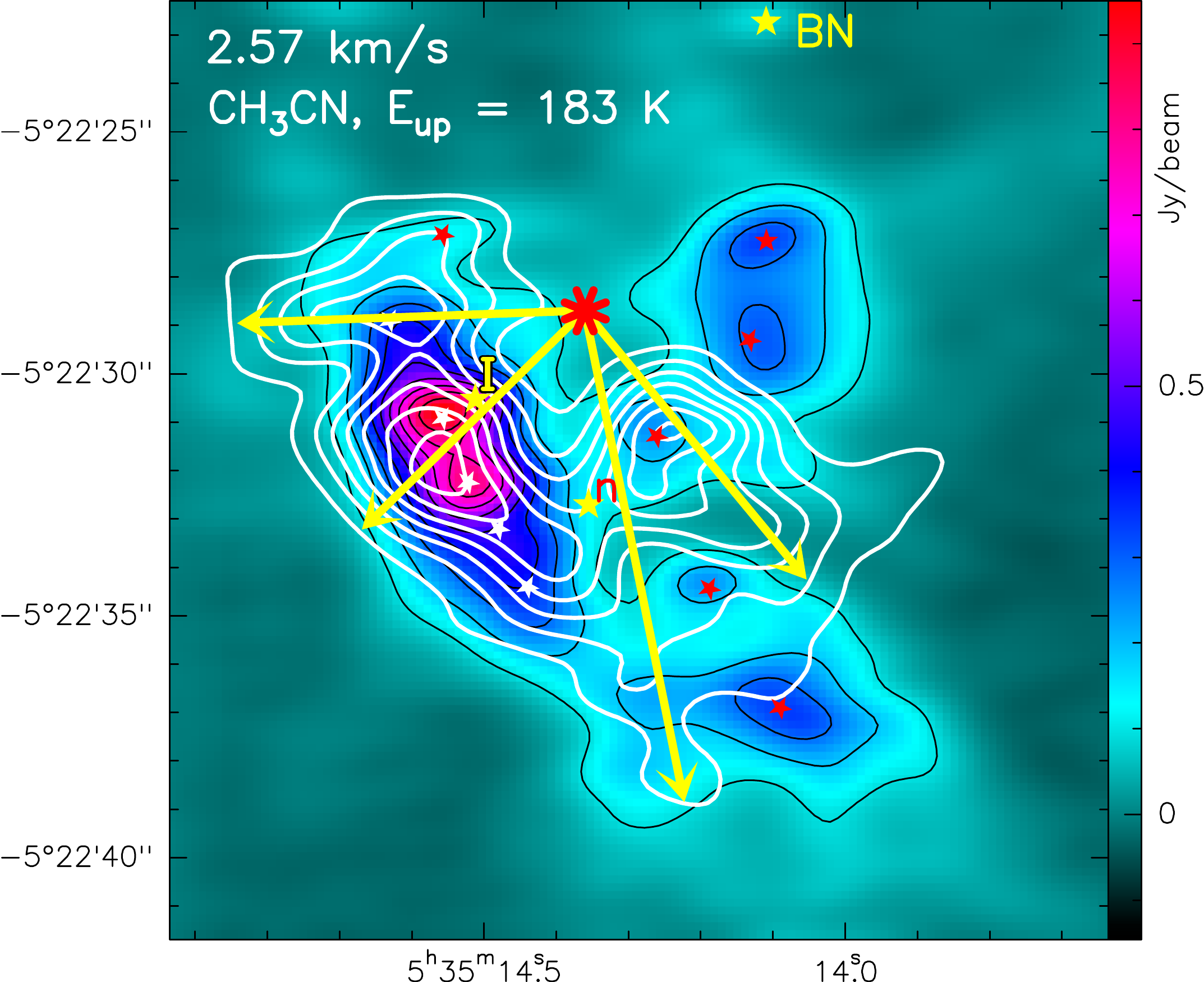}
 \includegraphics[scale = 0.248, trim=85 20 0 0,clip]{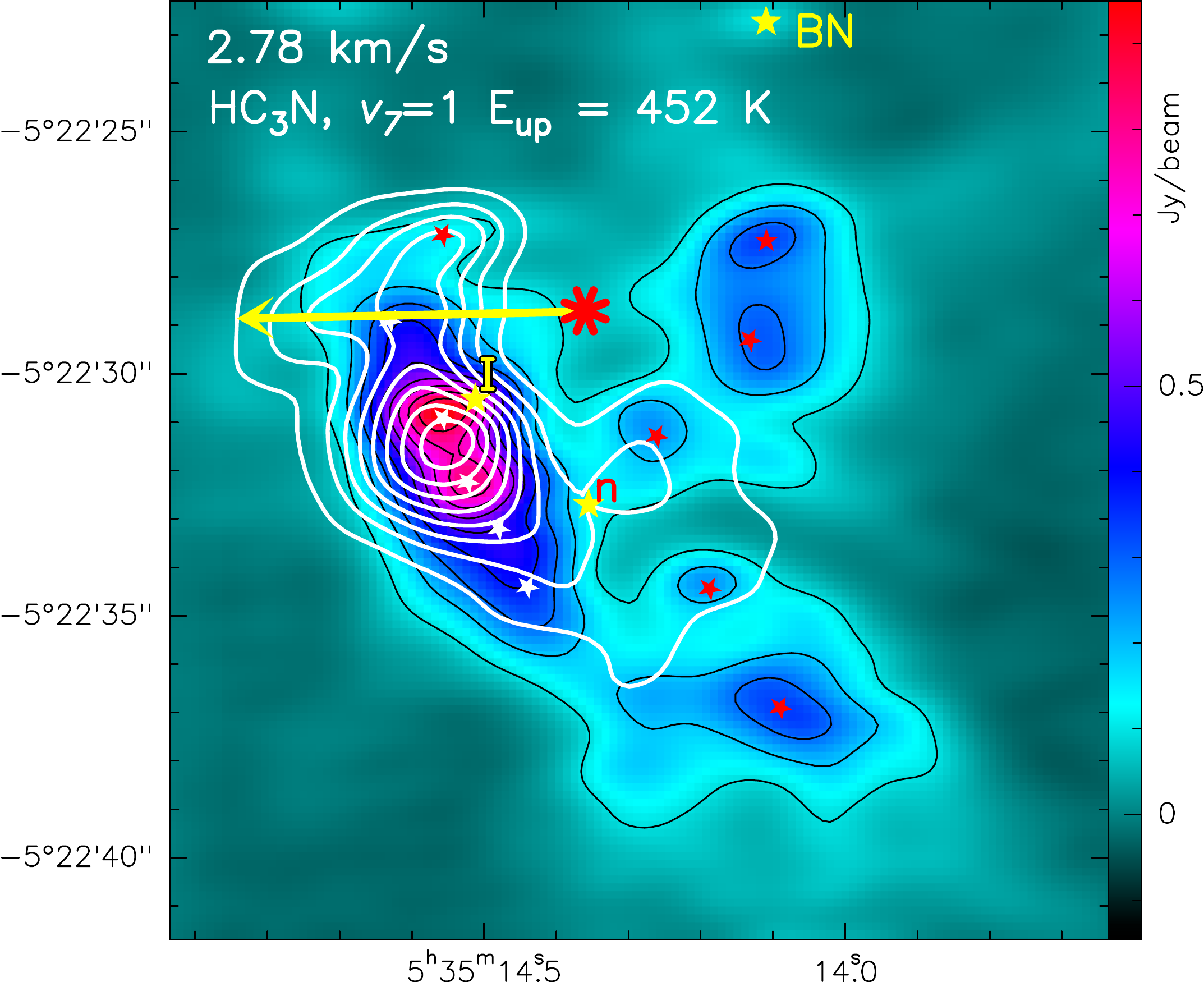}
 
 \includegraphics[scale = 0.248, trim=-40 20 0 0,clip]{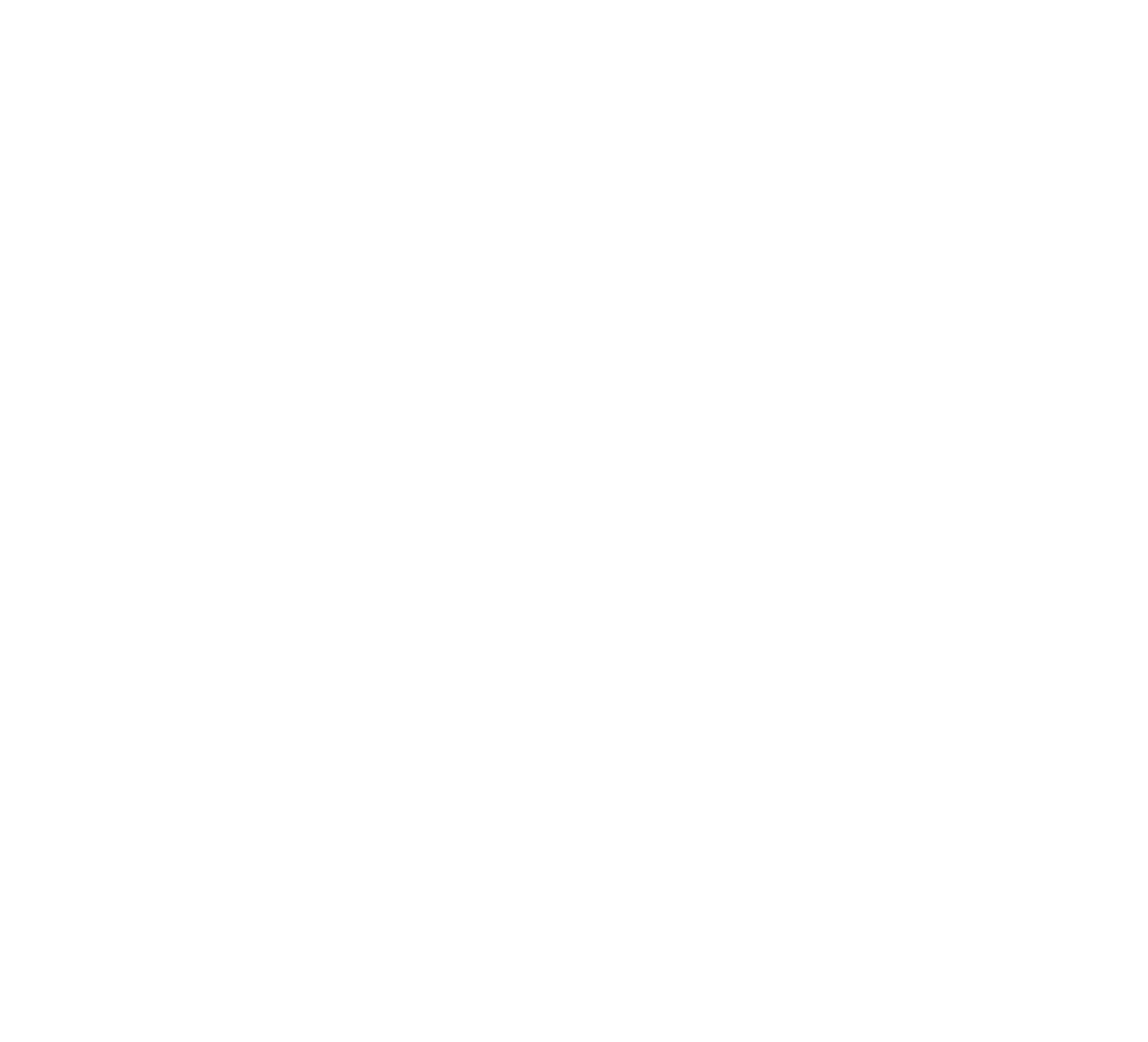}
 \includegraphics[scale = 0.248, trim=85 -10 48 0,clip]{{BNKL_H2CNH_cont+plan_28}.pdf}
 \includegraphics[scale = 0.248, trim=85 -10 48 0,clip]{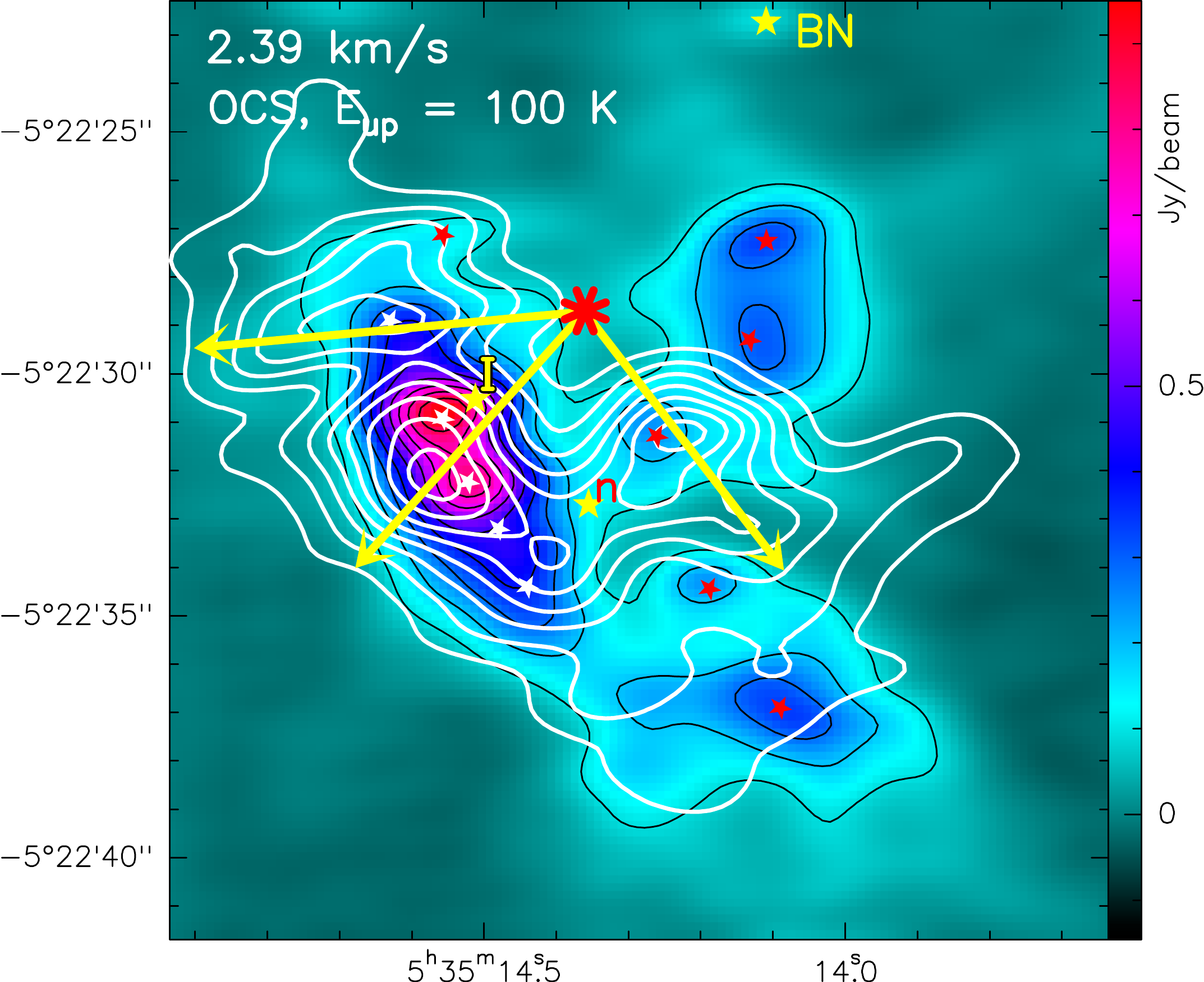}
 \includegraphics[scale = 0.248, trim=85 -10 0 0,clip]{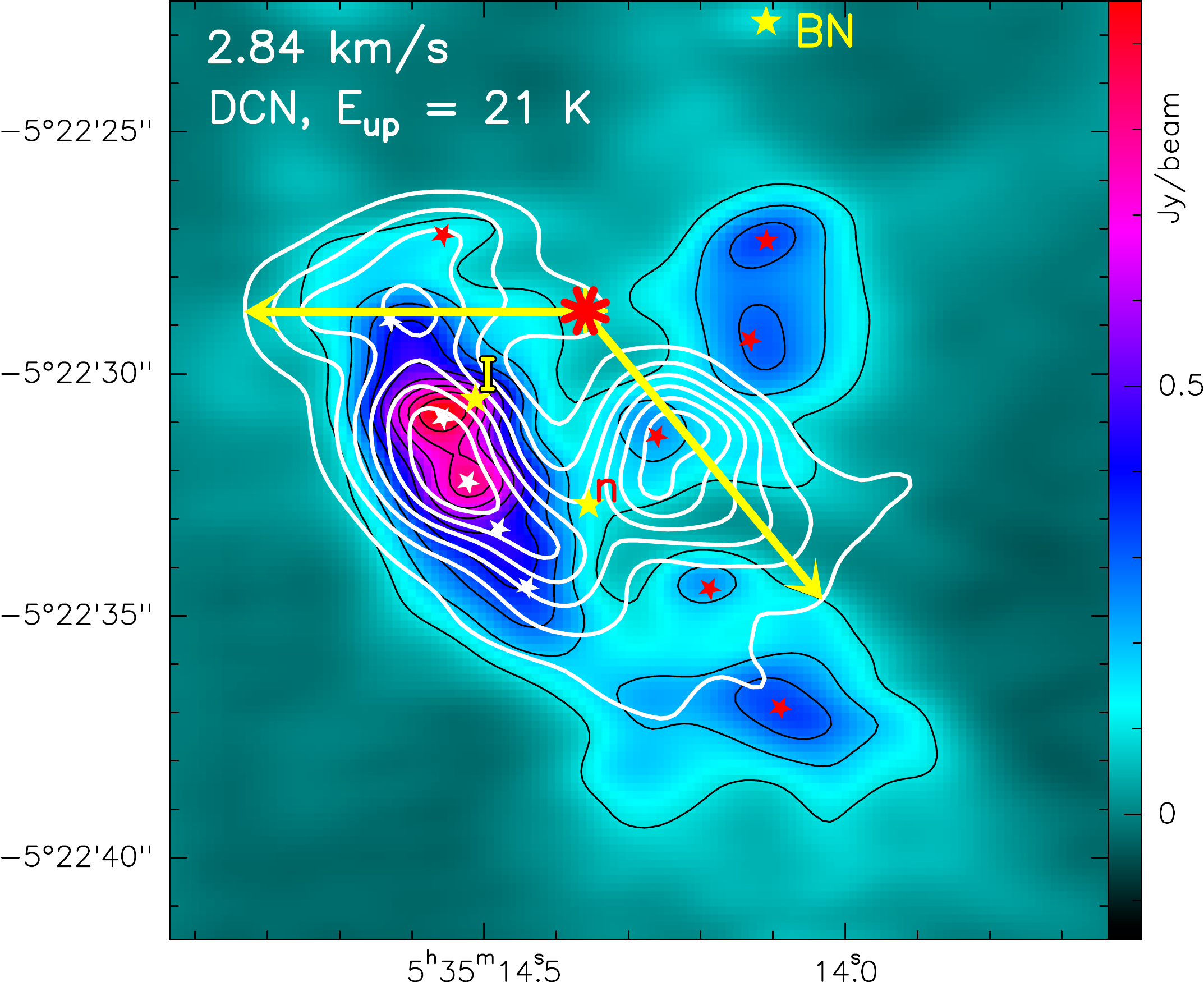}
 
 \includegraphics[scale = 0.248, trim=0 20 49 0,clip]{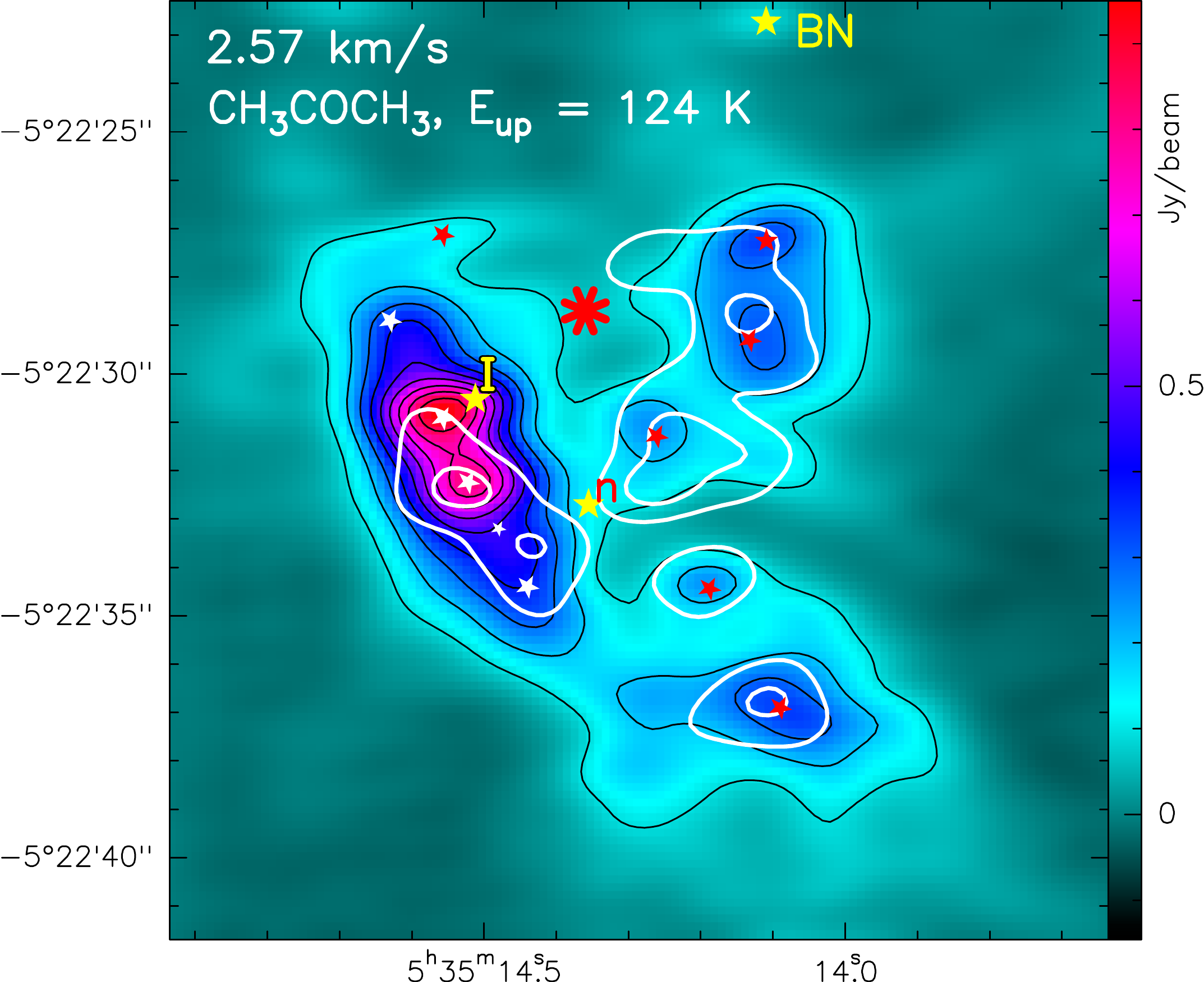}
 \includegraphics[scale = 0.248, trim=85 20 48 0,clip]{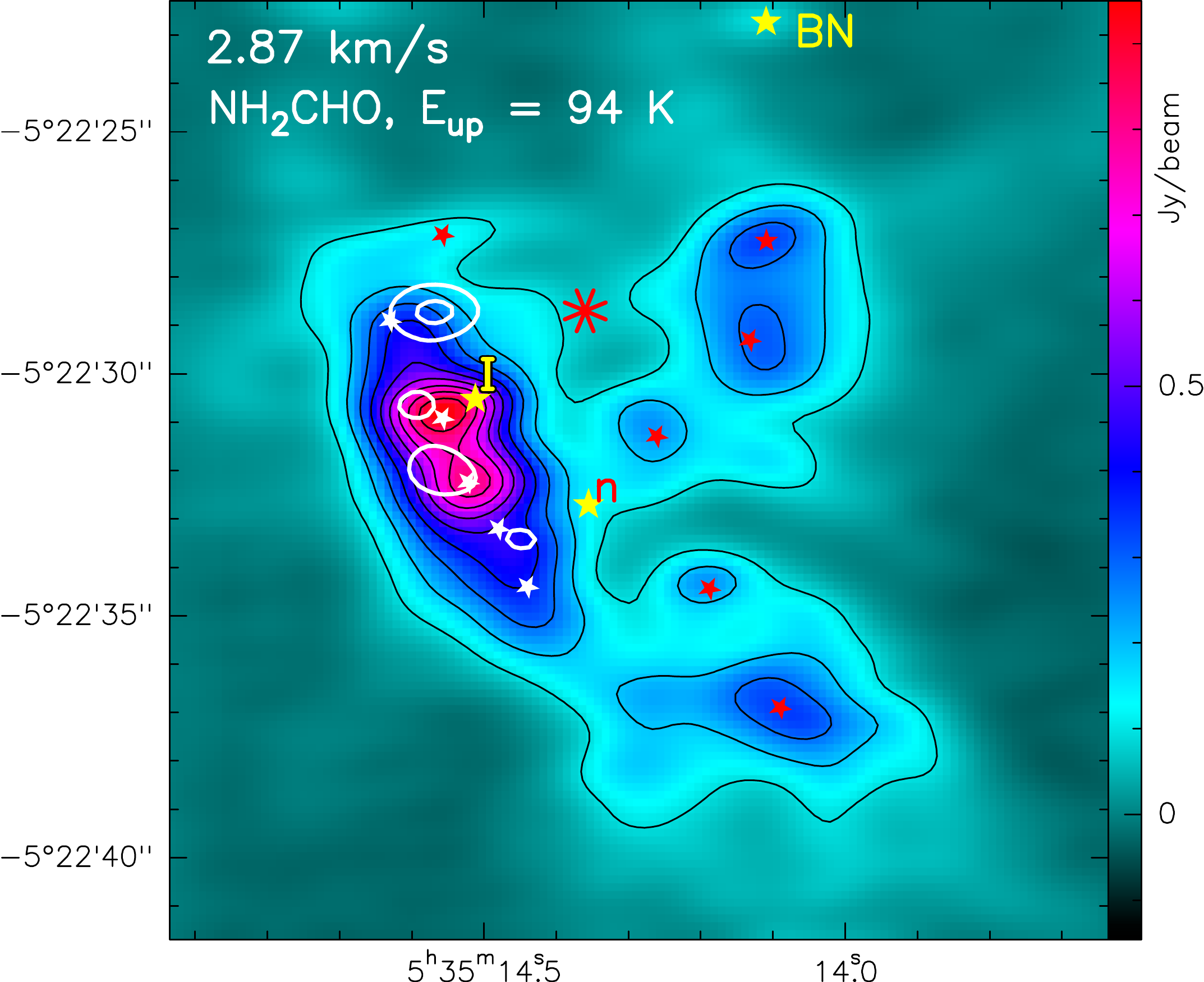}
 \includegraphics[scale = 0.248, trim=85 20 49 0,clip]{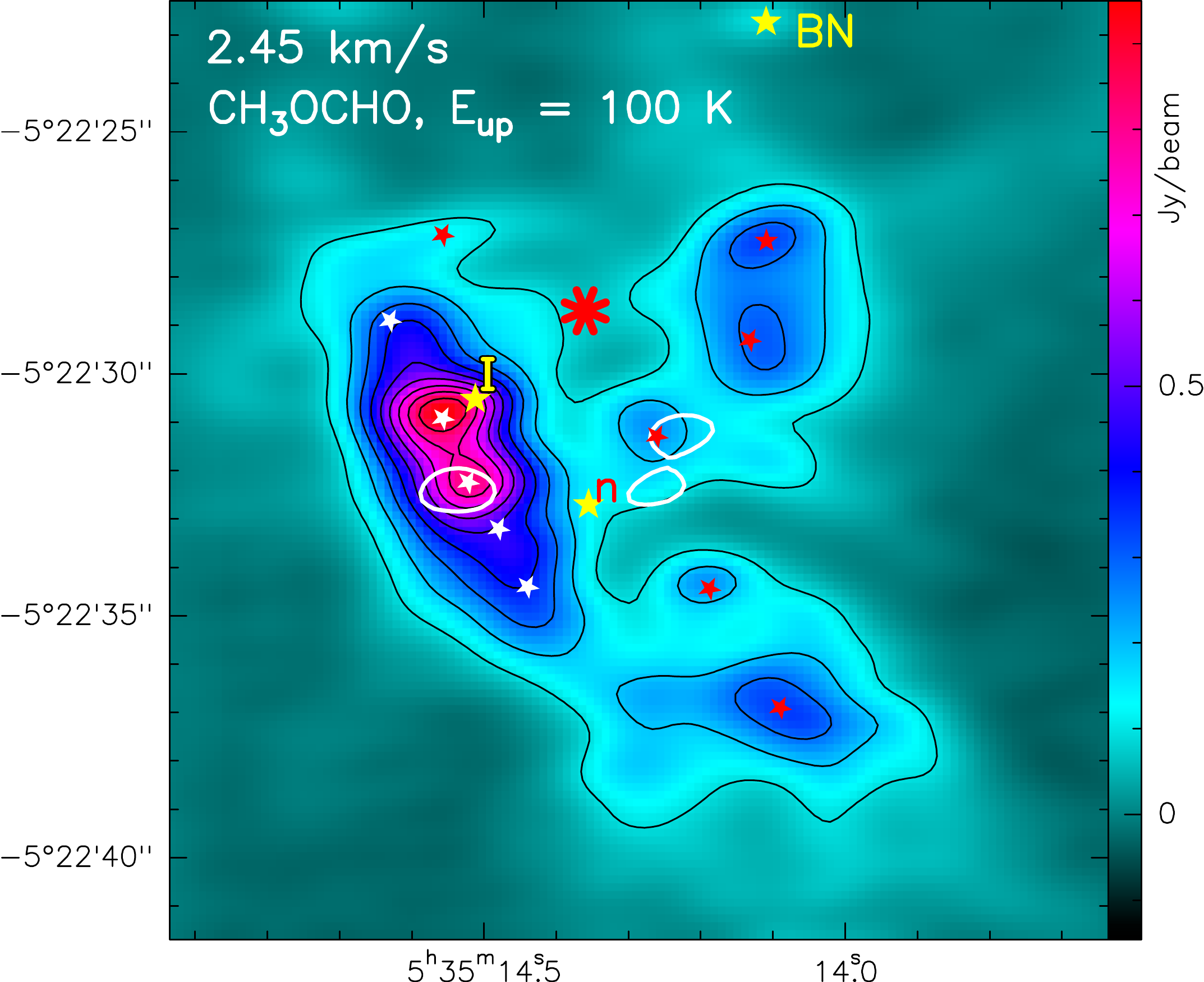}
 \includegraphics[scale = 0.248, trim=85 20 0 0,clip]{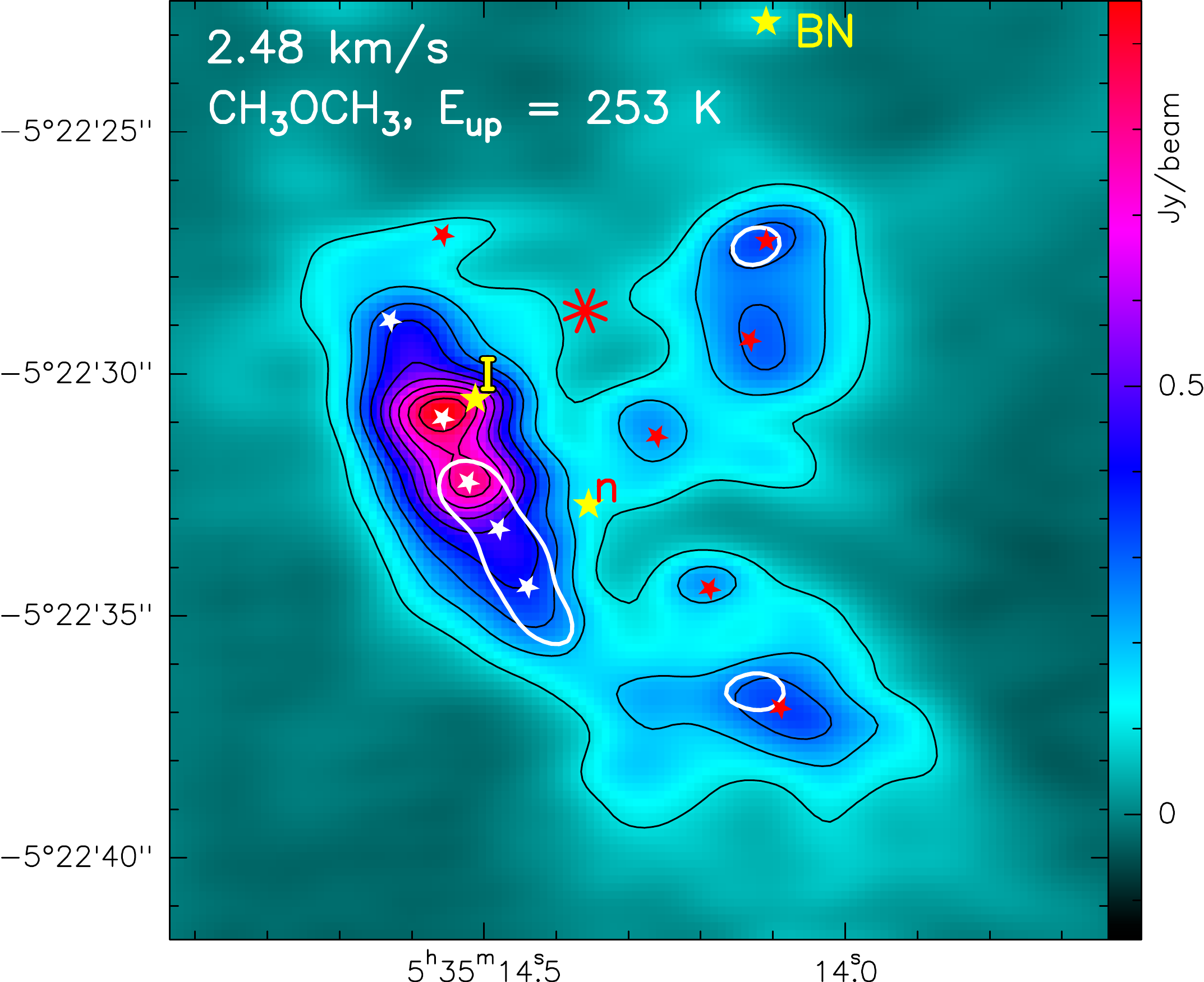}

 \includegraphics[scale = 0.248, trim=-200 20 0 0,clip]{carre_blanc.pdf}
 \includegraphics[scale = 0.248, trim=-200 20 0 0,clip]{carre_blanc.pdf}
 \includegraphics[scale = 0.248, trim=0 0 0 0,clip]{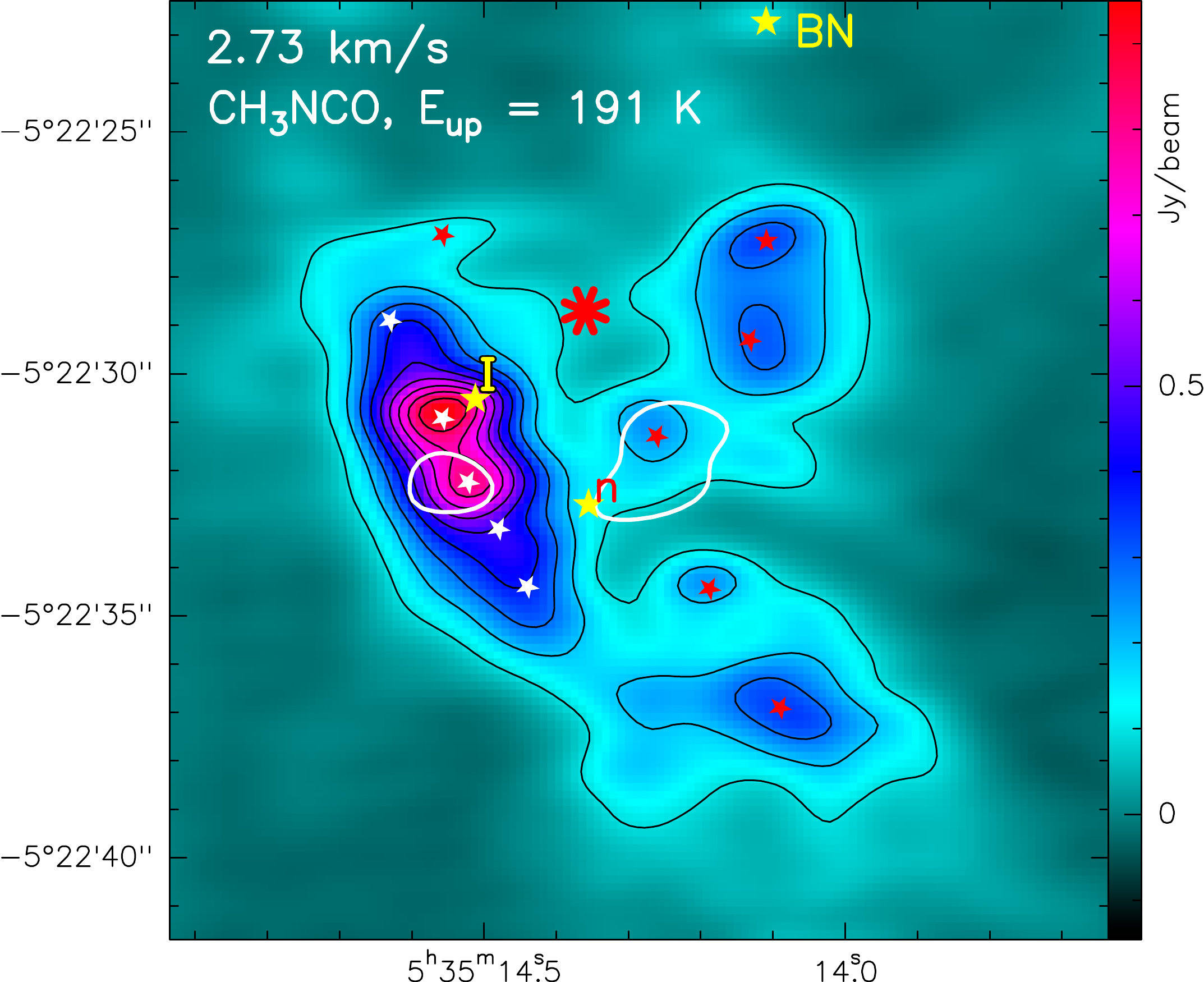}

  \caption{Selection of  species at 3 different velocities (here $\sim$2.5 \kmps).  There are no data for D$_2$CO and NH$_2$D
  due to line blending and for C$_2$H$_5$OH and c-C$_2$H$_4$O owing to lack of emission, i.e., maximum signal less than 10\% of the
  peak emission.  NO is partly masked, due to line blending. The top left corner indicates the channel velocity, the species, and
  its upper energy level. The white contours are 10 to 90\% of the peak emission of the strongest channel for that species. The
  yellow arrows starting from the explosion center  (red eight-pointed star) suggest possible displacement of gas linked to the
  explosive event, which occurred $\sim$550 years ago.  The species shown in the  top three rows show such extensions, while
  the two bottom rows do not. The molecular species transitions are listed in Table \ref{Tab:molfreq}. {All  are channel maps of 488 kHz width (0.59 to 0.69\,\kmps)}}
 
  \label{fig:2.5kmps}
 \end{figure*}
 
 \begin{figure*}[h!]
 \centering 
 \includegraphics[scale = 0.254,trim=0 20 49 0,clip]{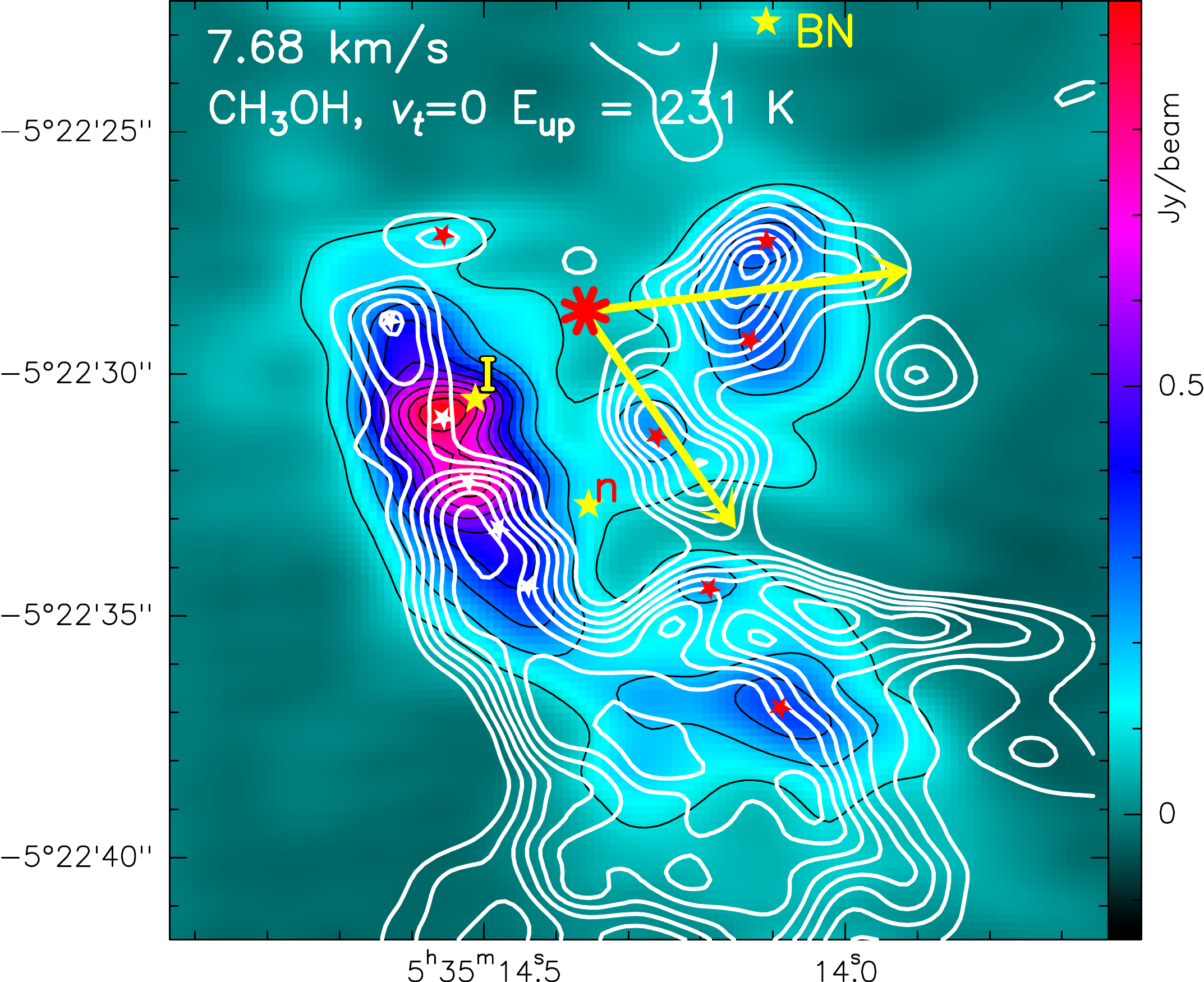}
 \includegraphics[scale = 0.254, trim=85 20 48 0,clip]{BNKL_H2CO_cont+plan_18.pdf}
 \includegraphics[scale = 0.254, trim=85 20 48 0,clip]{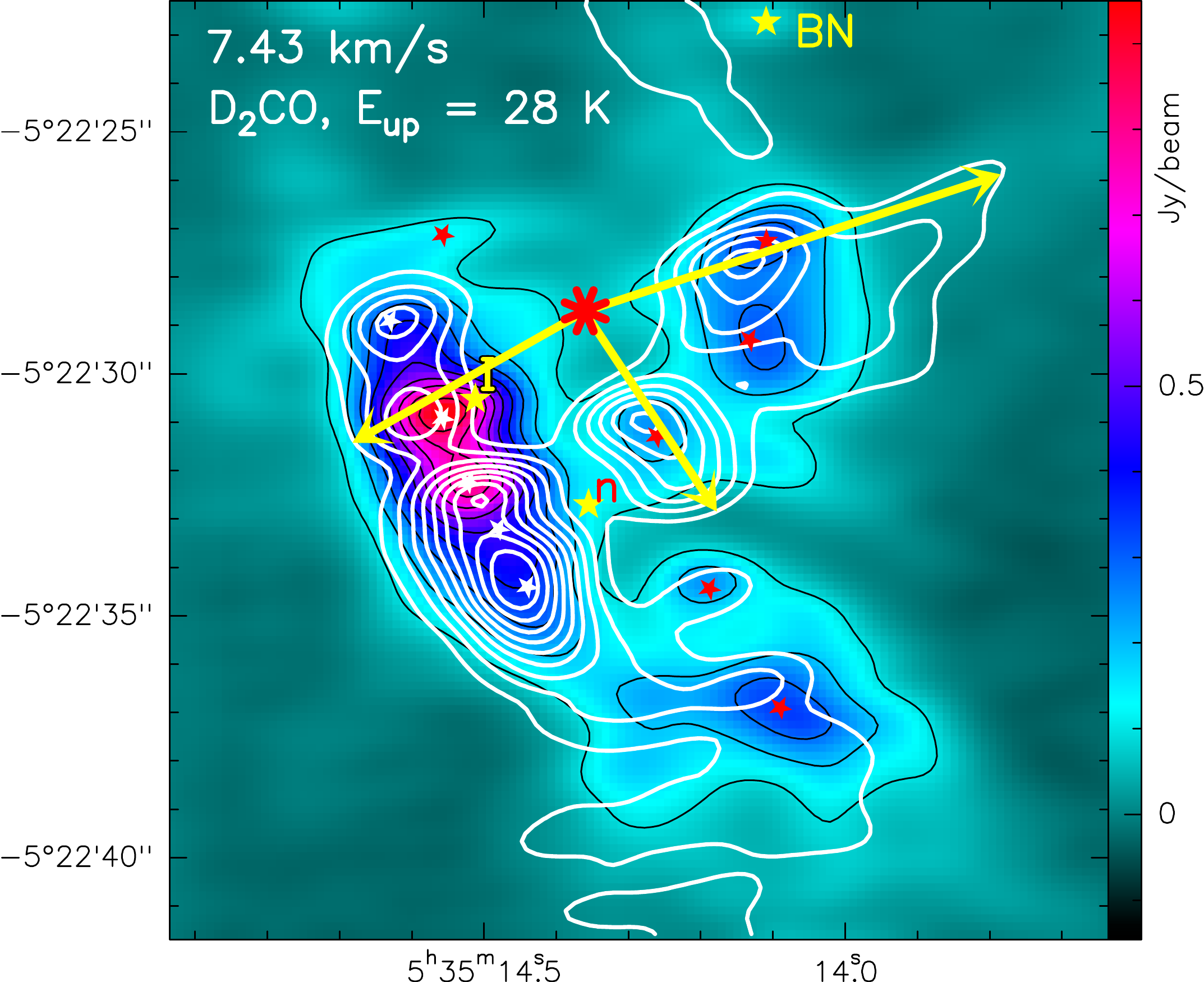}
  \includegraphics[scale = 0.254, trim=85 20 48 0,clip]{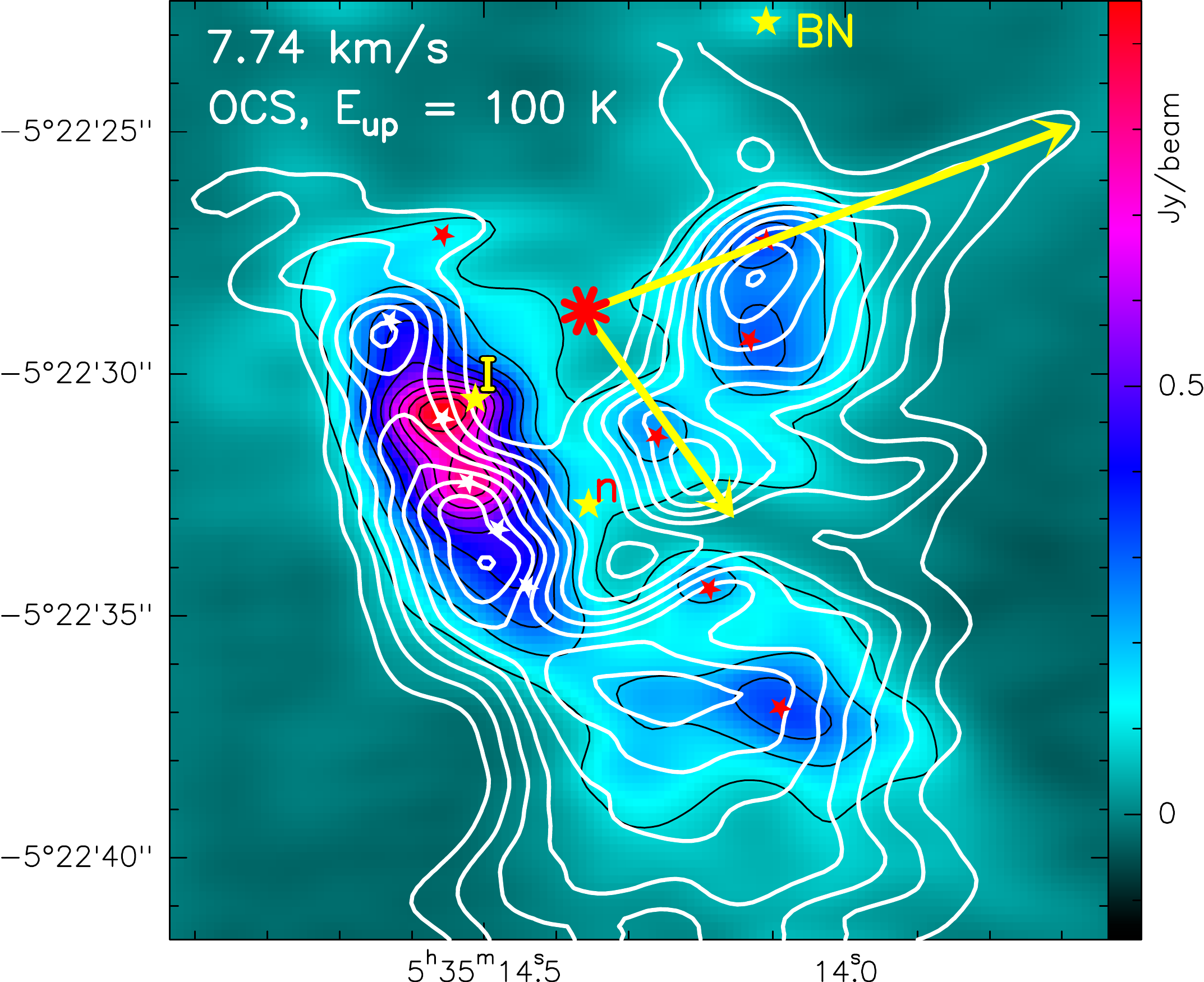}
 
 \includegraphics[scale = 0.254, trim=0 20 49 0,clip]{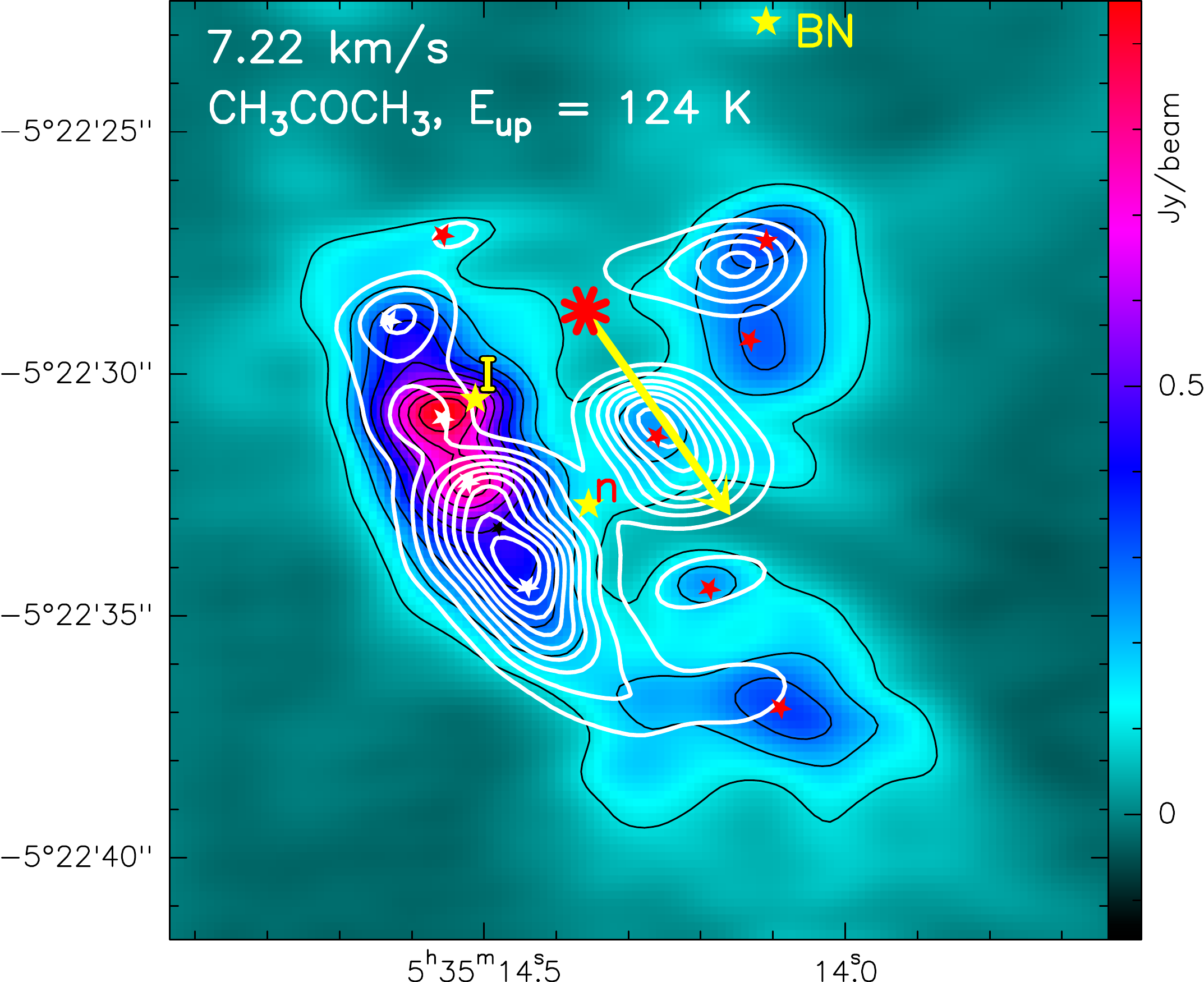}
 \includegraphics[scale = 0.254, trim=85 20 48 0,clip]{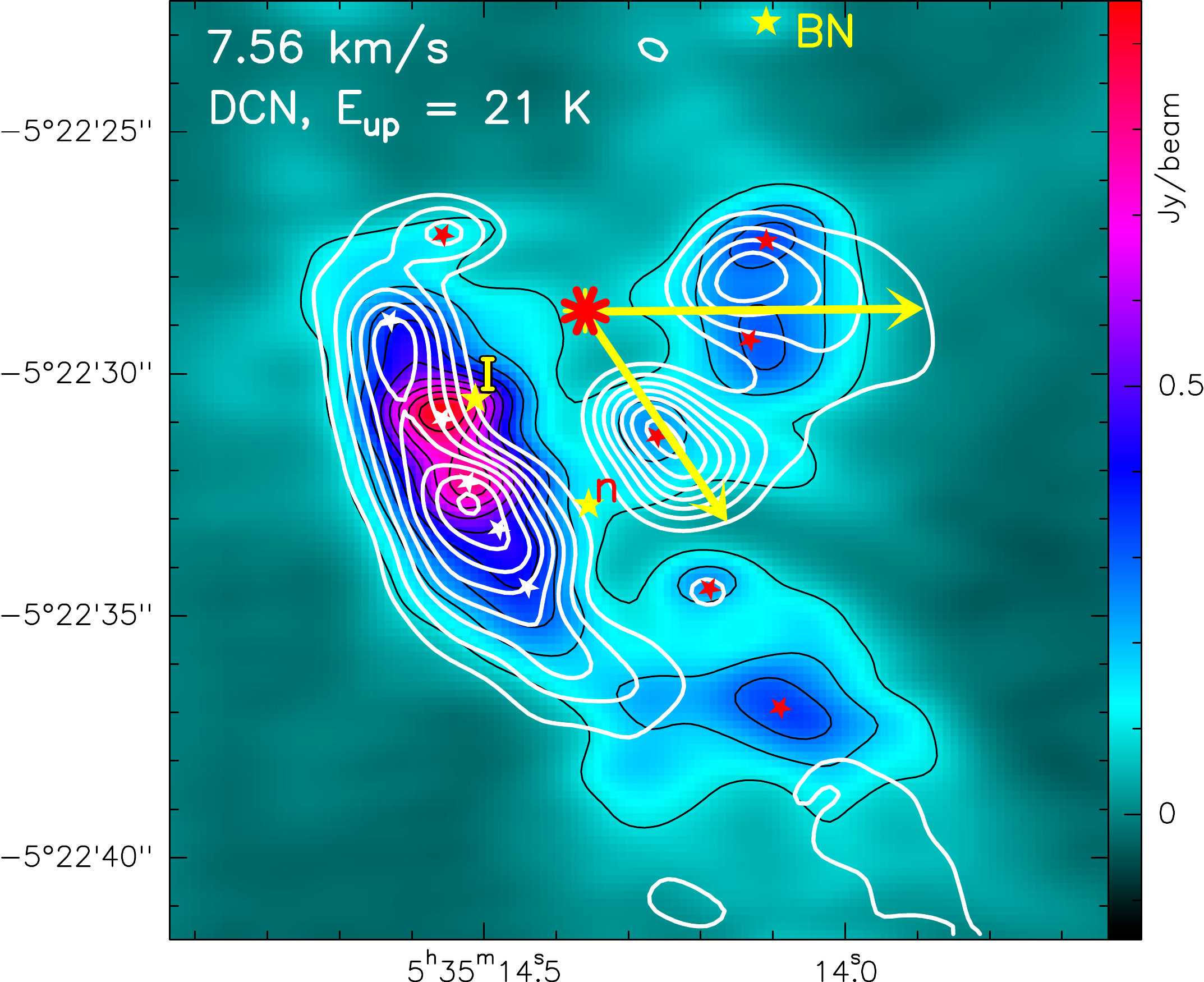}
 \includegraphics[scale = 0.254, trim=85 20 48 0,clip]{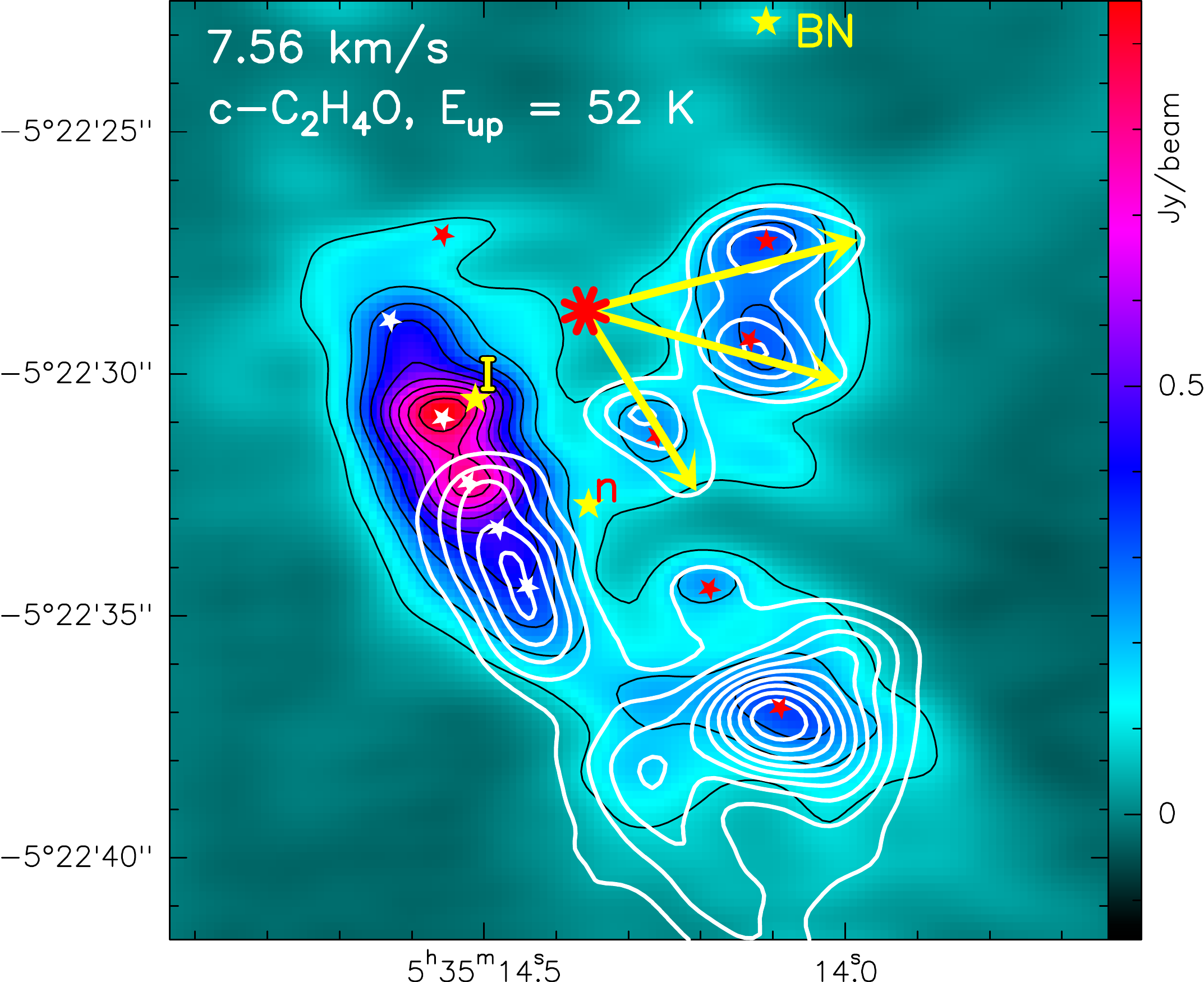}
\includegraphics[scale = 0.254, trim=85 20 49 0,clip]{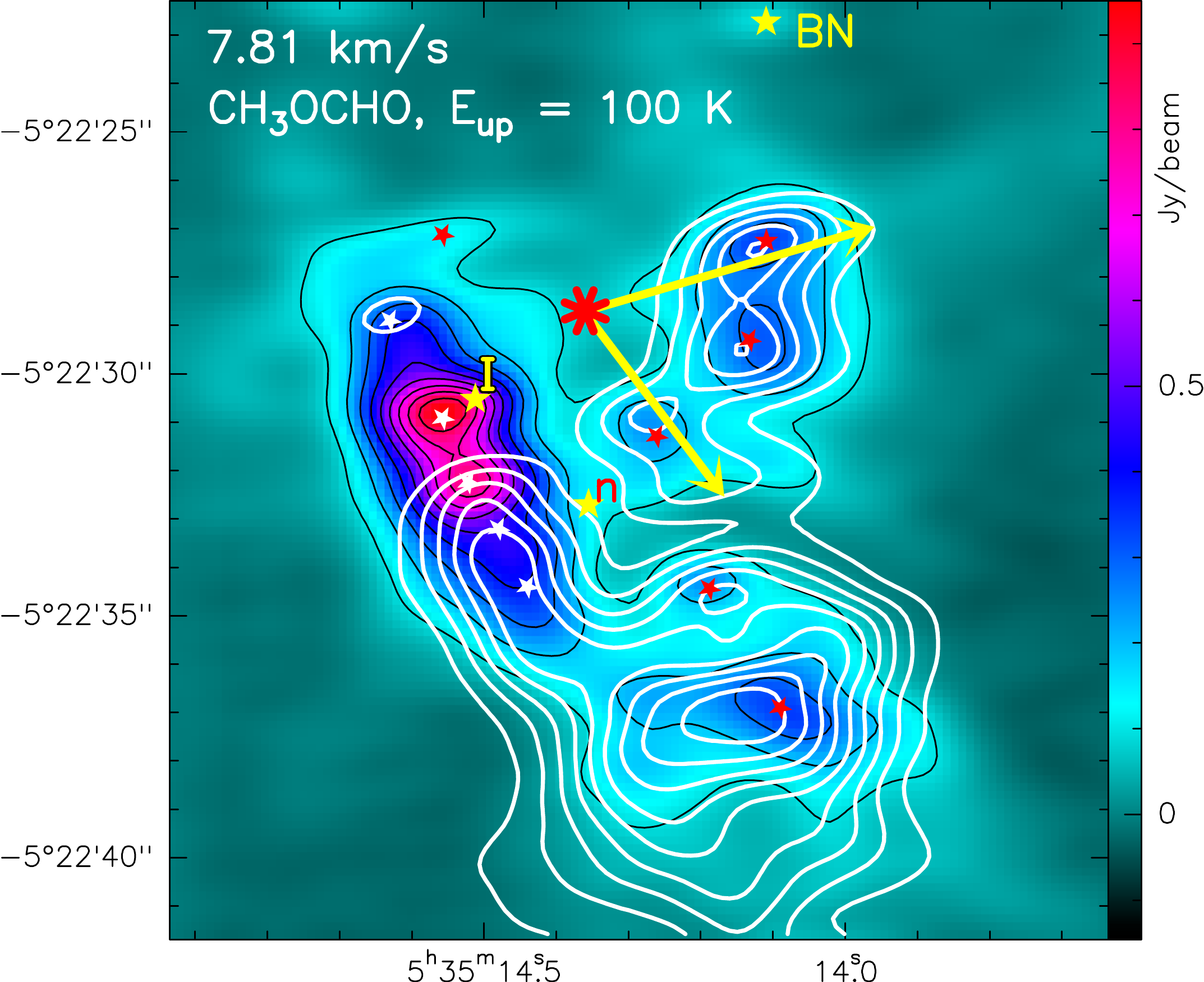}
 
  \includegraphics[scale = 0.254, trim=0 -10 49 0,clip]{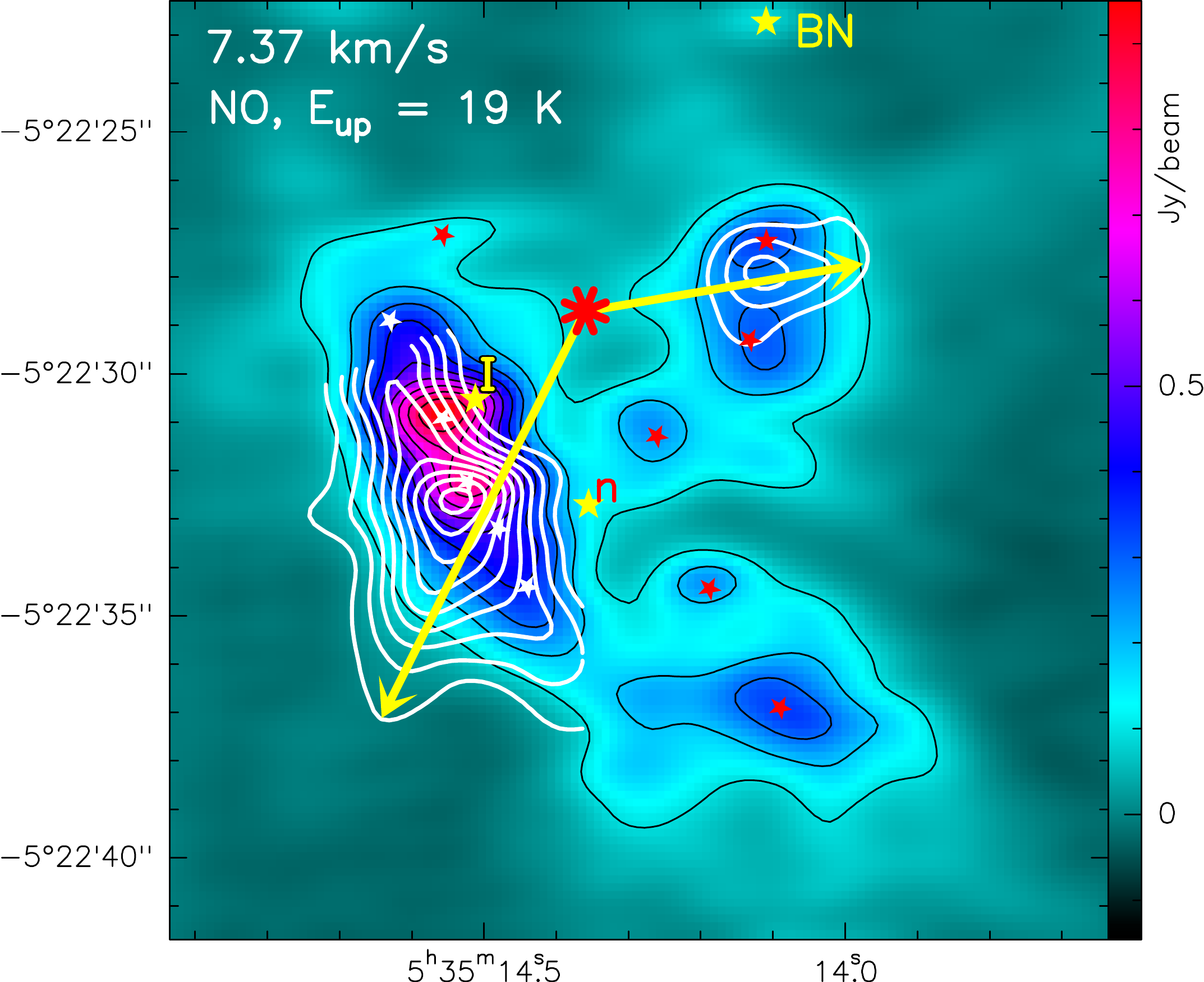}
 \includegraphics[scale = 0.254, trim=85 -10 49 0,clip]{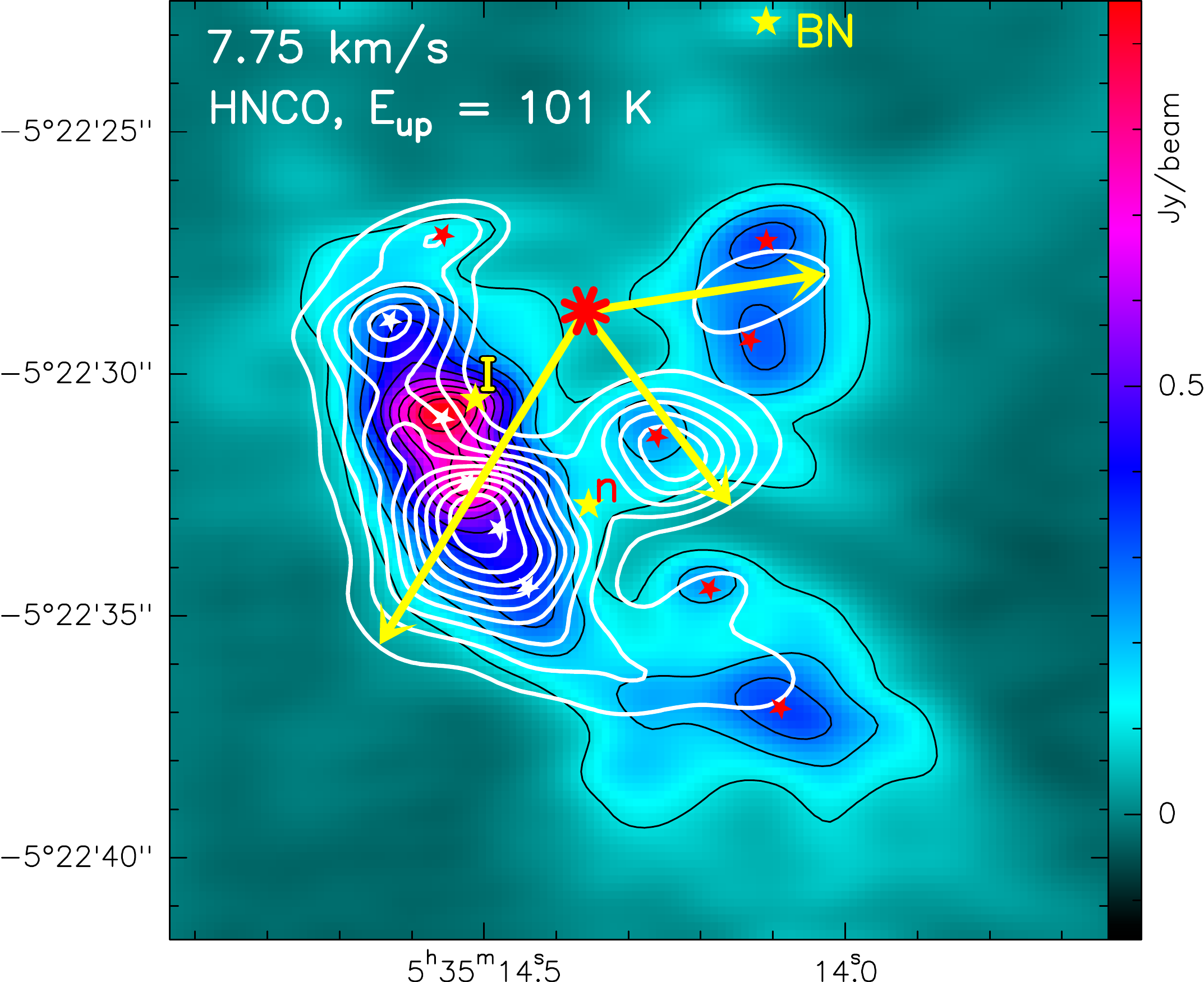}
  \includegraphics[scale = 0.254, trim=85 -10 48 0,clip]{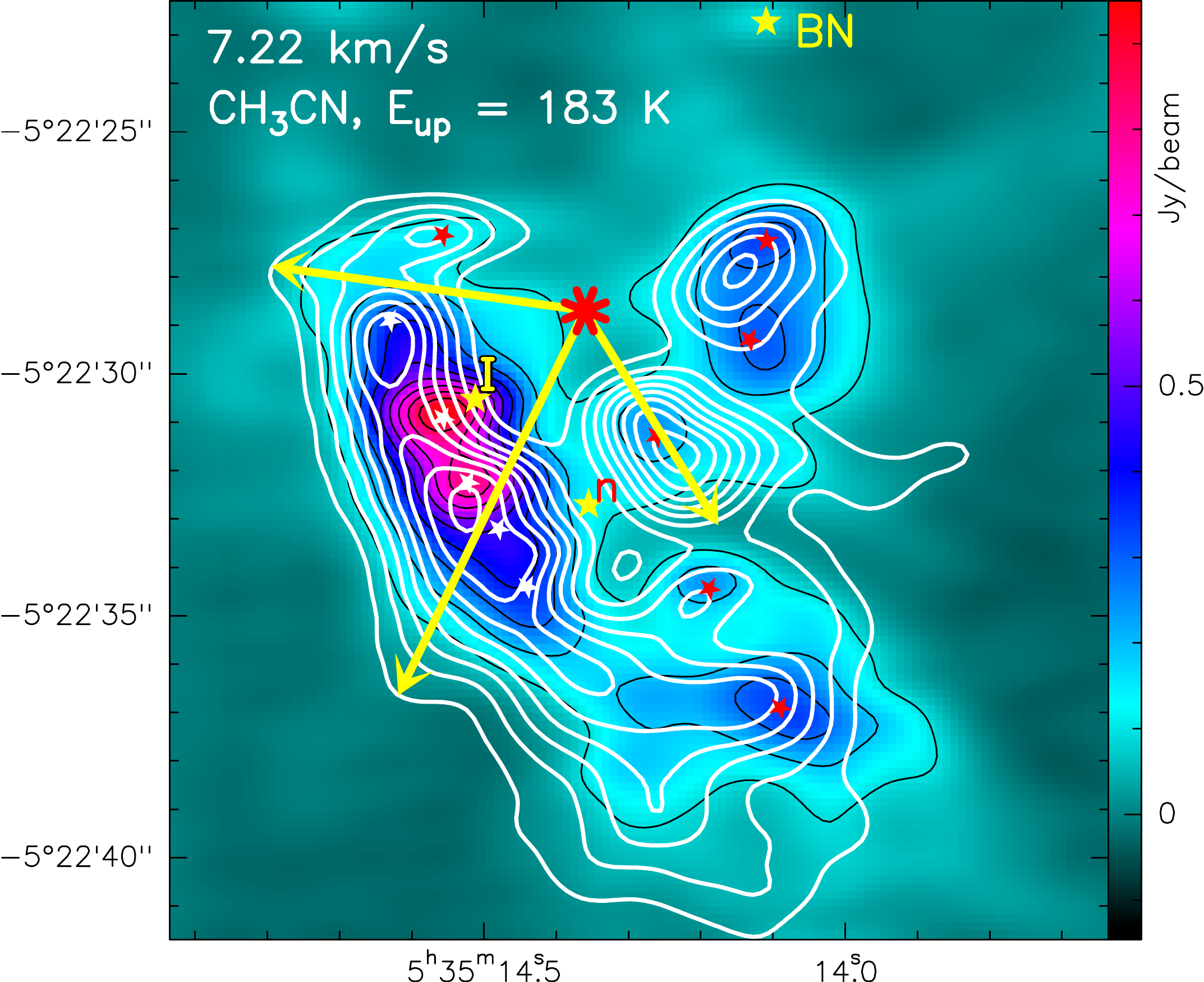}
\includegraphics[scale = 0.254, trim=85 -10 48 0,clip]{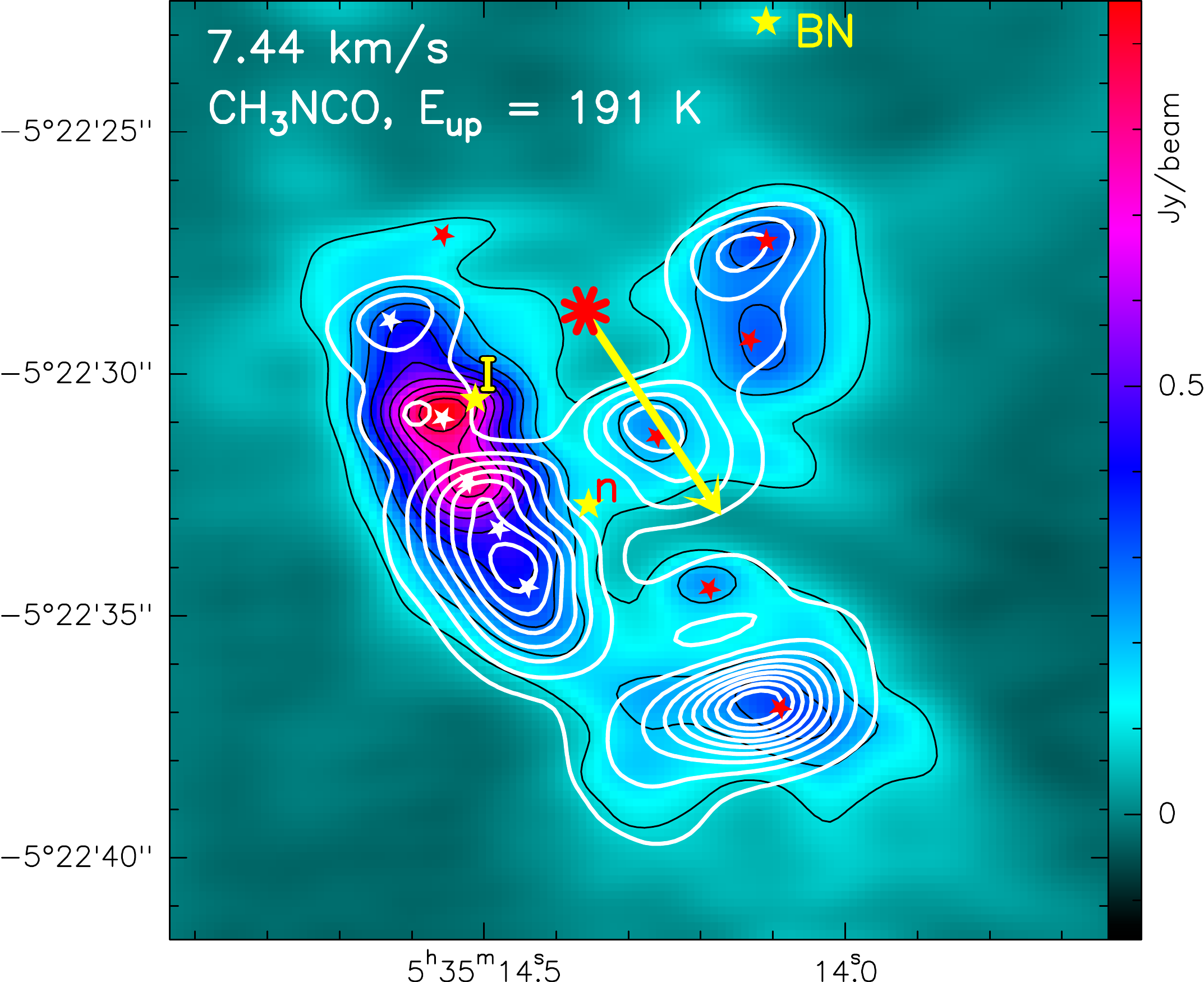}
 
 \includegraphics[scale = 0.254, trim=0 20 48 0,clip]{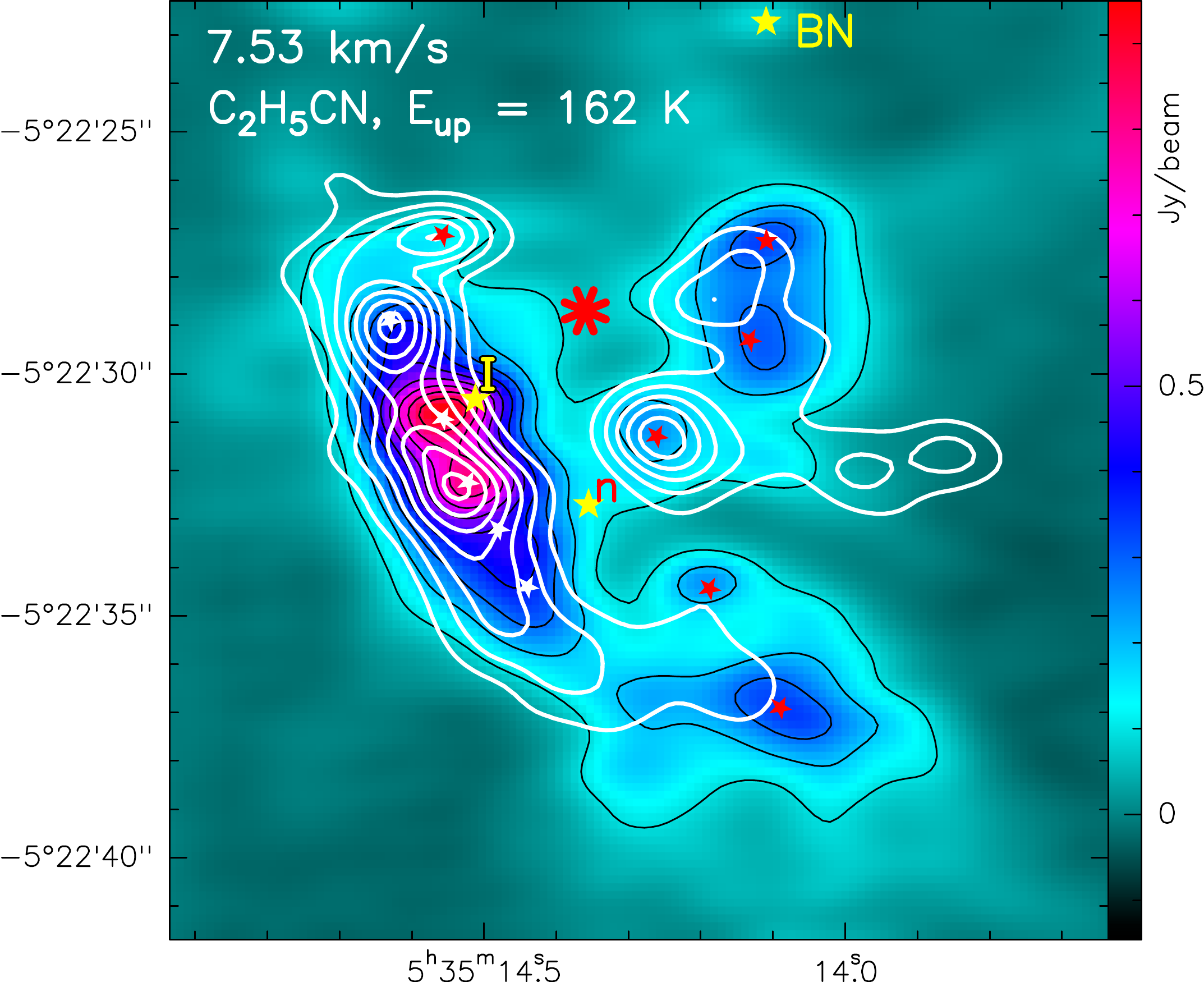}
 \includegraphics[scale = 0.254, trim=85 20 48 0,clip]{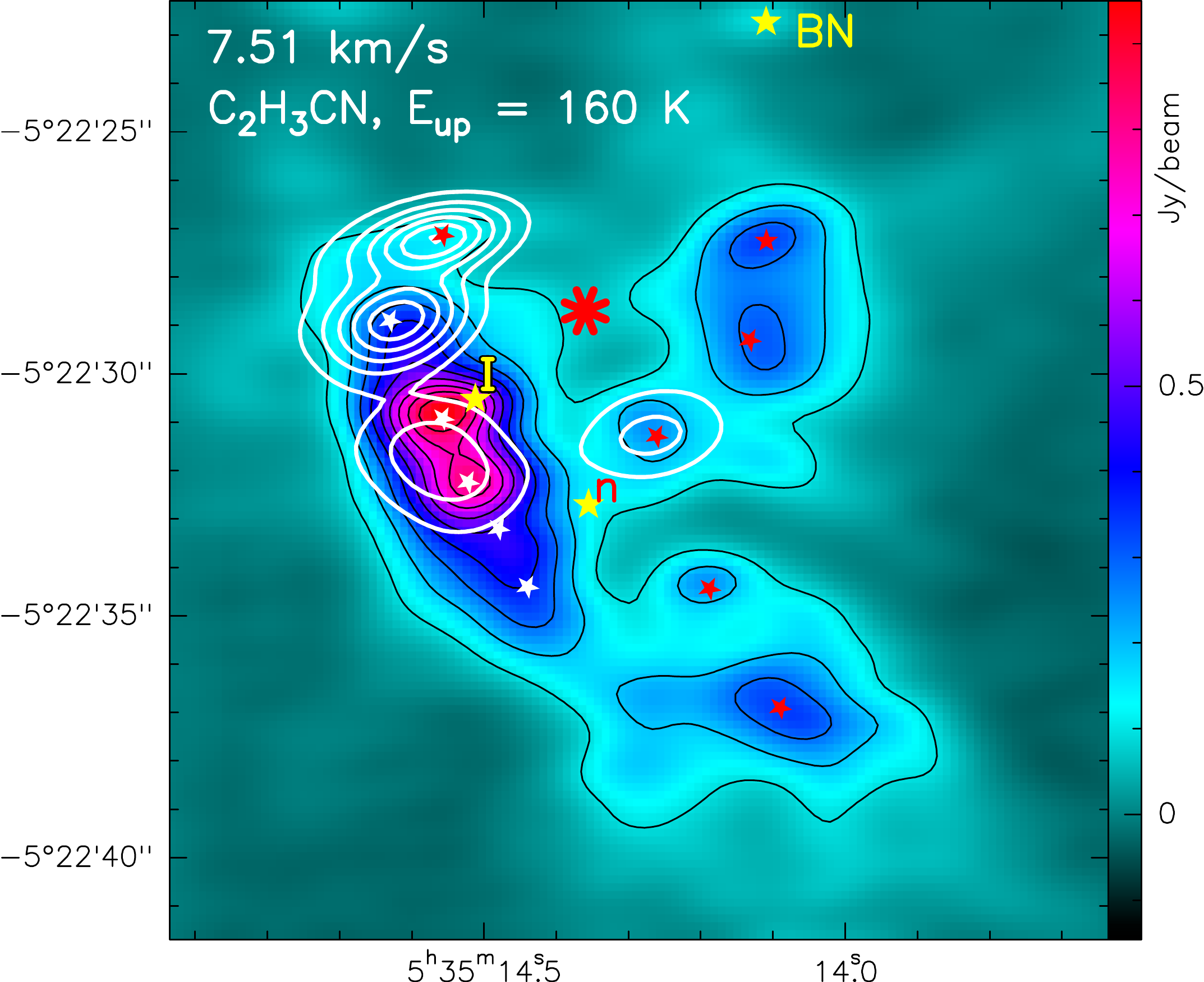}
 \includegraphics[scale = 0.254, trim=85 20 48 0,clip]{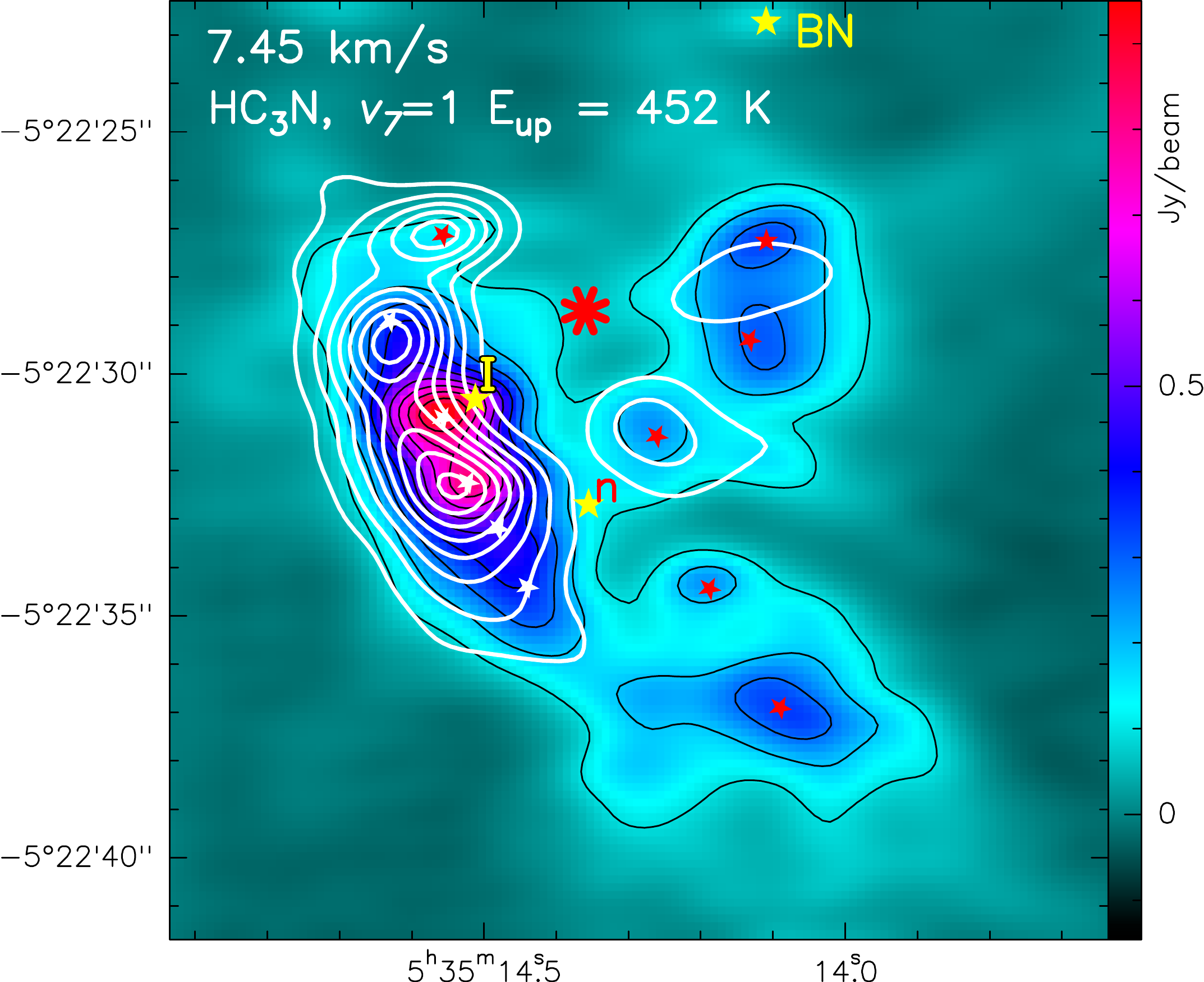}
 \includegraphics[scale = 0.254, trim=85 20 48 0,clip]{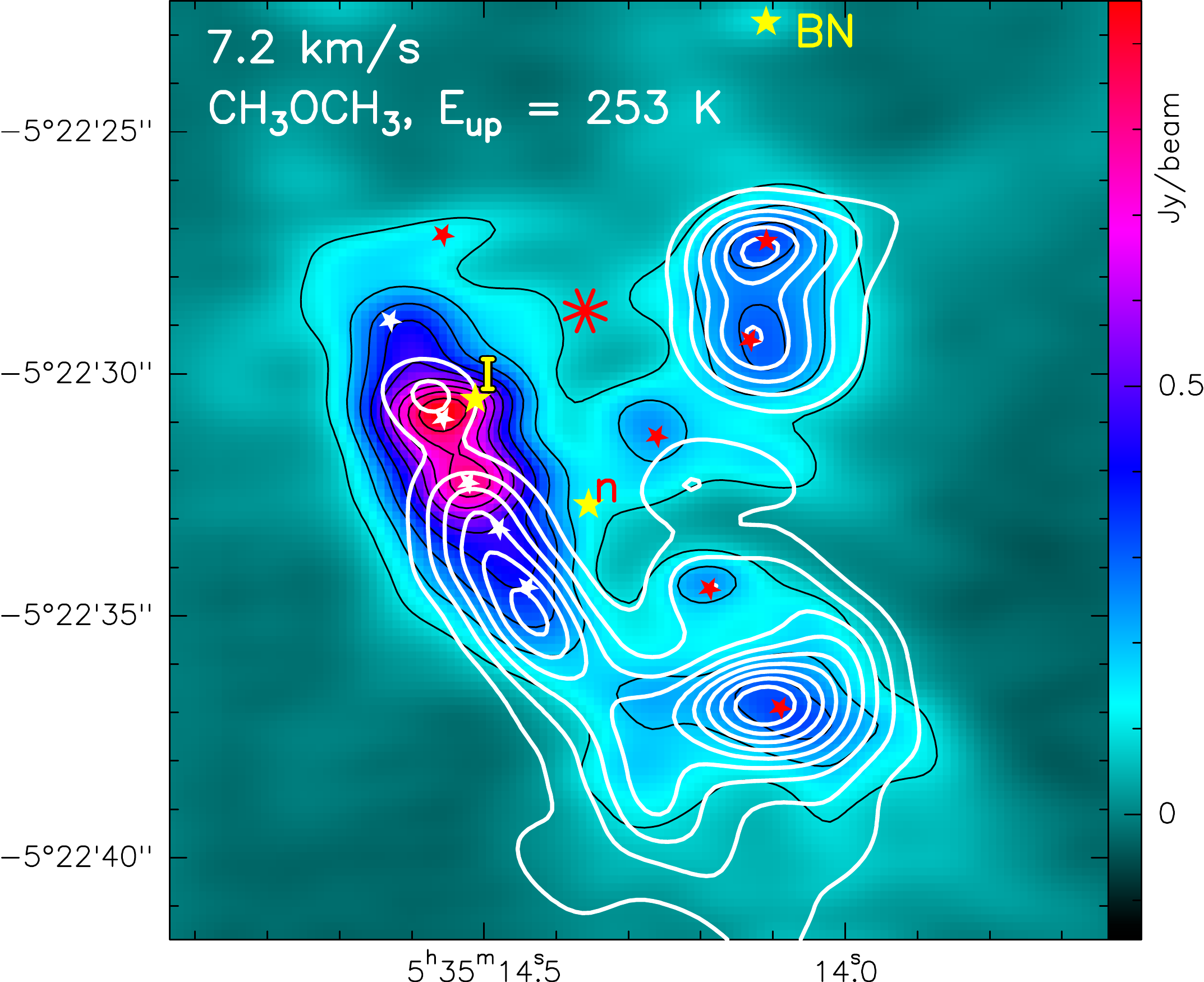}
 
  \includegraphics[scale = 0.254, trim=0 0 48 0,clip]{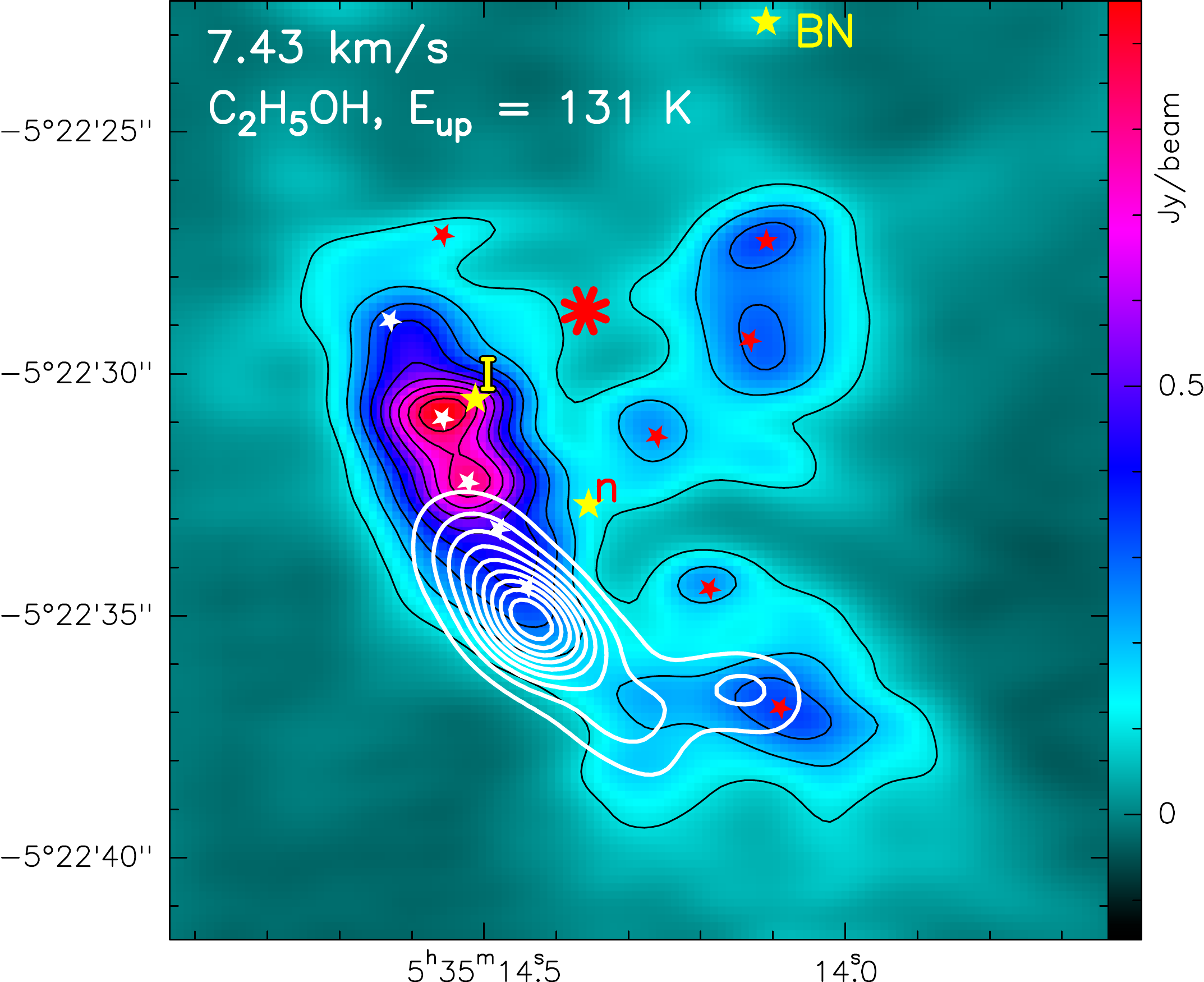}
  \includegraphics[scale = 0.254, trim=85 0 48 0,clip]{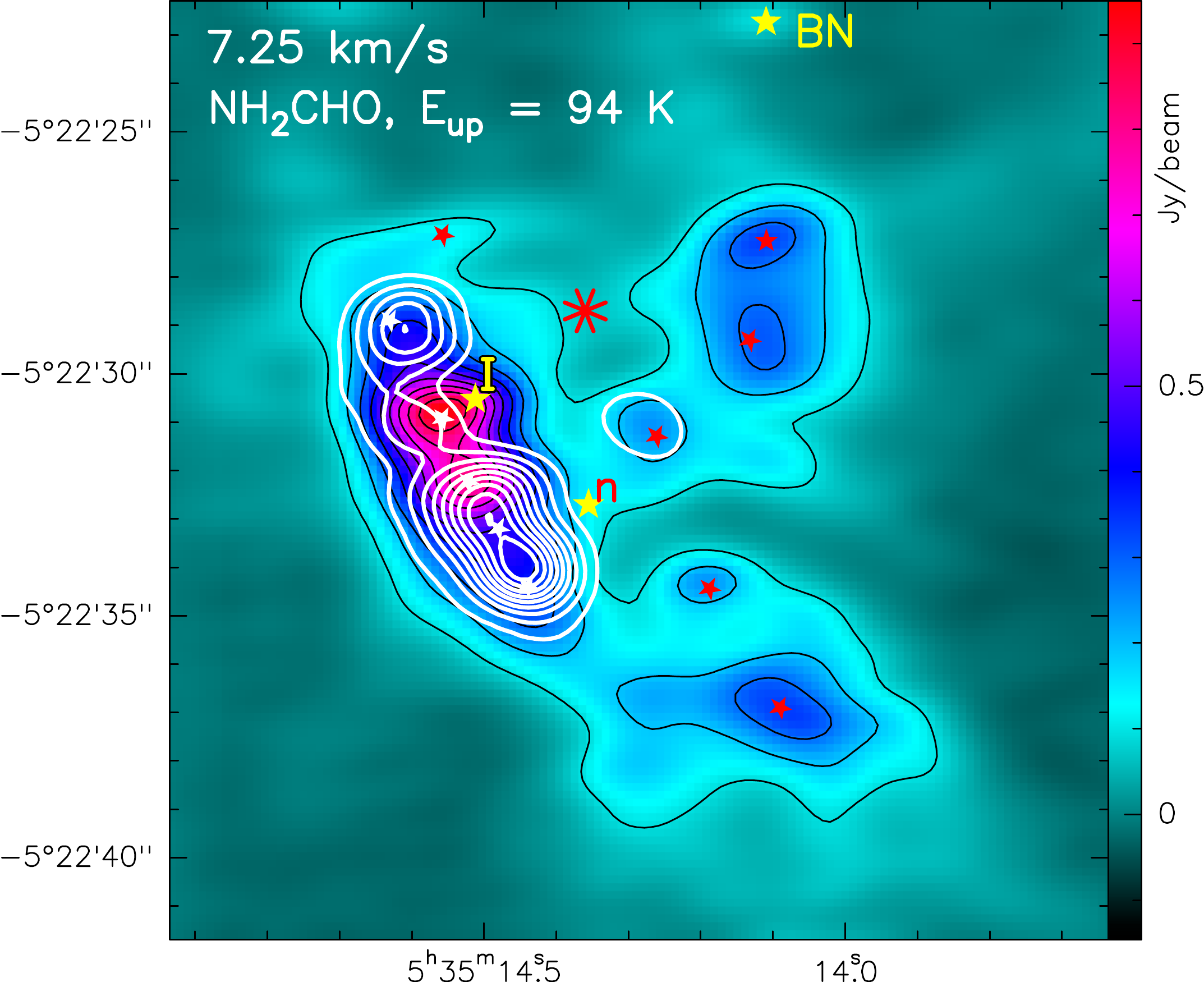}
   \includegraphics[scale = 0.254, trim=85 0 48 0,clip]{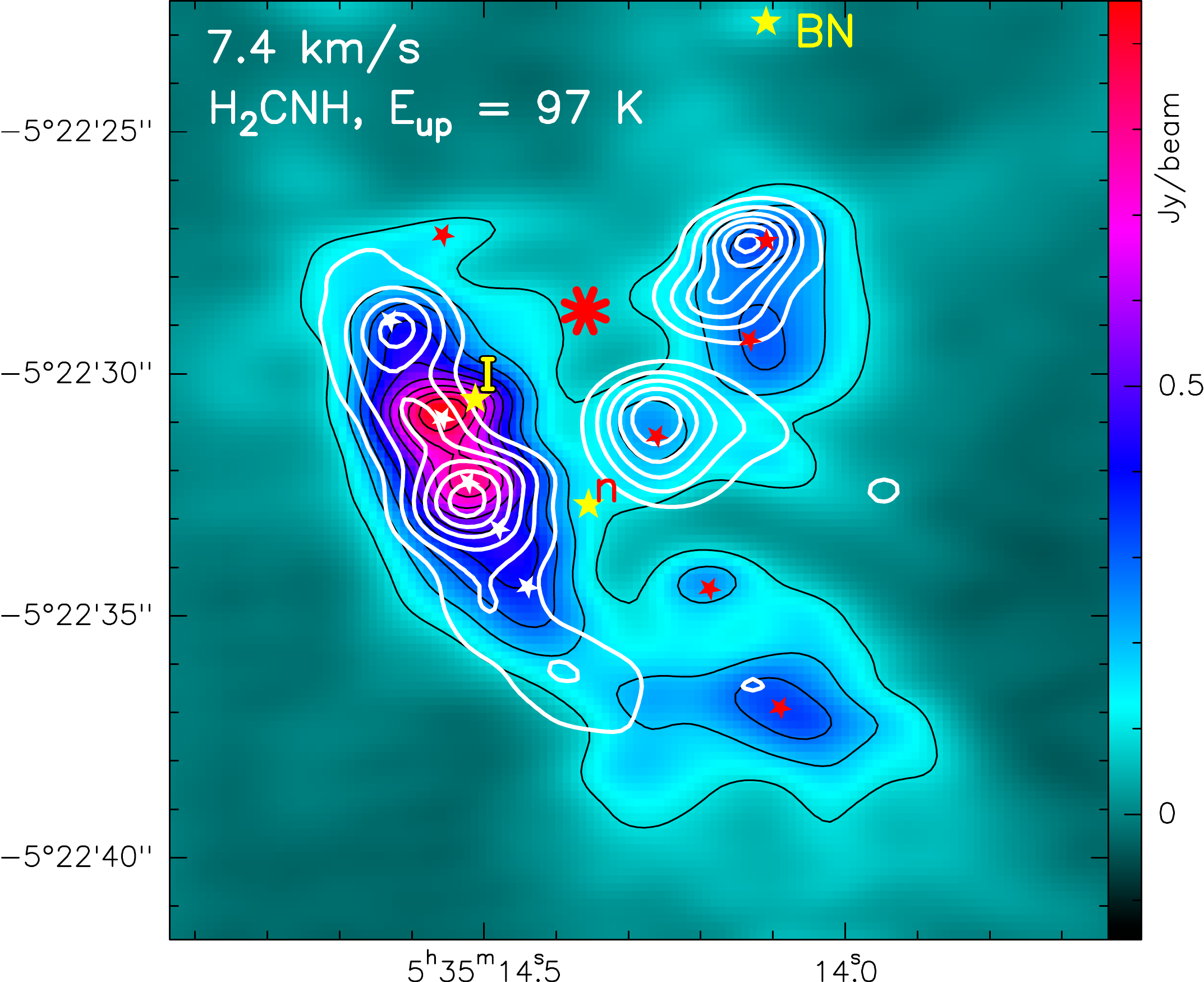}
   \includegraphics[scale = 0.254, trim=85 0 48 0,clip]{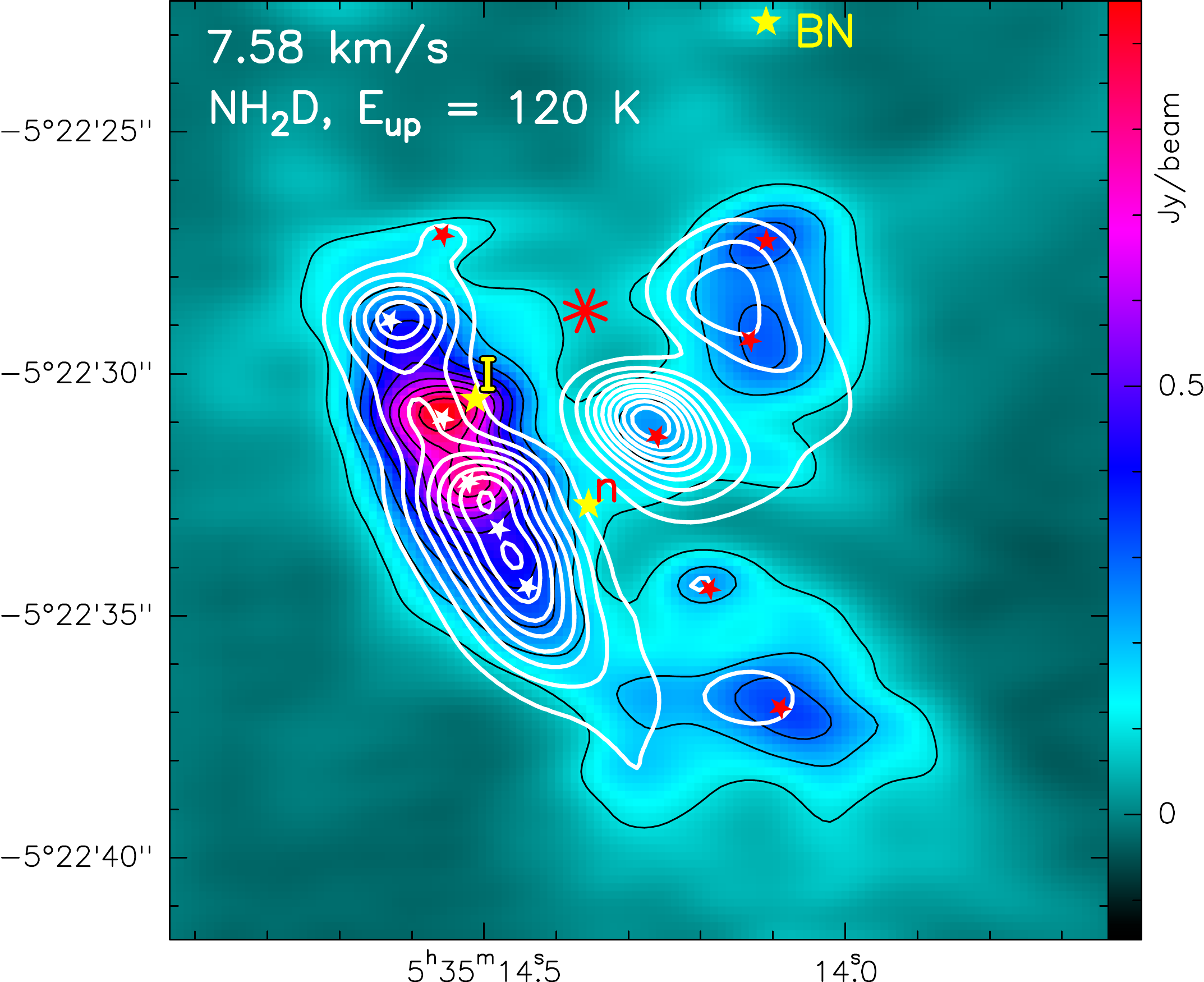}
   
  \caption{Selection of  species at 3 different velocities (here $\sim$7.5\,\kmps). The three top rows present species with
  extensions, the two bottom rows, without. NO is partly masked (where the emission is blended). }
  
  \label{fig:7kmps}
 \end{figure*}

 \begin{figure*}[h!]
 \centering 
 \includegraphics[scale = 0.254,trim=0 20 49 0,clip]{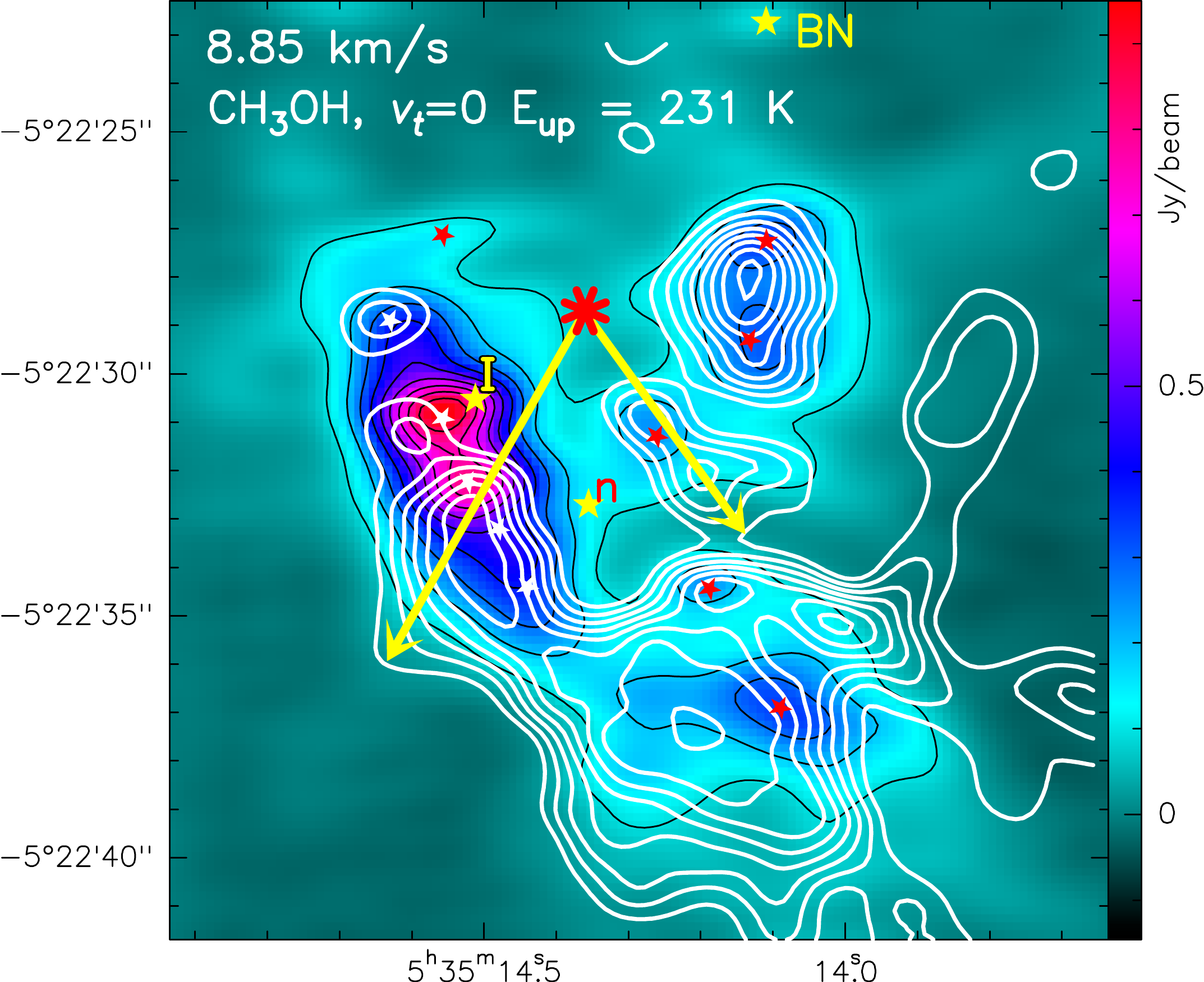}
 \includegraphics[scale = 0.254, trim=85 20 48 0,clip]{BNKL_H2CO_cont+plan_20.pdf}
 \includegraphics[scale = 0.254, trim=85 20 49 0,clip]{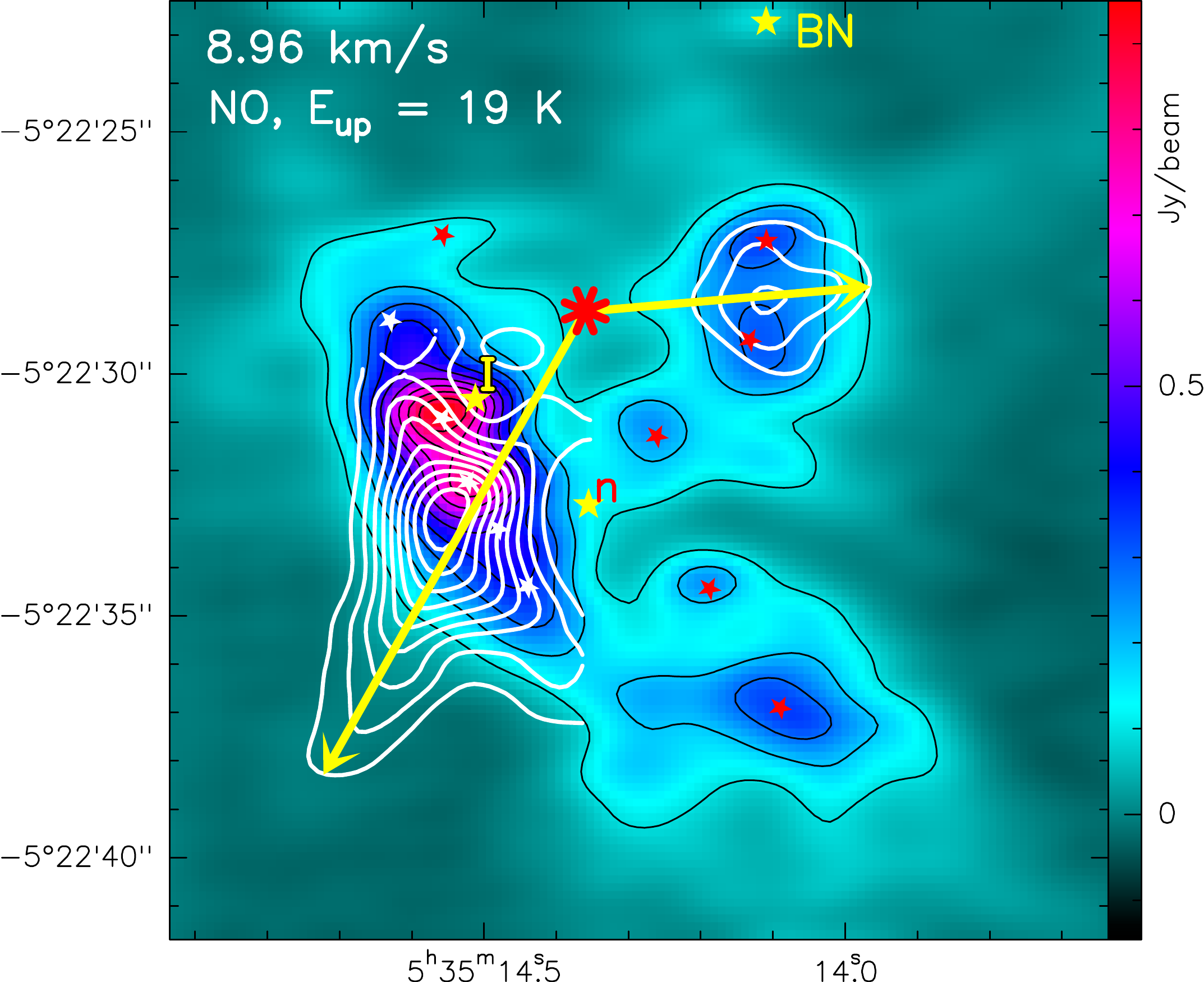}
   \includegraphics[scale = 0.254, trim=85 20 48 0,clip]{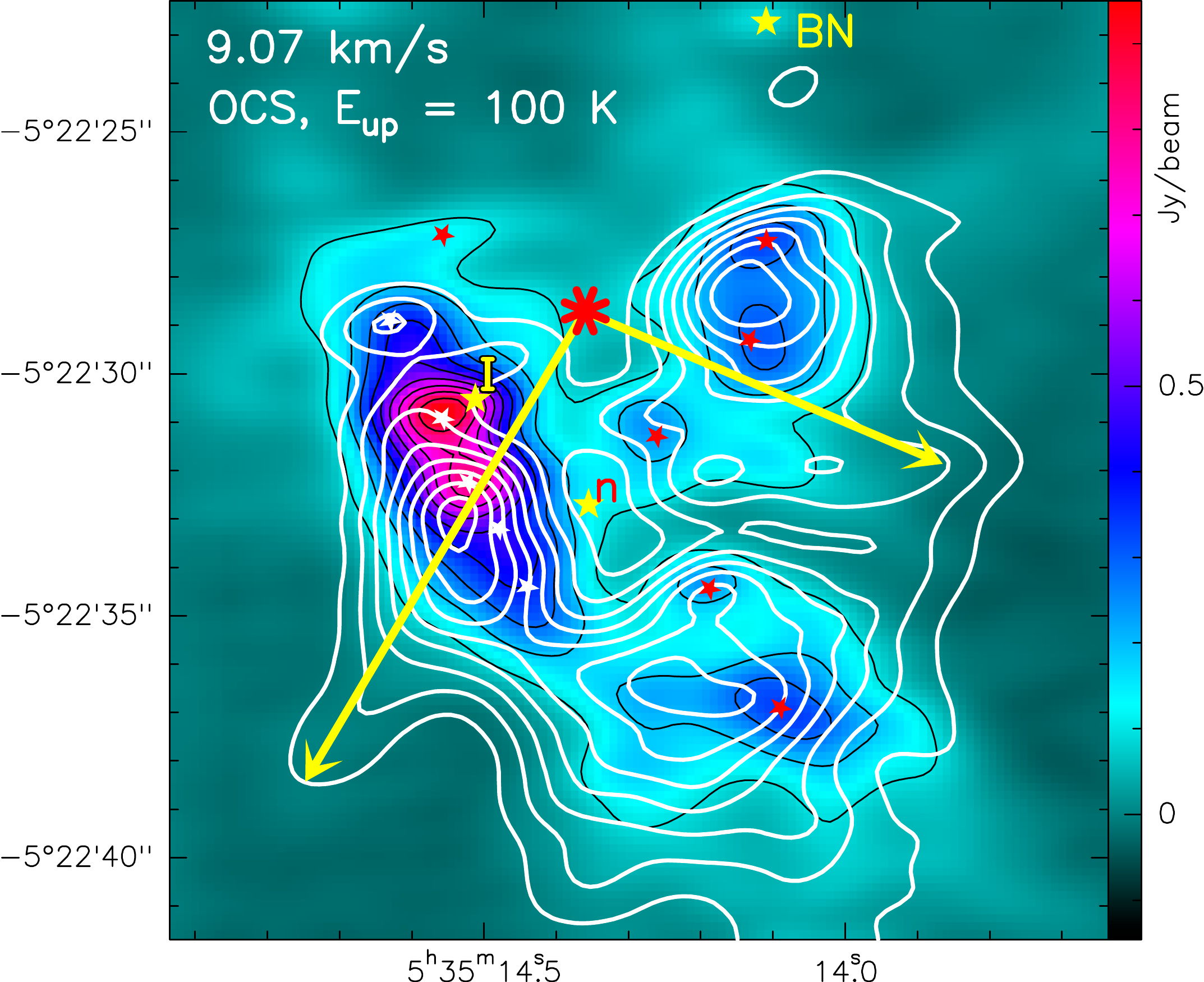}
  
    \includegraphics[scale = 0.254, trim=0 20 48 0,clip]{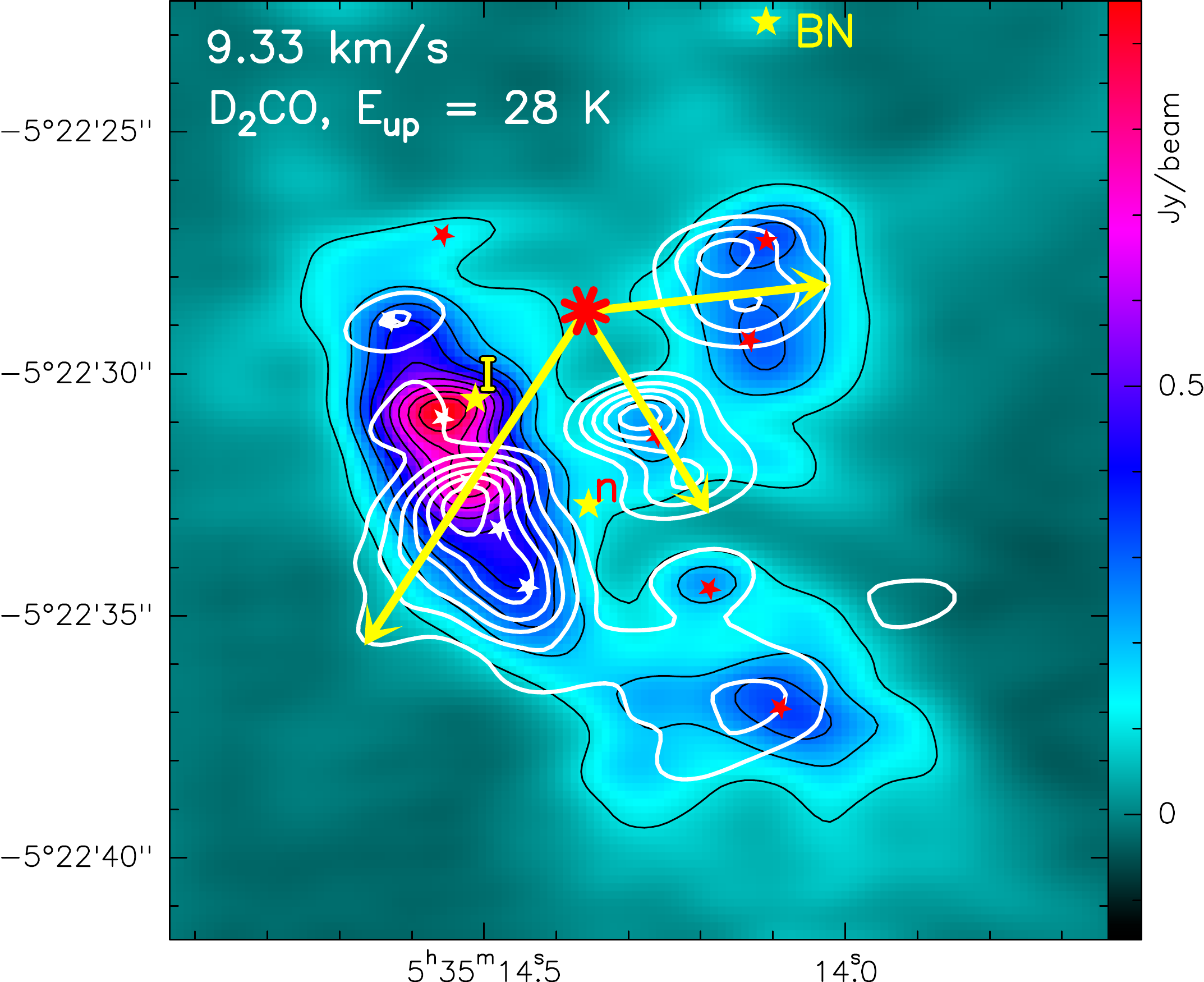}
   \includegraphics[scale = 0.254, trim=85 20 49 0,clip]{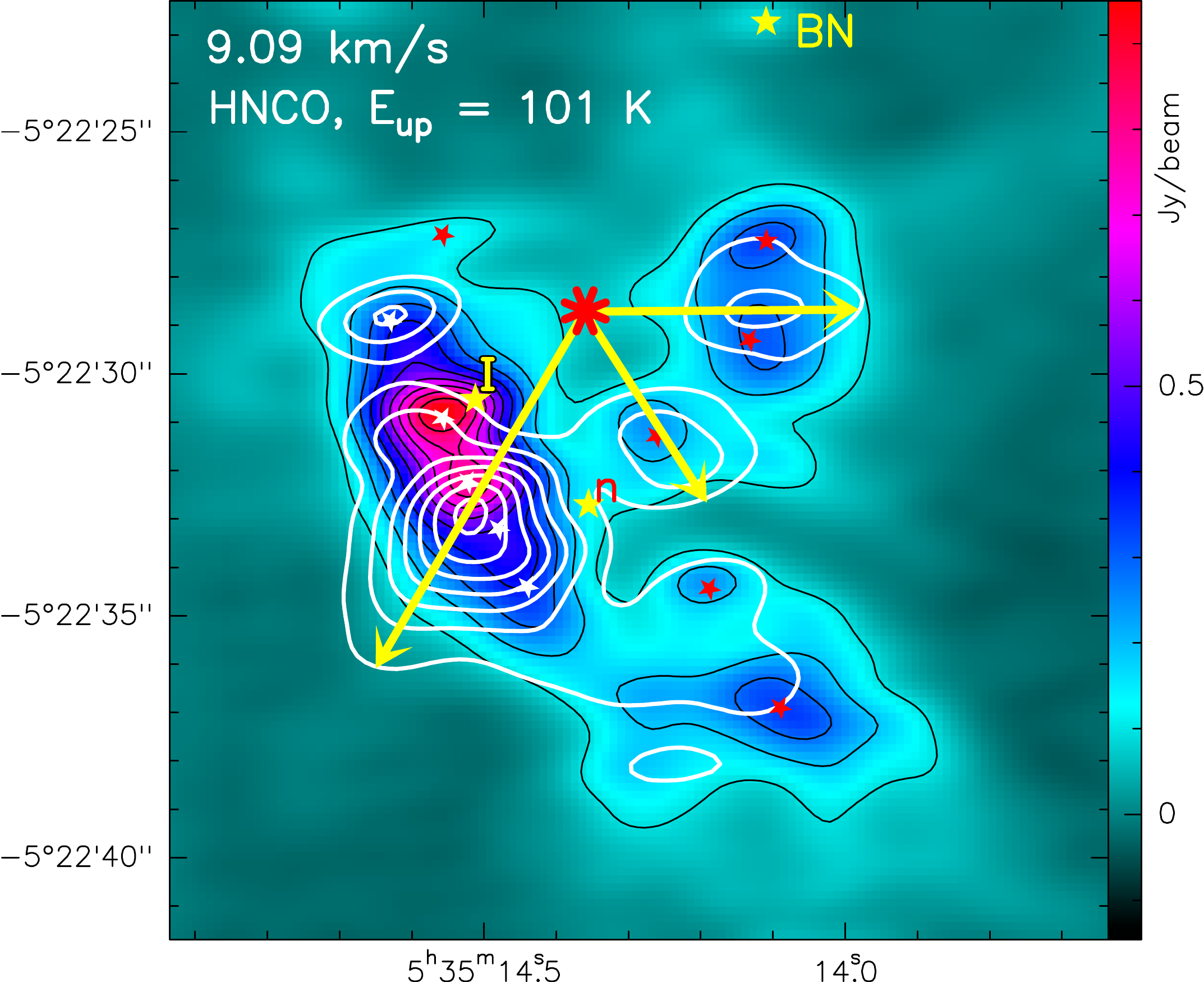}
 \includegraphics[scale = 0.254, trim=85 20 48 0,clip]{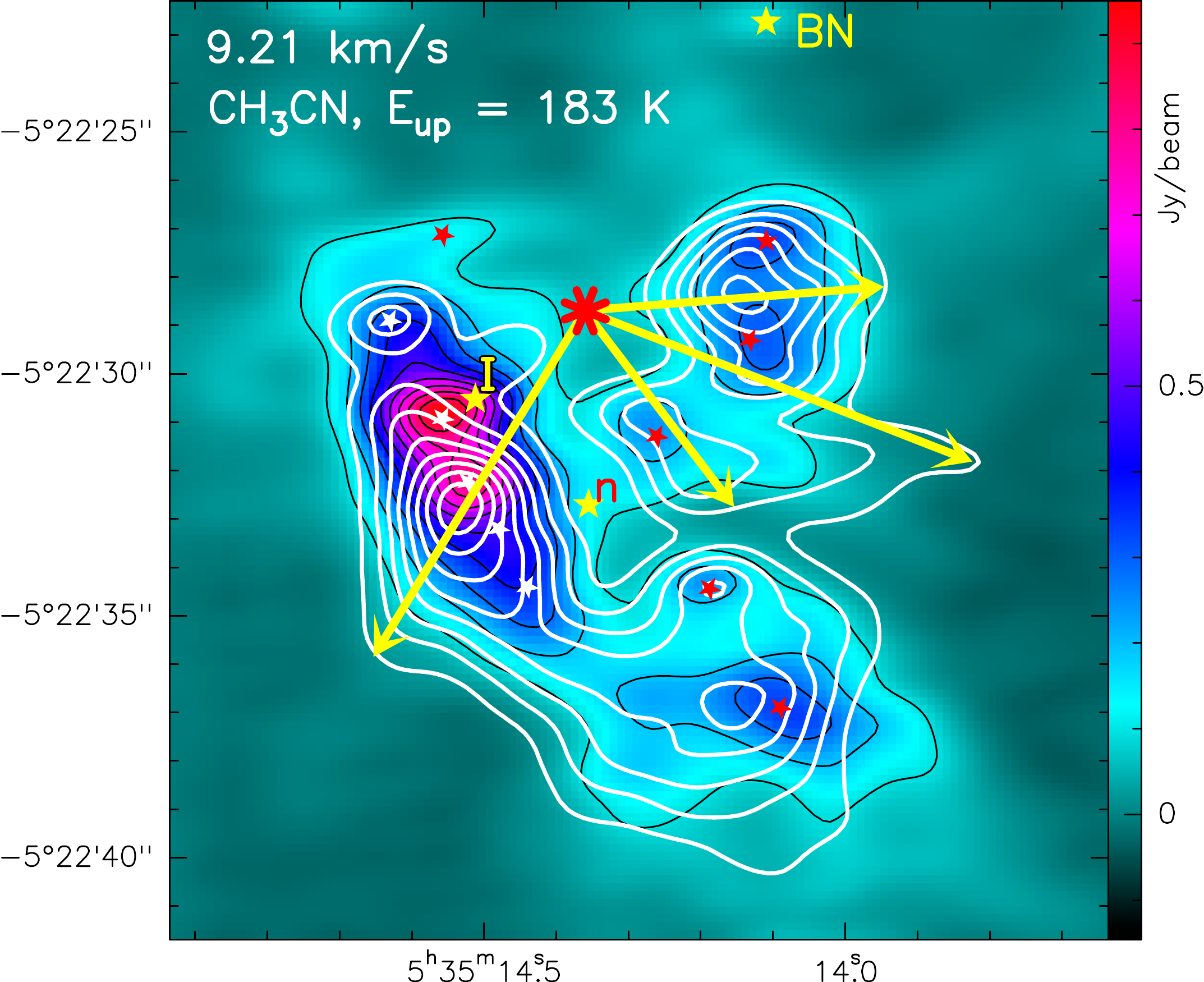}
 \includegraphics[scale = 0.254, trim=85 20 48 0,clip]{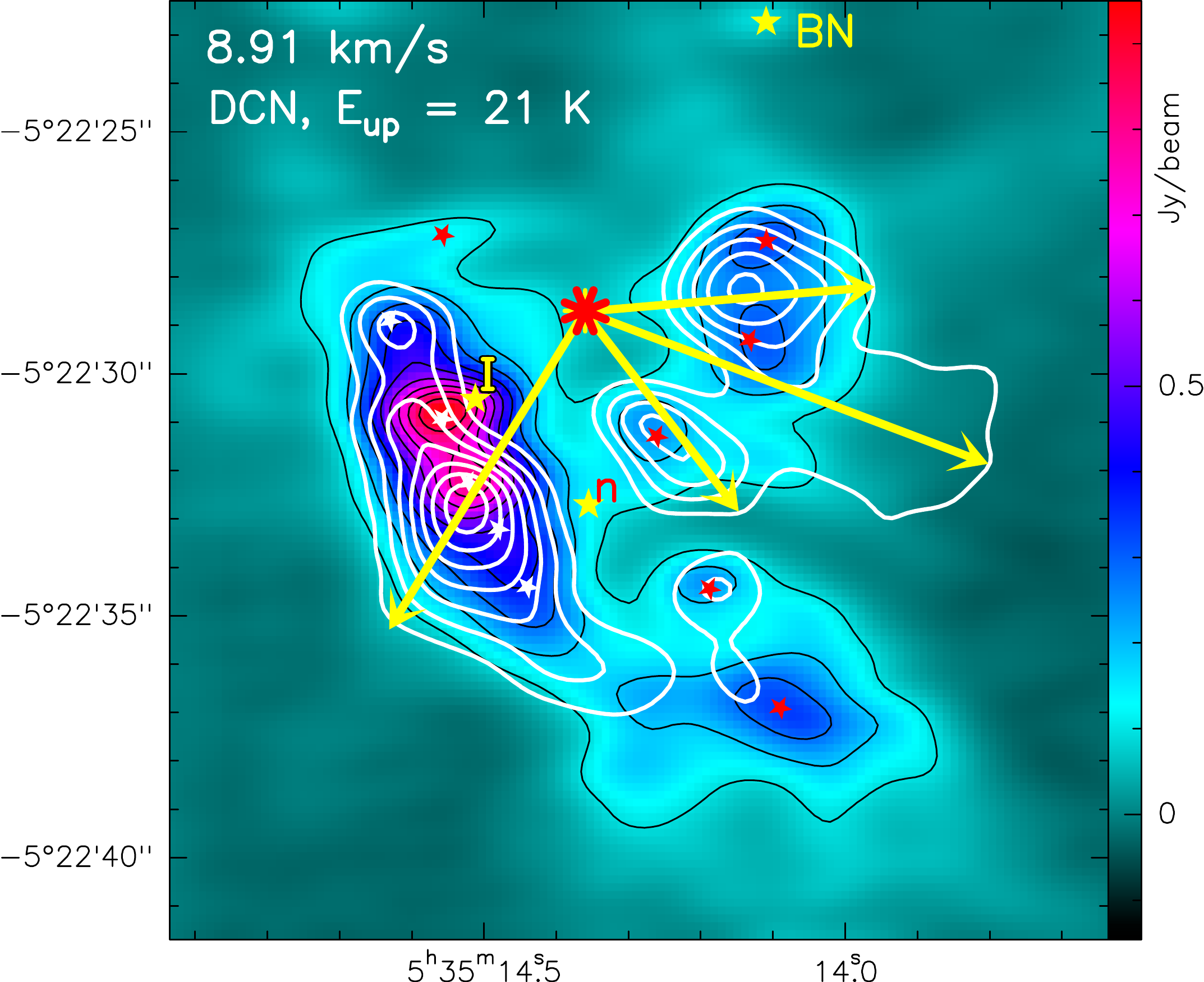}
 
 \includegraphics[scale = 0.254, trim=0 -10 48 0,clip]{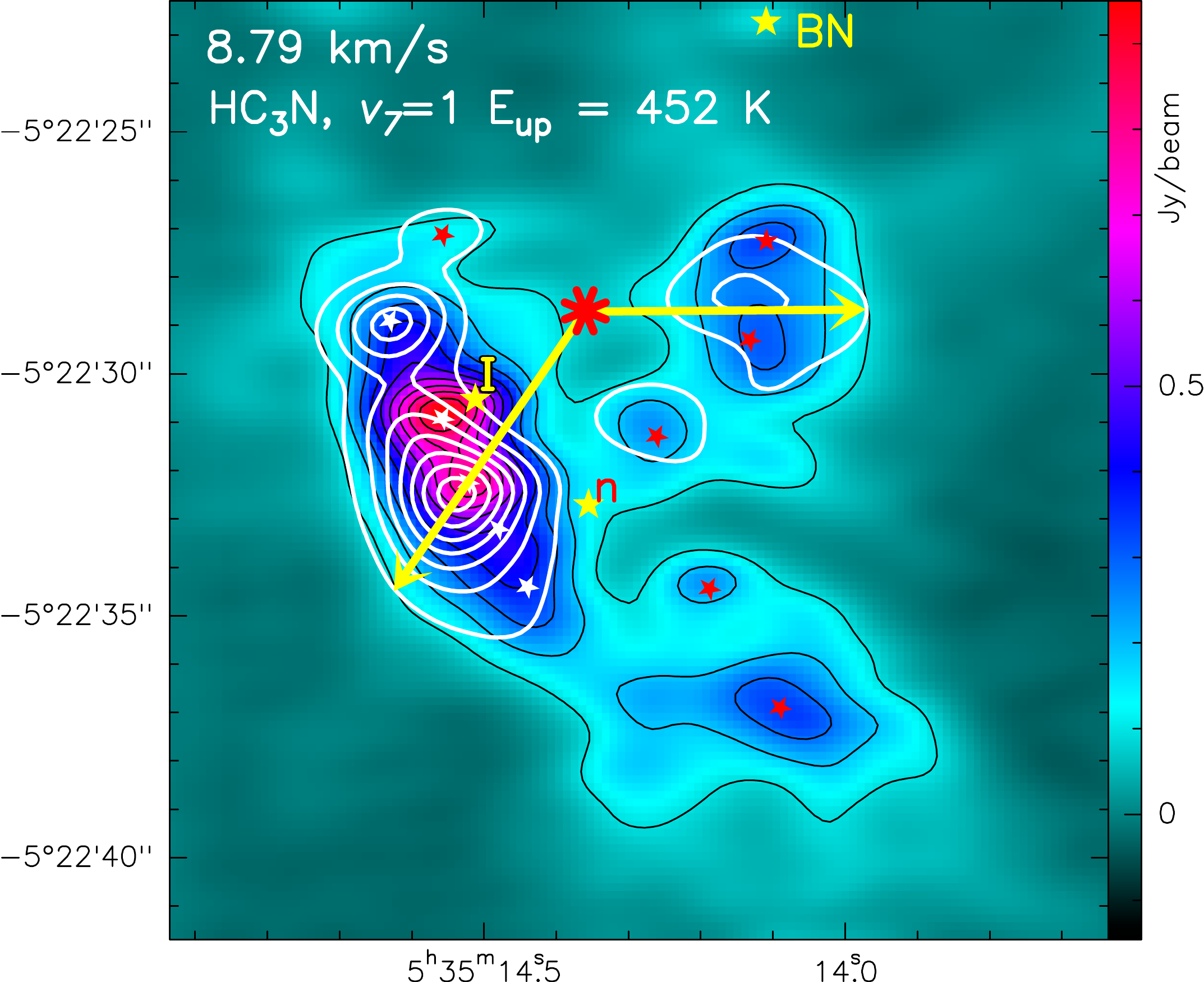}
 \includegraphics[scale = 0.254, trim=85 -10 48 0,clip]{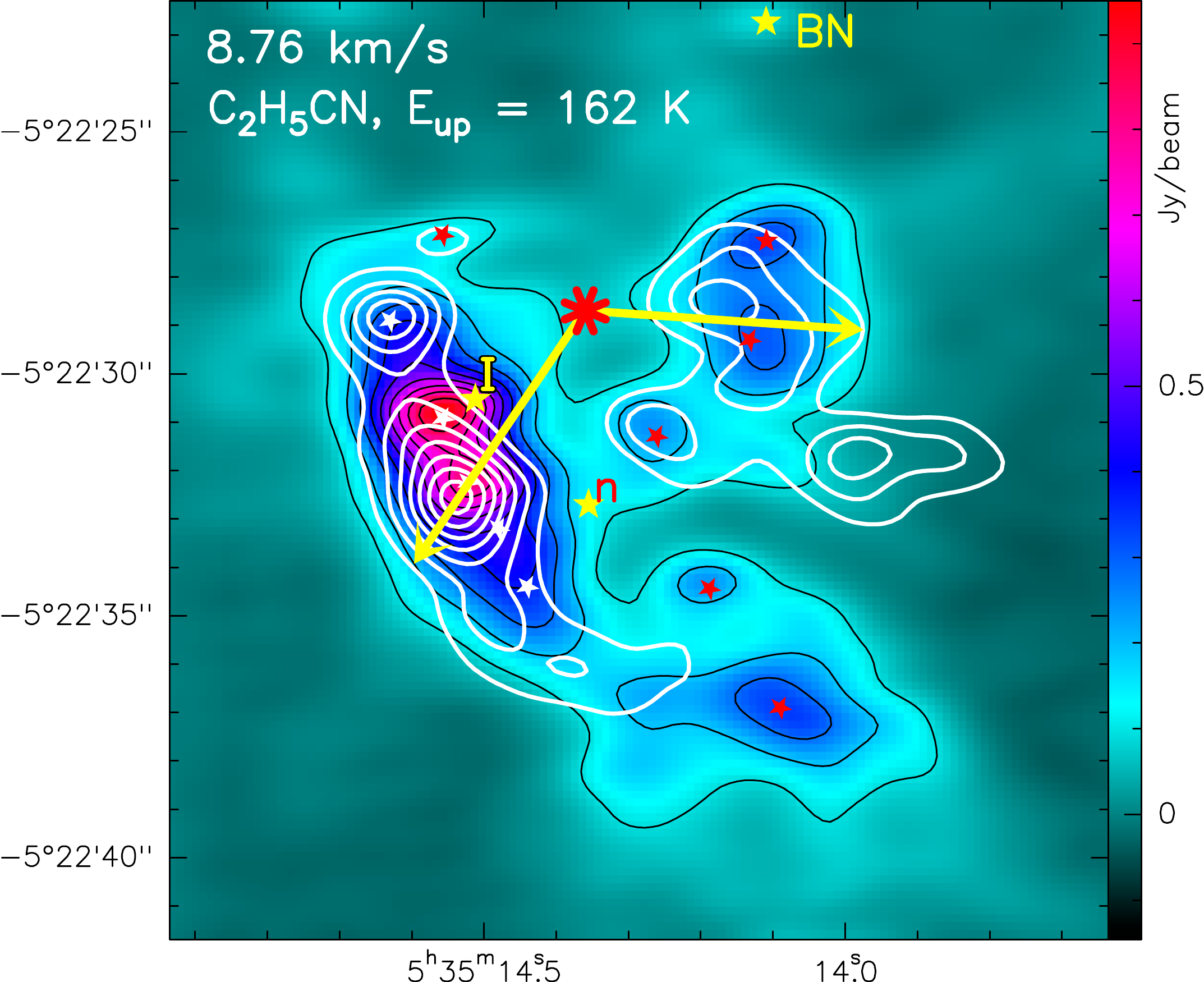}
 \includegraphics[scale = 0.254, trim=85 -10 49 0,clip]{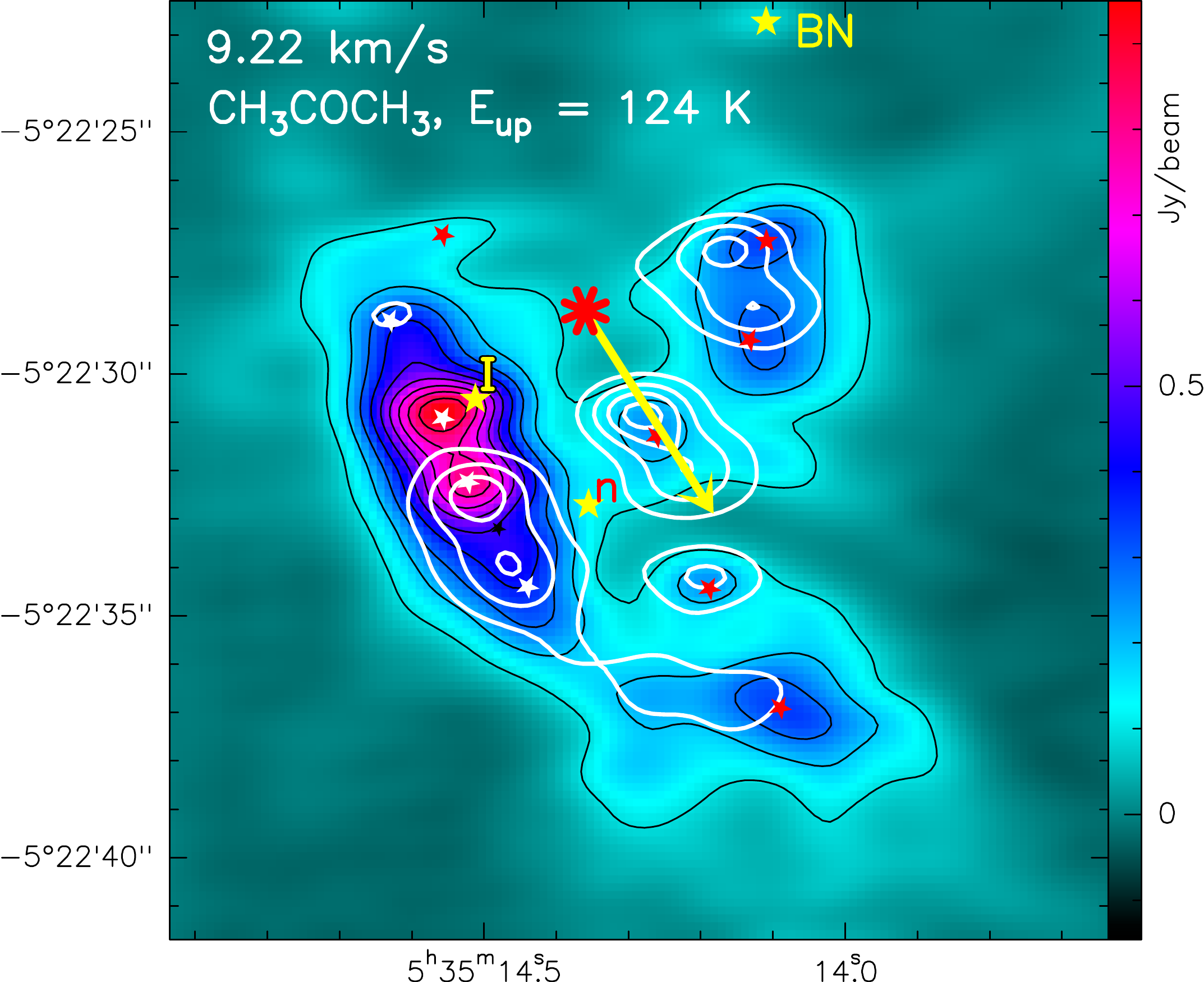}
 \includegraphics[scale = 0.254, trim=85 -10 48 0,clip]{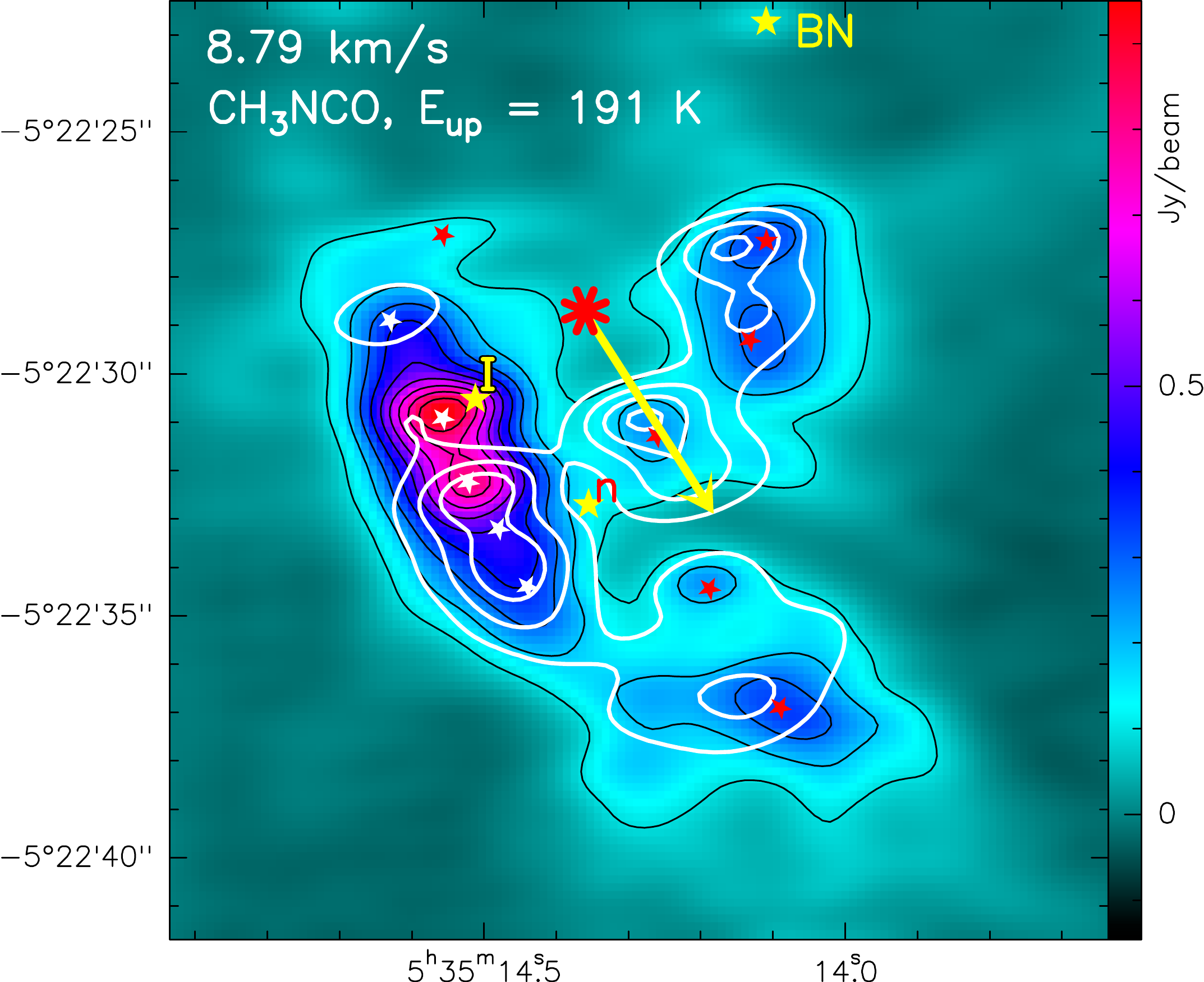}
 
   \includegraphics[scale = 0.254, trim=0 20 48 0,clip]{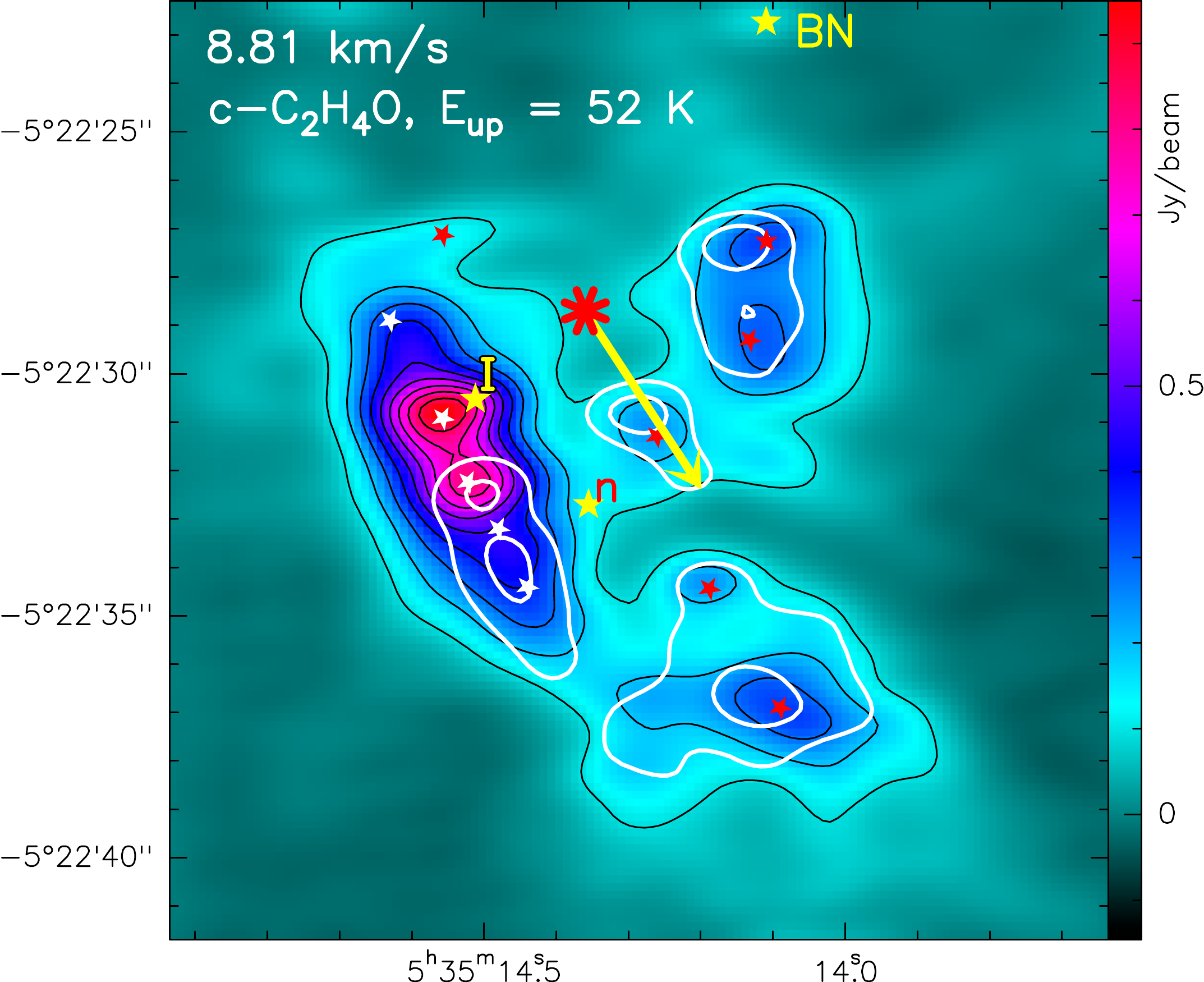}
\includegraphics[scale = 0.254, trim=85 20 48 0,clip]{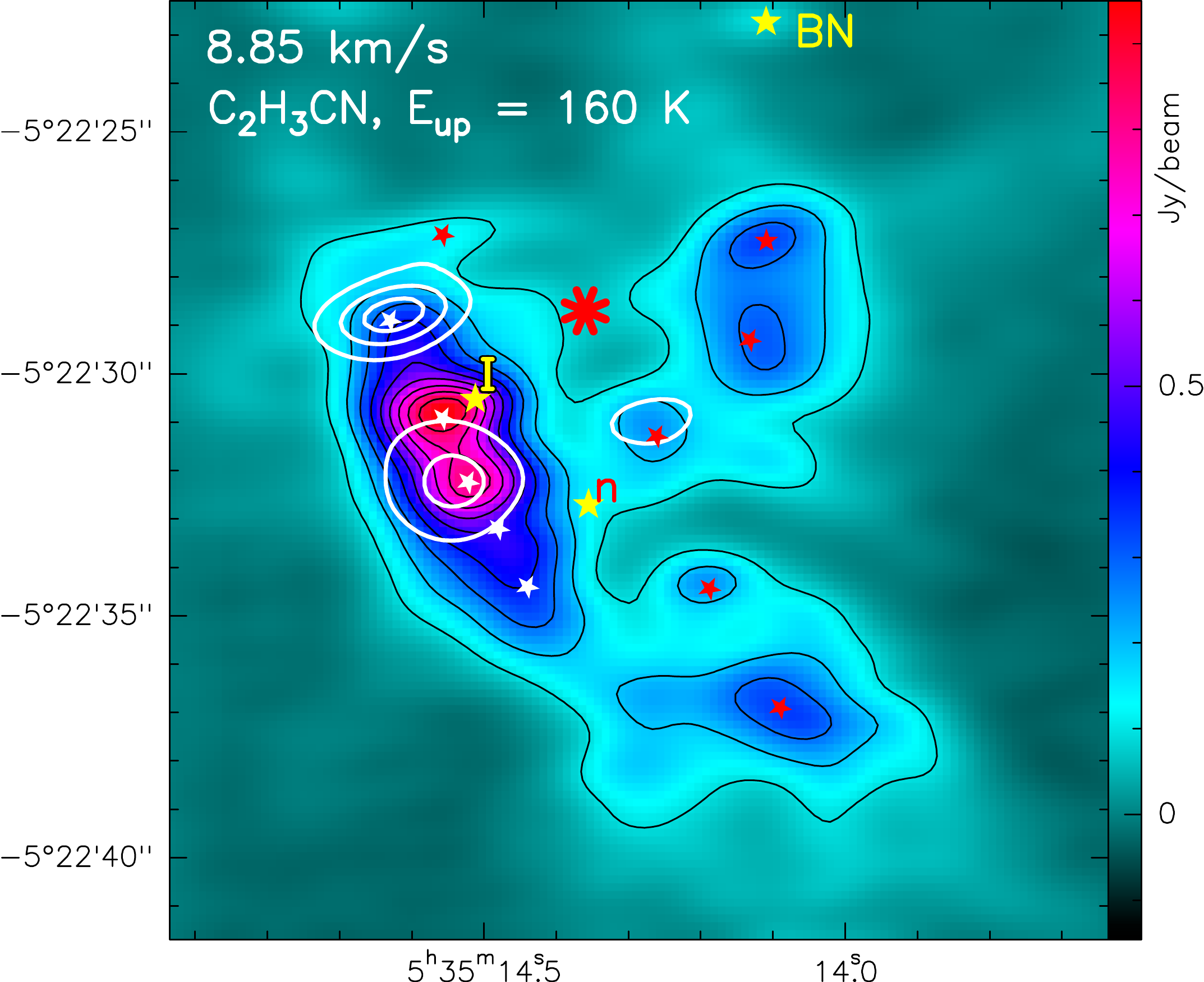}
   \includegraphics[scale = 0.254, trim=85 20 48 0,clip]{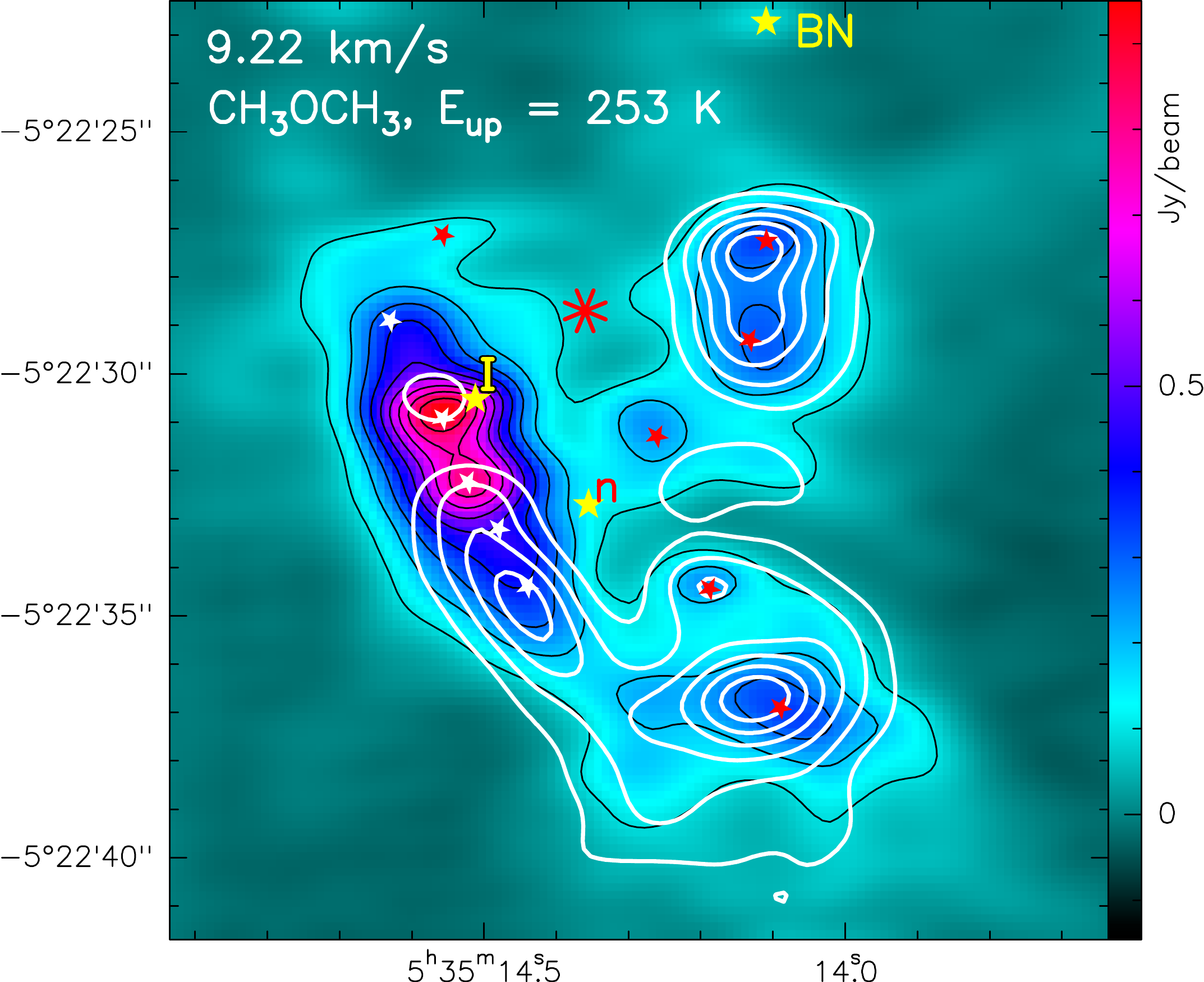}
 \includegraphics[scale = 0.254, trim=85 20 48 0,clip]{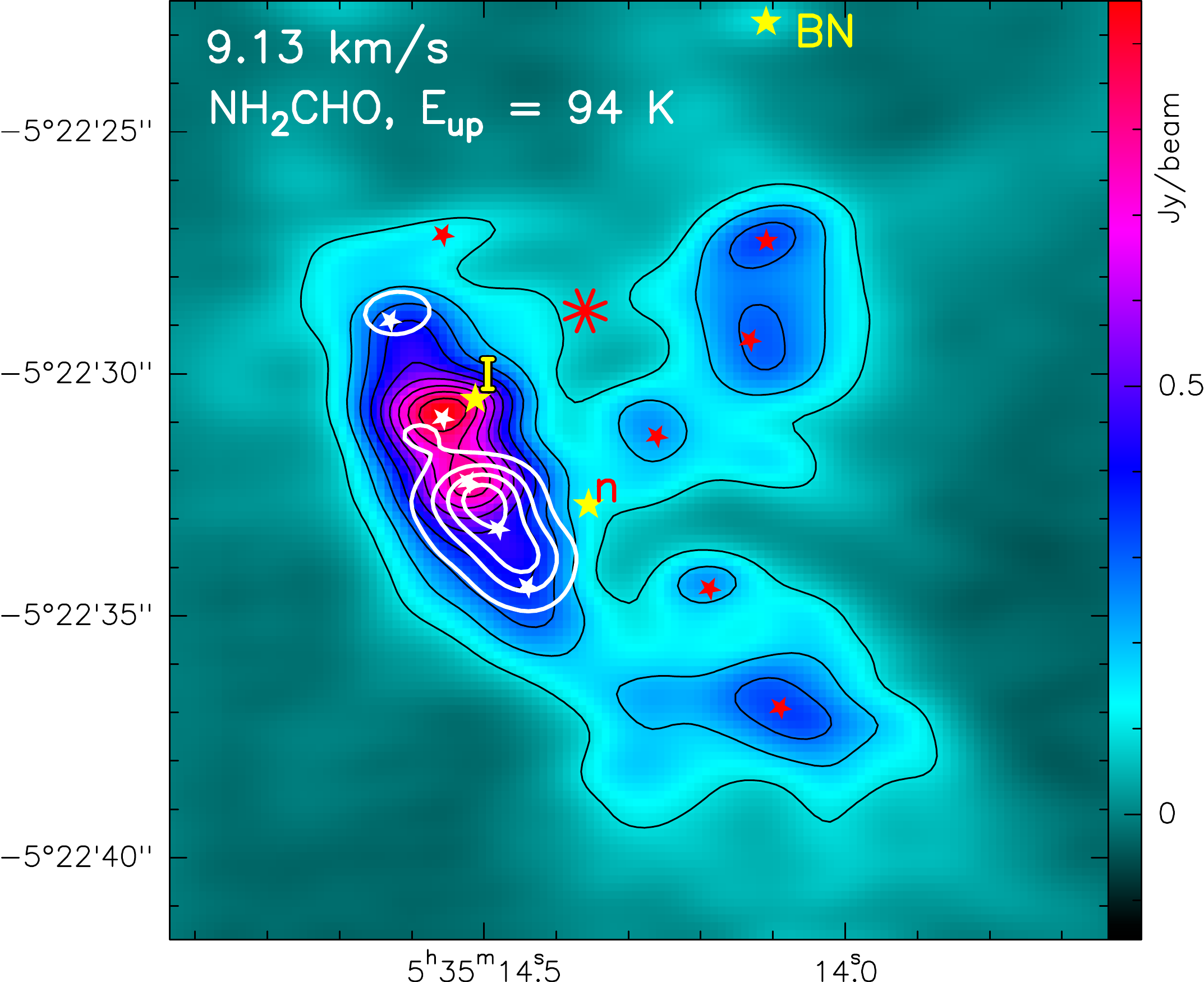}
   
  \includegraphics[scale = 0.254, trim=0 0 48 0,clip]{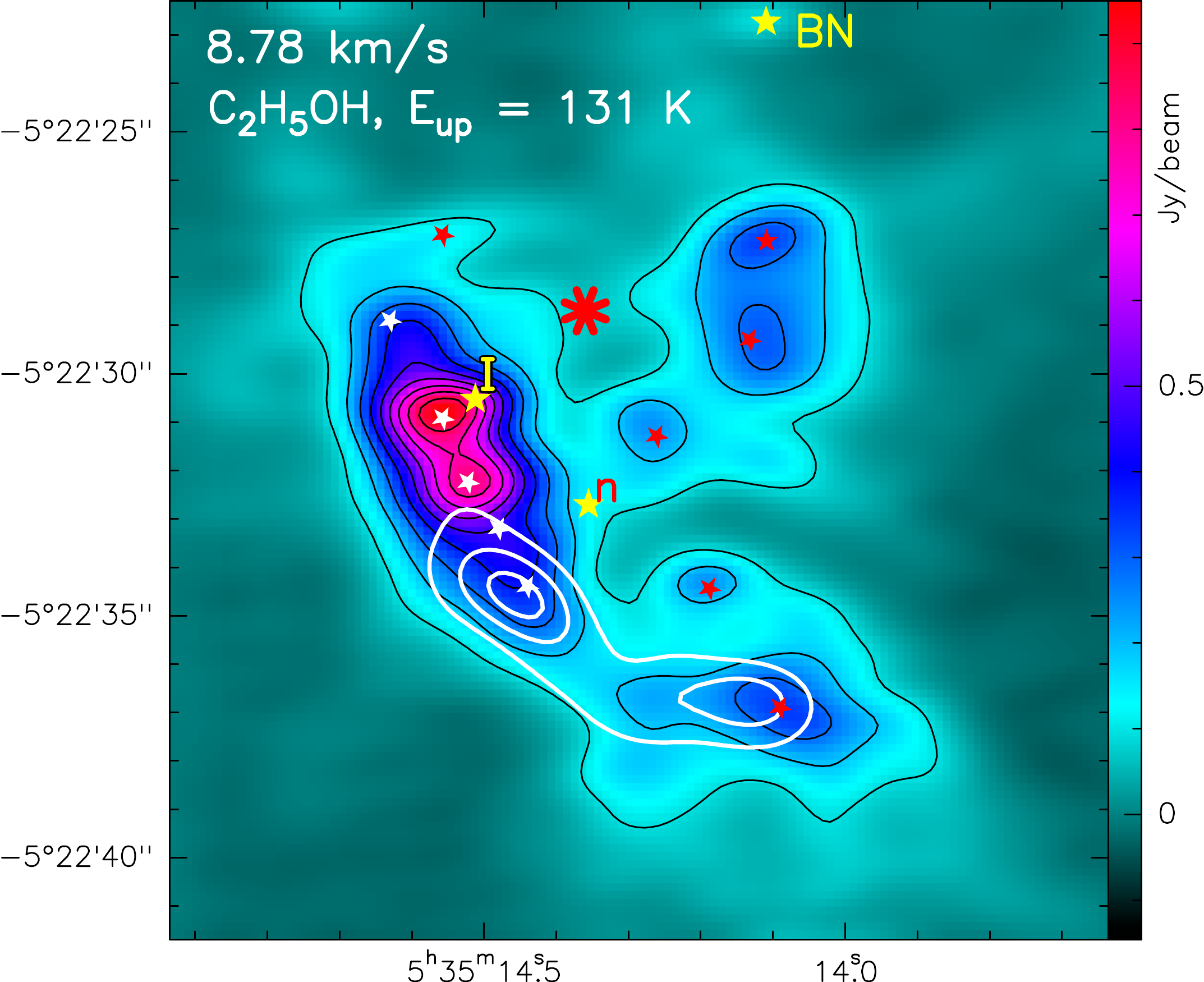}
 \includegraphics[scale = 0.254, trim=85 0 49 0,clip]{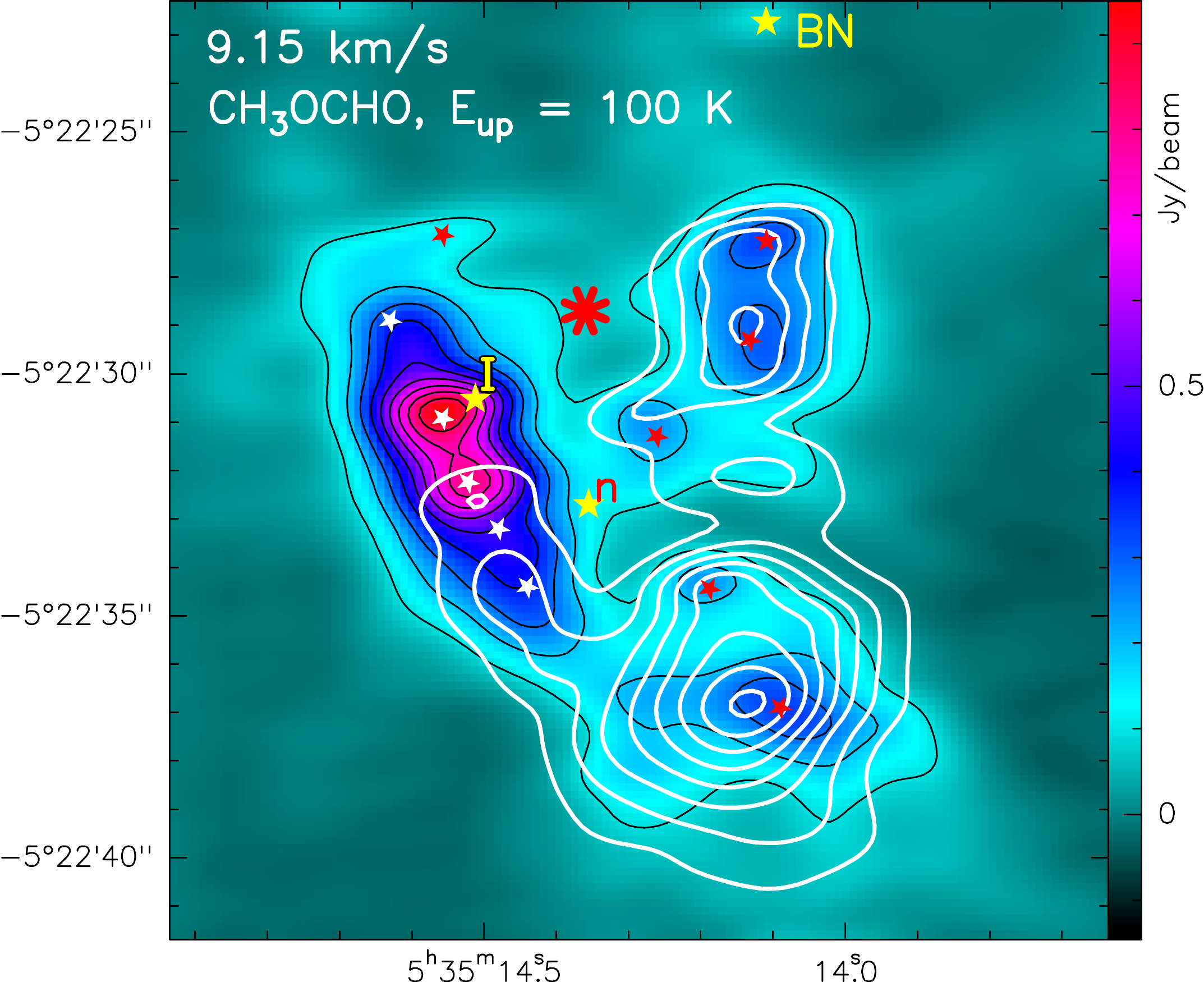}
 \includegraphics[scale = 0.254, trim=85 0 48 0,clip]{{BNKL_H2CNH_cont+plan_39}.pdf}
  \includegraphics[scale = 0.254, trim=85 0 48 0,clip]{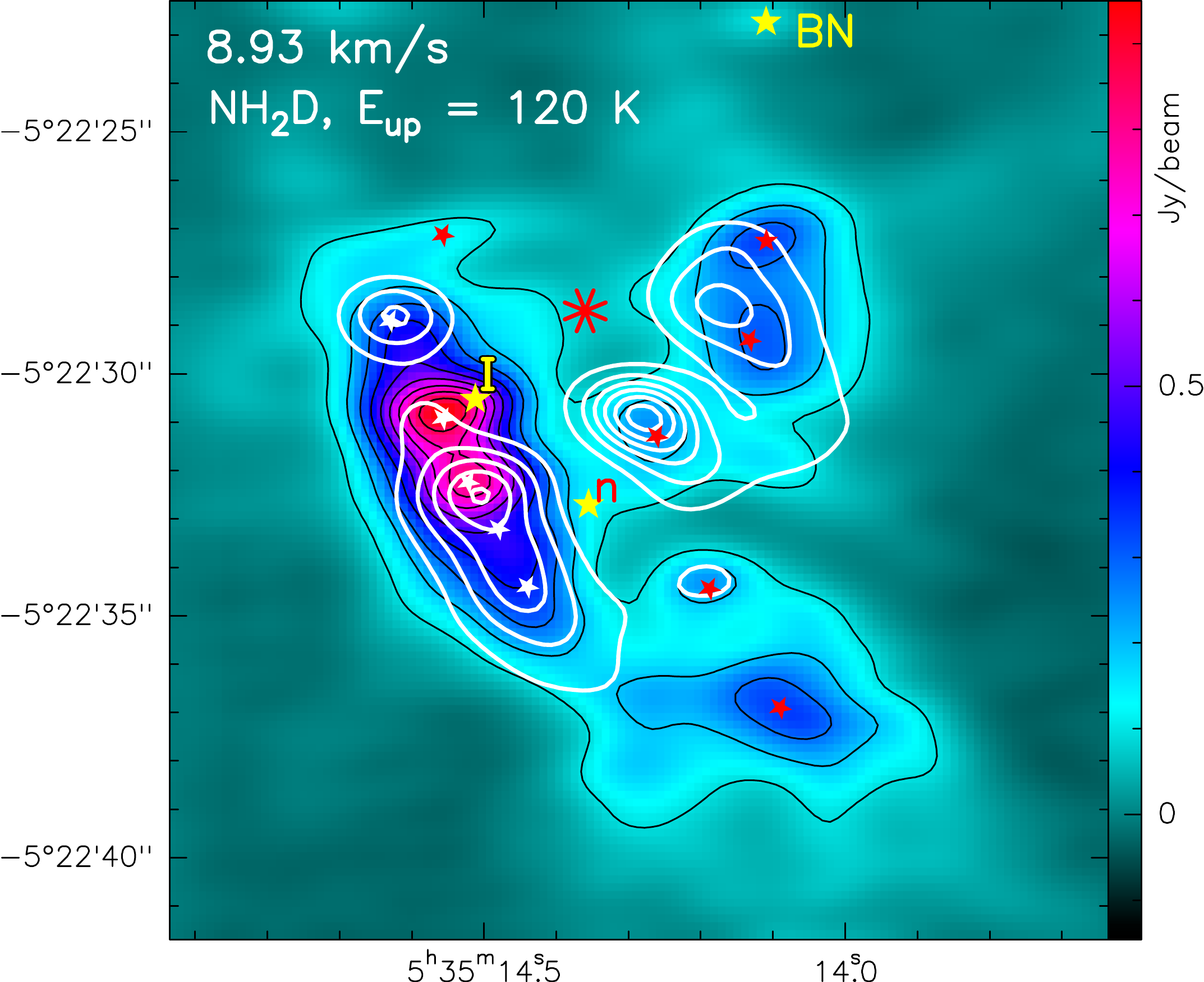}
   
   \caption{Selection of  species at 3 different velocities (here $\sim$9 \kmps). The three top rows present species with
     expansions, the two bottom rows, without (except possibly c-C$_2$H$_4$O). NO is partly masked (where the emission
     is blended).}
   
  \label{fig:9.0kmps}
 \end{figure*}
 
\end{document}